\documentclass[11pt]{article}
\usepackage{cite}
\usepackage{amsmath,amsfonts,amssymb}
\usepackage[small,bf,hang]{caption}
\usepackage{slashed}

\def\hybrid{
        \topmargin -20pt
        \oddsidemargin 0pt
        \headheight 0pt \headsep 0pt
        \textwidth 6.25in 
        \textheight 9.5in 
        \marginparwidth .875in
        \parskip 5pt plus 1pt \jot = 1.5ex}

\hybrid

 \csname
@addtoreset\endcsname{equation}{section}

\newcommand{\half} {{\tfrac{1}{2}}}

\newcommand{\nin}[1] {\underline{\phantom{h}}\hskip-6pt {#1}}

\def\moth{\mathsurround=0pt}
\newdimen\zo \zo=0pt

\def\tick{\leaders\hrule height 0.5ex depth 0pt \hskip 0.5pt}
\def\upboxfill{$\moth \setbox\zo\hbox{\tick}%
  \hskip 3pt\hbox to 0pt{$\tick$\hss}\hrulefill \hbox to 7.5pt{$\tick$\hss}$}

\def\dtick{\leaders\hrule height .34pt depth 0.5ex \hskip 0.5pt}
\def\downboxfill{$\moth \setbox\zo\hbox{\dtick}%
  \hskip 2pt\hbox to 0pt{$\dtick$\hss}\hrulefill \hbox to 2pt{$\dtick$\hss}$}

\def\ll{{\langle \hskip-1pt \langle}}
\def\rr{{\rangle \hskip-1pt \rangle}}

\def\bec{\begin{center}}
\def\ec{\end{center}}

\def\be{\begin{equation}}
\def\ee{\end{equation}}
\def\bea{\begin{eqnarray}}
\def\eea{\end{eqnarray}}
\def\ba{\begin{array}}
\def\ea{\end{array}}

\begin{document}

\begin{titlepage}
\rightline{January 2017} 
\rightline{\tt MIT-CTP-4875} 
\begin{center}
\vskip 2.5cm

{\Large \bf {$L_{\infty}$ Algebras and Field Theory }}   

 \vskip 0.5cm

  \vskip 2.0cm
 {\large {Olaf Hohm$^1$ and Barton Zwiebach$^2$}}
 \vskip 1cm

{\em $^1$ \hskip -.1truecm Simons Center for Geometry and Physics, \\
Stony Brook University, \\
Stony Brook, NY 11794-3636, USA \vskip 5pt }

{\em $^2$ \hskip -.1truecm Center for Theoretical Physics, \\
Massachusetts Institute of Technology\\
Cambridge, MA 02139, USA \vskip 5pt }

ohohm@scgp.stonybrook.edu, zwiebach@mit.edu\\

\vskip 2.5cm
{\bf Abstract}

\end{center}

\vskip 0.5cm

\noindent
\begin{narrower}

\baselineskip15pt

We review and develop the general properties of $L_\infty$ algebras
focusing on the gauge structure of the associated field theories.  
Motivated by the $L_\infty$ homotopy Lie algebra of closed string field theory
and the work of Roytenberg and Weinstein describing the Courant
bracket in this language we investigate the $L_\infty$ 
structure of general gauge invariant perturbative field theories. 
We sketch such formulations for non-abelian gauge theories,
Einstein gravity,  
and for double field theory.  We find that there is an $L_\infty$ algebra
for the gauge structure and a larger one for the full interacting field
theory.  Theories where the gauge structure is a strict Lie algebra 
often require  the full $L_\infty$ algebra for the interacting theory. 
The analysis suggests that $L_\infty$ algebras 
provide a classification of perturbative 
gauge invariant classical field theories.

\end{narrower}

\end{titlepage}

\tableofcontents

\newpage

\section{Introduction and summary}      

In bosonic open string field theory \cite{Witten:1985cc} the interaction of strings
is defined by a multiplication rule, a star product of string fields that 
happens to be associative.  While this  
formulation is advantageous for 
finding classical solutions of the theory, associativity is not strictly necessary
for the formulation of the string field theory; 
{\em homotopy} associativity is.  This more intricate structure has been
investigated in~\cite{Gaberdiel:1997ia} and has appeared when one incorporates
closed strings explicitly in the open string 
theory~\cite{Zwiebach:1997fe,Zwiebach:1990qj,Kajiura:2004xu}.
The homotopy associative $A_\infty$ algebra of Stasheff~\cite{stasheff} is the
mathematical structure underlying the general versions of the  
classical open string field theory sector.

The product of closed string fields is analogous to 
Lie brackets in that they
are graded commutative.  But a strict Lie algebra does not appear to allow 
for the formulation of closed string field theory.  Instead one requires a collection
of higher products satisfying generalized versions of Jacobi identities.
The classical string field theory is thus organized by a homotopy Lie algebra,
an $L_\infty$ algebra whose axioms and Jacobi-like identities were given explicitly in
\cite{Zwiebach:1992ie}.  The axioms and identities were later   
 given in different but equivalent
conventions in~\cite{Lada:1992wc}. These two formulations are related by a ``suspension", 
an operation where the degree of all vector spaces 
is shifted by one unit. 
An earlier mathematical motivation for homotopy Lie algebras is found in~\cite{sch-sta}.
The $L_\infty$ algebra describes the structure of classical string field theory: 
the collection of string field products is all one needs to write gauge  
transformations and field equations.  Equipped with a suitable inner product, one can
also write an action.  The interplay of $A_\infty$ and $L_\infty$ algebras feature in the
study of open-closed string field theory~\cite{Zwiebach:1997fe,Kajiura:2004xu,Munster:2011ij}.  $L_\infty$ algebras have recently
featured in a  study of massive
two-dimensional field theory~\cite{Gaiotto:2015aoa}.

The relevance of $L_\infty$ to closed string field theory, which is a field theory
for  an infinite number of component fields, suggests that it should also be relevant to 
arbitrary field theories, and this provided motivation for the present 
study.  In particular, in a recent work Sen has shown how to define consistent
truncations for a set of closed string modes~\cite{Sen:2016qap}.  For these degrees of freedom
one has an effective field theory organized by an $L_\infty$ algebra that can be derived
from the full algebra of the closed string field theory.  This again suggests 
the general relevance of $L_\infty$  to field theories.

A first look into the problem of identifying the $L_\infty$ gauge structure   
of some field theories was given by Barnich {\em et.al.}\cite{Barnich:1997ij}.  
The early investigation of non-linear higher-spin symmetries in 
Berends {\em et.al.} \cite{Berends:1984rq} eventually 
led to an analysis by Fulp {\em et.al.} \cite{Fulp:2002kk} 
of {\em gauge} structures that under some assumptions define 
$L_\infty$ algebras.  More recently, Yang-Mills-type gauge theories 
were fully formulated as $L_\infty$ algebras  by Zeitlin 
using the BRST complex of open string field theory~\cite{Zeitlin:2007vv,Zeitlin:2007fp,Zeitlin:2008cc}.

In an interesting paper,  
Roytenberg and Weinstein~\cite{roytenberg-weinstein} 
systematically analyzed the Courant algebroid in the language of $L_\infty$ algebras.  
The authors explicitly identified the
relevant vector spaces, the products, and proceeded to show that the Jacobi-like
homotopy identities are all satisfied.  As it turns out, in this algebra there is a bracket
$[ \, \cdot \,, \cdot \, ]$ that applied to two gauge parameters coincides with the
Courant bracket, and a triple product $[  \, \cdot \,, \cdot \, , \, \cdot \, ]$ that applied
to three gauge parameters gives a function whose gradient is the Jacobiator.  No higher
products exist in this particular $L_\infty$ algebra.  
The $L_\infty$ formulation of the C-bracket, the duality-covariant extension 
of the Courant bracket,  was considered by Deser and Saemann~\cite{Deser:2016qkw}. The authors used
 `derived brackets'  
 in their construction, making use of the results of 
 \cite{Deser:2014mxa,getzler,roytenberg}.  
 Since the Courant bracket underlies
the gauge structure of double field 
theory \cite{Siegel:1993th,Hull:2009mi,Hull:2009zb,Hohm:2010jy,Hohm:2010pp}, 
we were motivated by the above results  
to try to understand what would be the $L_\infty$ algebra of double field theory.

For the study of $L_\infty$ in field theory
a fact about $L_\infty$ in  closed string field theory 
was puzzling.
In this theory there is a triple product $[  \, \cdot \,, \cdot \, , \, \cdot \, ]$
and it controls several aspects of the theory. It enters in the quartic interactions of fields,
as the inner product of $\Psi$ with $[  \, \Psi \,, \Psi \, , \Psi \, ]$.  It 
appears in the gauge transformations as a nonlinear contribution
$\delta_\Lambda \Psi \sim \cdots +  [  \, \Lambda \,, \Psi \, , \, \Psi \, ]$.
 It makes the commutator of two gauge transformations $\delta_{\Lambda_1}$ 
 and $\delta_{\Lambda_2}$ a gauge transformation with a gauge parameter
 that includes a field-dependent term $[\Lambda_1 , \Lambda_2, \Psi]$.  Finally,
the triple product implies that the gauge transformations only close on shell, 
the extra term being $[\Lambda_1 , \Lambda_2, {\cal F}]$, where ${\cal F}$ is the string field equation.
 In closed string field theory all these peculiarities happen simultaneously
 because the triple product is
 non-vanishing; these facts are correlated.  Moreover, the definition of the product
 is universal, valid for arbitrary input string fields.
 
 These facts, however, are not correlated in ordinary field theories.  Yang-Mills, for example,
 has a quartic interaction in the Lagrangian 
 but the commutator of gauge transformations is a gauge
 transformation with a field-independent gauge parameter.  The same is true for 
 Einstein   gravity in a perturbative expansion around a background.  The gauge algebra
 of double field theory is field independent but 
 there is a triple bracket associated to the failure of the Jacobi identity for the Courant bracket. 
  The lack of correlation is quickly demystified by exploring such examples.  
 Gauge parameters, fields, and field equations appear
 in different vector spaces, according to some relevant grading.  In defining a triple
 product one must state its value for all possible gradings of the various inputs.  
 While in the string field theory one has a universal definition controlled by some 
 data about four-punctured Riemann spheres, the definition of products {\em in a given
 field theory} has to be done in a case by case approach as we vary the inputs.
The product is non-vanishing for inputs with certain gradings and can vanish
 for other sets of inputs.  The correlation observed in string field theory
 is not discernible in the various separate field theories.  
 
Another important feature is revealed by the explicit analysis: there are at least
two $L_\infty$ algebras associated to a field theory, one a subalgebra of the other.
There is an ${L_\infty^{\phantom{0}}}^{\hskip-7pt \rm gauge}$ for the gauge structure which is a subalgebra of the larger
${L_\infty^{\phantom{0}}}^{\hskip-7pt \rm full}$  
that includes the interactions associated with the action and the field equations:
\be
{L_\infty^{\phantom{0}}}^{\hskip-7pt \rm gauge}  \subset \  {L_\infty^{\phantom{0}}}^{\hskip-5pt \rm full} \,. 
\ee
Any  algebra for the gauge structure must include a vector space for gauge parameters.
If the algebra is field dependent, it must include a vector space for fields.  This algebra does
not include interactions.  When including interactions, the gauge algebra is supplemented
by a new vector space for field equations and a set of new products, including some defined
on fields and field equations.   One can easily have a theory where ${L_\infty^{\phantom{0}}}^{\hskip-7pt \rm gauge}$ is a Lie algebra but ${L_\infty^{\phantom{0}}}^{\hskip-5pt \rm full}$ has higher
products, because the theories have quartic or higher-order interactions.  
This is the case for Yang-Mills theory and for  Einstein gravity.
For a discussion of $L_\infty$ algebras associated with some class of
gauge algebras with field-dependent
structure constants see also \cite{Fulp:2002kk}.

  If ${L_\infty^{\phantom{0}}}^{\hskip-7pt \rm gauge}$ is field independent, there is
a third, intermediate algebra 
\be 
{L_\infty^{\phantom{0}}}^{\hskip-7pt \rm gauge+fields}\;, 
\ee 
that includes the off-shell realization
of the gauge transformations on fields but that does not include dynamics. 
Roytenberg and Weinstein~\cite{roytenberg-weinstein}, for example, 
computed  the field-independent algebra ${L_\infty^{\phantom{0}}}^{\hskip-7pt \rm gauge}$ for the Courant bracket.   We will review that work, translated in $O(d,d)$ covariant language, and 
extend it to consider the algebra ${L_\infty^{\phantom{0}}}^{\hskip-7pt \rm gauge+fields}$ and the
 algebra ${L_\infty^{\phantom{0}}}^{\hskip-5pt \rm full}$ for the interacting double field theory.

 Our work here points to 
 an intriguing possibility:  any gauge invariant perturbative field theory
 may be represented by an $L_\infty$ algebra that encodes all the information about the theory,
 namely, the gauge algebra and its interactions.  It is then possible that {\em  
 gauge invariant 
 perturbative field theories are classified by $L_\infty$ algebras.}   A few comments are
 needed here.  Associated to any gauge structure, field theories can differ by the interactions.
 Since the interactions define some of the products in 
 ${L_\infty^{\phantom{0}}}^{\hskip-5pt \rm full}$,
 that aspect of the theory is properly  
 incorporated.  Of course, field redefinitions establish equivalences
 between theories, and such equivalences must correspond to suitable isomorphisms of $L_\infty$
 algebras,  presumably in the way discussed for $A_\infty$ in~\cite{Gaberdiel:1997ia}. 
 We seem to be constrained to theories formulated in perturbative form, that is theories in which
 one can identify unambiguously terms with definite powers of the fields in the Lagrangian, 
 although there may be exceptions.   
 For Einstein 
 gravity, which in the standard formulation contains both the metric and its inverse,
 one must expand around a background to obtain a perturbative expansion in terms of the fluctuating
 field.  
 This formulation of gravity as $L_\infty$ is completely straightforward, as will be clear to the reader
 of this paper, but real insight would come only if some elegant explicit definition of the products
 could be given. 
 Constrained fields, such as the generalized metric of double field theory, are also problematic
 as the power of the field in any expression may be altered by use of the constraint. 
All in all, we do not attempt to show that {\em any} gauge invariant 
field theory
has a description as an $L_\infty$ algebra, although we suspect that the result is true for 
unconstrained fields.   
Perhaps a proof could be built using a different approach.   
Consider a perturbative theory
that can be formulated in the Batalin-Vilkovisky formalism with a master action $S$ satisfying
the classical Batalin-Vilkovisky master equation $\{ S, S\} =0$; see \cite{Henneaux:1989jq} for a
review of these techniques.  General arguments indicate that an $L_\infty$ structure 
can be systematically extracted from the master action~\cite{Alexandrov:1995kv}.

Since $A_\infty$ is the algebraic setup for open string field theory, one can ask
why is $L_\infty$, the setup for closed string field theory,  
 chosen for perturbative field theories.  We have no general answer   
but it appears that the $L_\infty$ setup is rather flexible.  We show that for
Chern-Simons theory both an $A_\infty$ and a $L_\infty$ formulation exists. 
The first formulation requires describing the Lie algebra in terms
of an associative algebra of matrix multiplication.  The second formulation
requires the use of a background metric in the definition of the products.
We have not investigated if other field theories have both formulations. 
See, however, the general discussion in~\cite{Movshev:2004aw}  
giving an $A_\infty$ setup to Yang-Mills theory.  
The formulation of gauge theories as $A_{\infty}$ 
algebras will also be investigated in \cite{MartinRocek}.

A fraction of the work here deals with the structure of $L_\infty$ algebras.
Following \cite{Zwiebach:1992ie} we discuss the axioms and main identities,
but develop a bit further the analysis.  We show explicitly 
that given an $L_\infty$ algebra
with multilinear products with $n\geq 1$ inputs,
one can construct consistent {\em modified} products
 with $n\geq 0$ inputs.  A product  $[ \cdot ]'$ without an input is just a special vector
 ${\cal F}$ in the algebra and that vector is in fact the field equation for a field $\Psi$
 in the theory defined by the original products.  The modified product with one
 input,  $[ B]' = Q' B$, 
 defines a linear operator $Q'$, 
 built from $\Psi$ and a $Q$ operator
 that squares to zero and defines the one-input original product.   
  We establish 
 the $L_\infty$ identity
 \be
 \label{bianchi-ft}
 Q' {\cal F} \ = \ 0 \,,
 \ee   
 which can be viewed as the {\em Bianchi identity} of the original theory, and $Q'$ 
 may be thought of 
 as a {\em covariant derivative}.  Indeed we also have $Q'^2 \sim {\cal F}$. 
 The modified products
 simplify the analysis of the gauge structure of the theory.  The gauge transformations
 take the form 
 \be
 \delta_\Lambda \Psi \ = \ Q' \Lambda\,, 
 \ee
  and the computation of the gauge algebra $[  \delta_{\Lambda_2}, 
 \delta_{\Lambda_1} ]$ can be simplified considerably.  We also compute
  the `gauge Jacobiator'  
  \be
  \label{gauge-jacobiator}
{\cal J} (\Lambda_1, \Lambda_2, \Lambda_3) \ \equiv \  
\sum_{\hbox{\tiny\rm cyc}}  \bigl[ \delta_{\Lambda_3}\,,  \, [  \delta_{\Lambda_2}\,, 
\, \delta_{\Lambda_1} ]  \bigr] \,.
\ee
The right hand side is trivially zero for any theory with well-defined gauge transformations;
this is clear by expansion of the commutators.  On the other hand, this vanishing
 is a nontrivial constraint on the form of the gauge algebra.  This constraint is satisfied by
 virtue of the identities satisfied by the higher products in the $L_\infty$ algebra.
 
 In the above approach, called the $b$-picture of the $L_\infty$ algebra, the signs
 in the field equation, gauge transformations, action, and gauge algebra are known.
 There is another picture, the $\ell$-picture of the algebra \cite{Lada:1992wc} 
 in which the signs of the Jacobi-like identities are more familiar.   The two pictures
 are related by suspension, a shift in the degree of the various spaces involved.
 We use this suspension to derive the form of field equations, gauge transformation,
 action, and gauge algebra in the $\ell$ picture.

Here is a brief summary of this paper.   
We begin in section 2 with a description
of the $L_{\infty}$ algebra in the conventions of the original closed string field theory
and discuss the gauge structure, particularly the closure of the algebra and the 
triviality of the Jacobiator. 
In sec.~3 we begin by defining 
the axioms of $L_\infty$ algebras in two conventions, one (the $\ell$-picture) that is 
conventional in the mathematics literature and one (the $b$-picture) that is conventional 
in string field theory and hence directly related to sec.~2. 
In sec.~3.3 we make some general remarks 
how to identify for a given field theory the corresponding structures of an $L_{\infty}$ algebra.   
Moreover, we explain how gauge covariance of the field equations and
closure of the gauge transformations imply that large classes of $L_\infty$
identities hold.
(Readers mainly interested in the applications to field theory can skip sec.~2 
and sec.~3.2.) 
These results will be applied in sec.~4  
 in order to describe Yang-Mills-like gauge theories, 
both for Chern-Simons actions in 3D and for general Yang-Mills actions.  
In sec.~5 we  
discuss the $L_{\infty}$ description of double field theory, which in turn is an 
extension of the construction by Roytenberg and Weinstein for the Courant algebroid. 
We finally compare these results with $A_{\infty}$ algebras by giving the $A_{\infty}$ description 
of Chern-Simons theory in sec.~6. We close with a summary and an outlook in sec.~7.

\section{$L_\infty$ algebra and gauge Jacobiator}\label{LinfJac}

In this section we review the definition of an $L_\infty$ algebra, 
state the main identities,\footnote{The identities for gauge invariance of the
classical theory first appeared in~\cite{Kugo:1989aa} and were re-cast as $L_\infty$ identities
in~\cite{Zwiebach:1992ie}.  Note that the structure of quantum closed string field theory
goes beyond $L_\infty$ algebras. } 
and introduce the field equation and the action.  
We then turn to a family of identities for modified products, giving 
the details of a result anticipated in \cite{Zwiebach:1992ie}.  
They correspond to the products that would arise after the expansion
of the string field theory action around a background that does not solve 
the string field theory equations of motion.\footnote{After expansion of
the string field theory around a background that satisfies the equations
of motion, the type of algebraic structure is not changed~\cite{Sen:1990ff}.}  
With these products, the Bianchi identities of string field theory
become clear and the modified BRST operator functions as a covariant 
derivative.   We elaborate on the types of gauge transformations,  
and the modified products simplify the
calculation of the gauge algebra.  We are also able to verify that the 
gauge Jacobiator for a general field theory described with an $L_\infty$ algebra
vanishes, as required by consistency.

\subsection{The multilinear products and main identity}

In an $L_\infty$ algebra we have a vector space $V$ graded by 
a degree, which is an integer.  
We will typically work with elements $B_i\in V$ of fixed 
degree.\footnote{In closed string field theory degree `deg' is related
to ghost number `gh' as deg $= 2- $gh.}   
 The degree enters in sign factors where,
for convenience, we omit the `deg' label.  Thus, for example: 
\be
 (-1)^{B_1 B_2}  \ \equiv \  (-1)^{{\small\hbox{deg} }(B_1) \cdot {\small\hbox{deg} } (B_2)}
 \,.  
\ee
In exponents, the degrees are relevant only mod 2.   
In an $L_\infty$ algebra we have multilinear products.
In the notation used for string field theory the multilinear products
are denoted by brackets $[ B_1, \ldots , B_n ]$ and are graded commutative
\be\label{Bcommutation}
[ \cdots B_i , B_j , \cdots ] = (-1)^{B_i B_j}  [ \cdots B_j , B_i , \cdots ]\,.
\ee
 All products are defined to be of intrinsic degree $-1$, meaning that 
the degree of a product of a given number of inputs
is given by   
\be
\hbox{deg}  \,(\,  [B_1, \, \ldots\,  , B_n]  ) \ = \   -1 + \sum_{i=1}^n  \hbox{deg} (B_i) \,. 
\ee 
The product with one input is sometimes called the $Q$ operator (for BRST)
\be
\label{sgbv}
[B] \ \equiv  \ QB\, . 
\ee
We also have a product $[ \cdot  ]$ with no input whose value is
just some special vector in the vector space.

\noindent
The $L_\infty$ relations can be written in the form~\cite{Zwiebach:1992ie}:
\be
\label{hla-sft}
\sum_{l, k \geq 0 \atop l+k = n} \sum_{\sigma_s}  \sigma( i_l, j_k)  
\bigl[ B_{i_1}, \, \ldots\,  , B_{i_l} \, [ B_{j_1}, \, \ldots \,,  B_{j_k}] \, \bigr]  \ = \ 0 \,,  \quad  n \geq 0\,. 
\ee
Here $n$ is the number of inputs (if $n=0$ we still get a nontrivial identity).   
The inputs $B_1, \ldots,  B_n$ are split into two sets:  a first set 
$\{ B_{i_1} \ldots B_{i_l} \}$ with $l$ elements and a second set 
$\{ B_{j_1} \ldots B_{j_k}\}$ with $k$ elements, where $l+k = n$.   
The first set is empty if $l=0$ and the second set is empty  if $k=0$.
The two sets do not enter the identity symmetrically: the second set   
has the inputs for a product nested inside a product that involves the first set of elements.
The set of numbers $\{ i_1, \ldots , i_l, j_1, \ldots, j_k\}$ is a permutation of the list $\{ 1, \ldots , n\}$.

The sums are over inequivalent splittings.  Sets with different values of $l$ and $k$ are inequivalent,
so we must sum over all possible values of $k$ and $l$.
Two splittings with the same values of $l$ and $k$ are equivalent if the first set $\{ B_{i_1} \ldots B_{i_l} \}$ contains the same elements, regardless of order. 
 The factor $\sigma(i_l, j_k)$ is the sign needed to rearrange the list
 $\{ B_*, B_1,  \ldots ,B_n\}$
into $\{ B_{i_1},  \ldots B_{i_l}, B_* ,\, B_{j_1}, \ldots B_{j_k} \}$:
\be
\{ B_*, B_1,  \ldots ,B_n\}  \quad \to \quad  \{ B_{i_1},  \ldots B_{i_l}, B_* ,\, B_{j_1}, \ldots B_{j_k} \}  \,, 
\ee
 using the degrees to commute the $B$'s
according to (\ref{Bcommutation}) 
and thinking of $B_*$ as an element of odd degree.  The element $B_*$ is needed to take into
account that the products are odd.

For classical string field theory, or for any field theory expanded around a  
classical solution, the value of the zeroth product $[ \cdot ]$ 
will be set equal to the zero vector:
\be
\label{zero-product} 
[ \, \cdot \, ] \ \equiv \  0 \,. 
\ee
Using the above rules for sign factors, we can write out 
the $L_{\infty}$ identities (\ref{hla-sft}).  
Note that in the absence of a zeroth product
$k >0$ and thus $n>0$ to get a nontrivial identity.  
For $n=1, 2, 3$ one gets:    
\be
\label{clksshrss}
\begin{split}
  0 \ &= \ Q ( Q B) \,,  \\
  0 \ &= \ Q [ B_1, B_2]  +  [QB_1 , B_2]  + (-1)^{B_1} \, [B_1, QB_2 ] \,, \\
 0 \ &=  \  Q [B_1 , B_2, B_3 ] \\
&\qquad  +
 [ QB_1 , B_2, B_3 ]  
 +  (-1)^{B_1}
[B_1 , Q B_2, B_3  ]  + (-1)^{B_1+ B_2} [B_1 , B_2, Q B_3 ]
 \\
 & \qquad +(-1)^{B_1}    [ B_1 , [B_2, B_3]\,  ] 
 \ + (-1)^{B_2 (1+B_1)} 
 [ B_2 , [B_1, B_3] \,  ] \\
 & \qquad + (-1)^{B_3 (1+B_1+B_2)} 
\,  [ B_3 , [\, B_1, B_2] ]\;.  \\
\end{split}
\ee

We will now discuss how to define in this language equations of motion and actions for a field theory.  
To this end and for brevity, we write 
products with repeated inputs 
as powers.  
When there is no possible 
confusion we also omit the commas between the inputs: 
\be
[ \Psi^3 ] \ \equiv \ [ \Psi, \Psi,  \Psi ] \,, \qquad    [ B \Psi^3] \ \equiv \  [ B, \Psi, \Psi, \Psi ] \,. 
\ee
Here $\Psi$, called the field,  is an element of 
degree zero:  
\be
\hbox{deg}  \, \Psi \ = \ 0 \,.
\ee
If $\Psi$ had been of odd degree, the above products would vanish
by the graded commutativity property.

Given a set of products satisfying the $L_\infty$ conditions and a Grassmann even
field $\Psi$ we 
introduce a field equation ${\cal F}$ of degree minus one:  
\be
\label{calF-def}
{\cal F} \ = \, \sum_{n=0}^\infty  {1\over n!}\  [\Psi^n] \, = \, 
 \, Q \Psi  + \tfrac{1}{2}[\Psi^2]  + \tfrac{1}{3!} [\Psi^3 ]  + \ldots  \, = \ Q\Psi + \tfrac{1}{2}[ \Psi, \Psi] 
 + \tfrac{1}{3!}[\Psi, \Psi, \Psi] + \ldots \ .
\ee
Again, we used that the term with $n=0$ vanishes, as it involves a product with no input.
The field equation  
${\cal F}$ is of  degree minus one because $\Psi$ is of degree zero
and all products are of degree minus one.  Certain infinite sums appear often when dealing with gauge
transformations and make it convenient to define
modified, primed products:
\be
 [  A_1
 \ldots A_n ]' \ \equiv \ \sum_{p=0}^\infty \,  {1\over p!} \, 
 [  A_1
 \ldots A_n \, \Psi^p ]\,,   \qquad  n \geq 1 \,. 
\ee
Thus, for example, 
\be\label{primeQ}
\begin{split} 
[ A]' \ \equiv \ Q' A  \ = \ & \  QA  + [A \Psi] + \tfrac{1}{2} [A \Psi^2] + \ldots \;,  \\[1.0ex]
 [  A_1 \ldots A_n ]' \ = \ & \  [  A_1 \ldots A_n ]+ 
 [  A_1 \ldots A_n \Psi ]+ \tfrac{1}{2}  [  A_1 \ldots A_n \Psi^2 ] \, + \, 
 \ldots \;. 
\end{split}
\ee
The variation of those products is rather simple:
\be
\label{var-primed-products}
\delta  [  A_1\ldots A_n ]' \ = \  [  \delta A_1 \ldots A_n ]'
+ \ldots +  [  A_1\ldots \delta A_n ]'
+  [  A_1\ldots  A_n \delta \Psi]' \,.
\ee
The identification of $[A]'$ with $Q' A$ is natural given
(\ref{sgbv}). The variation of the field equation takes the form of a modified product.
We have
\be
\label{varF}
\delta {\cal F} \ = \ Q' (\delta \Psi) \,, 
\ee
which is readily established:
\be
\delta {\cal F} \ = \    \delta \sum_{k=0}^\infty {1\over k!}
 [\Psi^k]
\ = \   \sum_{k=1}^\infty {k\over k!}
 [\Psi^{k-1}\delta \Psi] \ = \  \sum_{k=0}^\infty {1\over k!}
 [\Psi^k\delta \Psi]  \ = \ \ [ \,\delta \Psi\,  ]'\,.  \ee

\medskip
\noindent
{\bf Inner product and action:}  
The action exists if there is a suitable inner product $\langle \cdot \,, \cdot \rangle$.
One requires that  
\be
\label{gr-comm-ip}
\langle  A, B \rangle \ = \ (-1)^{(A+1)(B+1)}  \langle B, A \rangle\,,
\ee
and that the expression 
\be
\langle \, B_1,  [B_2 , \ldots , B_n] \rangle \,,
\ee
for $n \geq 1$ is a multilinear {\em graded-commutative} function of all the arguments. 
From the above one can show, for example, that 
\be
\label{jnctns}
\langle QA , B\rangle \ = \ (-1)^A \langle A , QB \rangle \,. 
\ee
The action is  given by
\be
\label{jnbtflrgs}
S \ = \ \sum_{n=1}^\infty \, {1\over (n+1)!}\langle \Psi \,, \, [\Psi^n]\,  \rangle \,.  
\ee 
A short calculation shows that that under a variation $\delta \Psi$ one has 
\be
\label{vary-action}
\delta S  \ = \ \langle \delta \Psi, {\cal F} \rangle \,,
\ee
confirming that ${\cal F}=0$ is the field equation corresponding to the action.

\subsection{A family of identities} 

In this subsection we will establish a number of identities that will be useful below when 
computing the Jacobiator. 
The products $[ \ldots ]'$ can in fact be viewed as a set
of products satisfying a simple extension of the $L_\infty$ identities.
To see this and to get the general picture we consider a few examples.

Consider (\ref{hla-sft}) when all $B$'s 
are of  even degree.  
 The sign factor is then always equal to $+1$ and we have
\be
\label{hla-sft1}
\sum_{l , k\geq 0 \atop l+k = n} \sum_{\sigma_s}  \bigl[ B_{i_1} \ldots B_{i_l} \, [ B_{j_1} \ldots  B_{j_k}] \, \bigr]  \ = \ 0 \,,  \quad  n \geq 0\,    \quad  B_k \ \hbox{even}  \ \forall k \,. 
\ee
 For $l$ and $k$ fixed
there are $n! / (l! k!)$ inequivalent splittings of the inputs;  this is the number of terms
in the sum $\sum_{\sigma_s}$.  Assume 
now  that all the $B$'s are the same: $B_1 = B_2 = \ldots = B$, so that all those terms are equal.
  We then have
\be
\label{hla-sft3}
\sum_{l, k \geq 0 \atop l+k = n} {1\over l!\,  k!}
\bigl[ B^l \, [ B^k] \, \bigr]  \ = \ 0 \,,  \quad  n \geq 0\,,    \quad  B \ \hbox{even}   \,, 
\ee
where we have taken out an overall factor of $n!$ from the numerator. 
If we now take $B = \Psi$ and sum over $n$ this identity becomes 
\be
\label{hla-sft4}
\sum_{n \geq 0} \sum_{l, k \geq 0 \atop l+k = n} {1\over l!\,  k!}
\bigl[ \Psi^l \, [ \Psi^k] \, \bigr]  \ = \ 0       \,. 
\ee
Reordering the double sum we have
\be
\label{hla-sft5}
 \sum_{l, k \geq 0 } {1\over l!\,  k!}
\bigl[ \Psi^l \, [ \Psi^k] \, \bigr]  \ = \  \sum_{l \geq 0 } {1\over l!}
\bigl[ \Psi^l \, {\cal F}  \, \bigr]  \ = \ 0         \,, 
\ee
where we summed over $k$ in the second step and used  (\ref{calF-def}). 
Recalling (\ref{primeQ}), the sum over $l$ finally gives 
\be
 Q'  {\cal F }  \ = \ 0 \,.  
\ee
If we view $Q'$ as the analogue of the covariant derivative $D$ and ${\cal F}$ as the analogue 
of the non-abelian field strength $F$ in Yang-Mills theory, then this identity is the analogue of the 
Bianchi identity $DF=0$.

Let us consider a second $L_\infty$ identity again based on (\ref{hla-sft}) but with $n+1$ inputs
\be
B_1 = A\, ,  \ B_2 = \ldots  = B_{n+1} \, = \, B \,, \quad  B \ \hbox{even} \,. 
\ee
There are two possible classes of splittings, both of which involve separating the $n$ copies of
$B$ into a set with $l$ elements and a set with $k$ elements, with $l+k = n$.  These are
\be
\{ A B^l\} \,,  \,  \{B^k\}   \quad  \hbox{and}   \quad   \  \{ B^l\}  \,, \,  \{A B^k\} \,. 
\ee
The sign factors arise from reordering $B_* A B^l B^k$ into $A B^l B_*B^k$ for the first sequence, 
giving a sign $(-1)^A$, and into $B^l B_*AB^k$ for the second sequence, giving no sign.
We thus have 
\be
\sum_{l, k \geq 0 \atop l+k = n} {n!\over l!\,  k!} 
\bigl(  \, (-1)^A [ A B^l [ B^k]]  + [ B^l [ A B^k]] \bigr) \ = \ 0  \,. 
\ee
For $n=0$ all terms vanish.  This is a fine identity but an alternative form 
is also useful.  We cancel the $n!$ in the numerator and sum
over $n$:
\be
\label{vmbb}
\sum_{n\geq 0} \sum_{l, k \geq 0 \atop l+k = n} {1\over l!\,  k!} 
\bigl(  \, (-1)^A [ A B^l [ B^k]]  + [ B^l [ A B^k]] \bigr) \ = \ 0 \,, 
\ee
which becomes, reorganizing the sums and letting $B= \Psi$
\be
\label{vmbv}
\sum_{l, k \geq 0 } {1\over l!\,  k!} 
\bigl(  \, (-1)^A [ A \Psi^l [ \Psi^k]]  + [ \Psi^l [ A \Psi^k]] \bigr) \ = \ 0  \,.
\ee
Performing the sum over $k$ gives
\be
\label{vmkhrv}
\sum_{l, \geq 0 } {1\over l!} 
\bigl(  \, (-1)^A [ A \Psi^l {\cal F}]  + [ \Psi^l Q'A ] \bigr) \ = \ 0  \,.
\ee
Doing the sum over $l$ now gives
\be
\label{vmksshrbv}
 (-1)^A [ A  {\cal F}]' + Q' ( Q' A) \ = \ 0  \,. 
\ee
Since ${\cal F}$ is odd, this result is equivalent to 
\be
\label{vmdlcsbtt}
\,   Q' (Q'A)  \, + \,  [ \, {\cal F} \, A \, ]'\ = \ 0 \, . 
\ee
Again, if we view $Q'$ and ${\cal F}$ as the analogues of covariant derivative $D$ and field strength $F$ in 
Yang-Mills theory, then this relation is the analogue of $D^2=F$. 

Consider another $L_\infty$ identity, again based on (\ref{hla-sft}), but with $n+2$ inputs
and two string fields $A_1, A_2$:
\be
B_1 = A_1\, ,  \ B_2= A_2 \,,   B_3 = \ldots  = B_{n+2} \, = \, \Psi \,. 
\ee
This time the above procedure yields:
\be
 0 \ = \ Q' [ A_1, A_2]'  +  [Q'A_1 , A_2]'  + (-1)^{A_1} \, [A_1, Q'A_2 ]'
 + [ \, {\cal F}A_1 A_2 ]'\,. 
\ee
Comparing the above and (\ref{vmdlcsbtt}) with the first two equations in (\ref{clksshrss}) the pattern becomes clear.  We are obtaining for the primed products
the same $L_\infty$ identities with one extra term.  In fact, the extra term corresponds to having
a zeroth product, as in (\ref{zero-product}), that this time is nonzero:  
\be
[ \, \cdot \, ]' \ \equiv \  {\cal F} \,. 
\ee
As noted in \cite{Zwiebach:1992ie} the identities (\ref{hla-sft}) indeed give, for $n=0,1,2,3$
\be
\label{jyndlcsbtt}
\begin{split}
&  0 \ = \ Q' {\cal F}  \,,  \\
&  0 \ = \ Q' ( Q' A) + [{\cal F} A]' \,,  \\
&  0 \ = \ Q' [ A_1 A_2]'  +  [Q'A_1  A_2]'  + (-1)^{A_1} \, [A_1 Q'A_2 ]' + [ {\cal F} A_1 A_2]'  \,, \\
& 0 \ =  \  Q' [A_1  A_2, A_3 ]' \\
& \quad\ +
 [ Q'A_1 A_2 A_3 ]'  
 +  (-1)^{A_1}
[A_1  Q' A_2 A_3  ]'  + (-1)^{A_1+ A_2} [A_1  A_2 Q' A_3 ]'
 \\
 &\quad\ +(-1)^{A_1}    [ A_1  [A_2 A_3]'\,  ]' 
 \ + (-1)^{A_2 (1+A_1)} 
 [ A_2  [A_1 A_3]' \,  ]' \\
 &\quad\ + (-1)^{A_3 (1+A_1+A_2)} 
\,  [ A_3  [\, A_1 A_2]' ]' \ + \  [ {\cal F}  A_1 A_2 A_3]'  \,.
\end{split}
\ee 
We have thus demonstrated that the modified products satisfy an extended
form of the $L_\infty$ identities, one that includes a nontrivial zeroth product. 
These identities simplify some of the work that was done in \cite{Zwiebach:1992ie}
and allow us to examine the Jacobiator.   The above identities would be needed
to construct a classical field theory around a background that is not a solution
of the field equations.   

The inner product interacts nicely with the modified products.  One can
quickly use (\ref{jnctns}) and multi-linearity to show that, for example,
\be
\label{jnsmllyns}
\langle Q'A , B\rangle \ = \ (-1)^A \langle A , Q'B \rangle \,. 
\ee

\subsection{Gauge transformations and algebra}

We now describe the gauge transformations and their gauge algebra. Here our primed
identities are very helpful.  We also discuss trivial or equation-of-motion symmetries
of two types.  We introduce the notion of trivial gauge parameters, and that of 
extended gauge transformations.

\medskip
\noindent
{\bf Standard Gauge transformations:}   These take a very simple form in terms of the
new product: they are simply the result of $Q'$ acting on the gauge parameter $\Lambda$,
an element of degree $+1$. 
Indeed
\be
\label{vmlvsagdlck}
\delta_\Lambda \Psi \ =  \  [\Lambda]' \ = \ Q' \Lambda  \ = \ Q \Lambda  + [\Lambda \Psi]  + \tfrac{1}{2} [\Lambda \, \Psi^2 ]  + \ldots  \ . 
\ee
The key constraint is that the  field equations must be gauge covariant.  This requires that
the gauge transformation of ${\cal F}$ vanishes when ${\cal F}=0$.  With the help of the new
products this is now a trivial computation.  Using (\ref{varF}) and the second of (\ref{jyndlcsbtt})
\be
\label{jnlvltts}
\delta_\Lambda {\cal F}  \ = \ Q' (\delta_\Lambda \Psi)  \ = \ Q' (Q' \Lambda) 
\ = \  [ \Lambda \,  {\cal F}\,]'  \,.   
\ee
We see that covariance holds.  Writing out the result more explicitly,
\be
\label{jnlvlsckbltts}
\delta_\Lambda {\cal F}  
\ = \  [  \Lambda\,{\cal F}]  \, + [ \Lambda {\cal F}  \Psi \,] + \tfrac{1}{2}  [ \Lambda  {\cal F}\Psi^2 \,] + \ldots      \,,  
\ee
makes it clear that the bare  field appears on the right-hand side.  
The action is, 
of course, gauge invariant:
\be
\delta S \ = \  \langle \delta_\Lambda \Psi , {\cal F} \rangle \ = \  
\langle Q' \Lambda , {\cal F} \rangle \ = \ -\langle  \Lambda ,Q'{\cal F} \rangle \ = \ 0 \,,
\ee
making use of (\ref{jnsmllyns}) and the first identity in (\ref{jyndlcsbtt}).

\medskip
\noindent
 {\bf Equations-of-motion symmetries}:   
These are transformations
that vanish when using the equations of motion and are invariances
of the action.    For example, $\delta \Psi  = [\chi, {\cal F}]$, for even $\chi$ is 
a trivial gauge transformation.  It vanishes
on-shell and leaves the action invariant because 
\be
\delta S \ = \ \langle \delta \Psi , {\cal F} \rangle \ = \ \langle\,  {\cal F}\,, \,  [\chi, {\cal F}]\,  \rangle 
 \ = \  \langle\,  \chi\,, \,  [{\cal F}, {\cal F}]\,  \rangle \ = \ 0\,,  \ee
because ${\cal F}$ is Grassmann odd.  Two types of 
equations-of-motion symmetries will play
a special role, one parameterized by a Grassmann even single string field $\chi$ of ghost number zero and 
another parameterized by two gauge parameters $\Lambda_1, \Lambda_2$.  
They are:
\be
\begin{split}
\delta_{\chi}^{{}^T} \Psi \ \equiv \ & \   [ \chi\,  {\cal F} ]' \ = \ 
-Q' (Q'\chi) \,, 
\\[1.0ex]
 \delta^{{}^T}_{\Lambda_1, \Lambda_2  } \Psi \ \equiv \ & \   [ \Lambda_1 \Lambda_2 {\cal F} ]' \,, 
\end{split}
\ee
using (\ref{vmdlcsbtt}) in the first line. 
The  second type of 
equations-of-motion symmetries 
 shows up in the commutator  
of two standard gauge transformations, as we will discuss  now.

\noindent
{\bf Gauge algebra:} We claim that the standard gauge transformations form an algebra
that includes the 
equations-of-motion symmetries of the second type.  Indeed, assuming the
gauge parameters $\Lambda_1$ and $\Lambda_2$ 
are field independent we have\footnote{In \cite{Zwiebach:1992ie}  the 
equations-of-motion symmetry 
on the right hand side has a wrong sign.}  
\be
\label{jnhwlvly}
\bigl[ \delta_{\Lambda_2} \,, \, \delta_{\Lambda_1} \bigr] 
\ = \  \delta_{[\Lambda_1\, \Lambda_2]'}\ + \  \delta^{{}^T}_{\Lambda_1, \Lambda_2 } \,.
\ee
With the help of our identities, the proof of this claim is much simplified.   Using the variation
formula (\ref{var-primed-products}) we find
\be
\delta_{\Lambda_2}  \, \delta_{\Lambda_1} \Psi \ = \  \delta_{\Lambda_2} [ \Lambda_1 ]' \ = \ [\Lambda_1 
\delta_{\Lambda_2} \Psi ]'  \ = \ [  \Lambda_1 
Q' \Lambda_2]'\,.
\ee
Thus, it follows that 
\be
\bigl[ \delta_{\Lambda_2} \,, \, \delta_{\Lambda_1} \bigr]  \Psi  \ = \ [  \Lambda_1 Q' \Lambda_2]' - [  \Lambda_2 Q' \Lambda_1]'\,. 
\ee
The third identity in (\ref{jyndlcsbtt}) gives
\be
0 \ = \ Q' [ \Lambda_1 \Lambda_2]'  +  [Q'\Lambda_1  \Lambda_2]'  - \, [\Lambda_1 Q'\Lambda_2 ]' + [ {\cal F} \Lambda_1 \Lambda_2]'  \,. 
\ee
As a result
\be
\bigl[ \delta_{\Lambda_2} \,, \, \delta_{\Lambda_1} \bigr]  \Psi  \ = \ Q' [ \Lambda_1 \Lambda_2]'
+ [ \Lambda_1 \Lambda_2 {\cal F}]' \ = \ \delta_{[ \Lambda_1 \Lambda_2]'} \Psi
+ \delta^{{}^T}_{\Lambda_1, \Lambda_2  } \Psi  \,,
\ee
which is what we wanted to prove.

\medskip
\noindent
{\bf Trivial gauge parameters}:   A field-dependent 
parameter $\Lambda$ is said to be trivial
if $\Lambda = Q'\chi$ for some Grassmann even $\chi$.
A standard transformation with a trivial gauge parameter is a 
equations-of-motion symmetry of the first kind:
\be
\delta_{ Q'\chi}  \Psi \ = \ Q' (Q' \chi)  \ = \ -  [{\cal F} \,  \chi  ]' \ = \ -\delta_{\chi}^{{}^T} \Psi \,.
\ee

\medskip
\noindent
{\bf Extended gauge transformations}:  They are the sum of a standard
gauge transformation with parameter $\Lambda$  and a 
equations-of-motion symmetry of the first kind with parameter $\chi$ of ghost-number zero:
\be
\delta^{\,{}^E}_{\Lambda, \chi } \,\Psi \ \equiv \ \delta_\Lambda \,\Psi   \, + \, 
\delta_{\chi}^{{}^T} \Psi \ = \ Q' \Lambda \, + \, 
[\chi\, {\cal F}\,  ]' \,. 
\ee

\medskip
\noindent
{\bf Null transformations}:  These are extended gauge transformations
that give no variation of the  field.  Indeed, if $\Lambda = Q' \chi$ the
transformation $\delta^{\,{}^E}_{\Lambda, \chi } $ of the string field
vanishes:
\be
\begin{split}
\delta^{\,{}^E}_{Q'\chi, \,\chi } \,\Psi \ \equiv  \ 
 \delta_{Q'\chi } \Psi + \, \delta^{{}^T}_\chi  \Psi  
\ =  \ Q' (Q'\chi)   \, + \, 
[\chi\, {\cal F}\,  ]' \ = \ 0\;, 
\end{split}
\ee
because $\chi$ is Grassmann even.  We say that $\chi$ generates the null transformation 
$\delta^{\,{}^E}_{Q'\chi, \,\chi }$. 
 Since null transformations give no variation of the
fields we declare they are identically zero: $\delta^{\,{}^E}_{Q'\chi, \,\chi }  =  0$.

\subsection{Gauge Jacobiator}

Given a set of gauge transformations one can consider the gauge algebra,
as we did above. In addition, one can consider the `gauge Jacobiator'  ${\cal J}$,  
\be
{\cal J} (\Lambda_1, \Lambda_2, \Lambda_3) \ \equiv \  
\sum_{\hbox{\tiny\rm cyc}}  \bigl[ \delta_{\Lambda_3}\,,  \, [  \delta_{\Lambda_2}\,, 
\, \delta_{\Lambda_1} ]  \bigr] \,,
\ee
a definition inspired by that of the Jacobiator of a bracket.  Here the cyclic sum
involves the sum of three terms in which we cycle the three indices $1,2,3$.   
The Jacobiator ${\cal J}$,
if non-zero, would be a gauge transformation because gauge transformations
close.  But  in the above, the 
brackets are simply commutators, and if the gauge transformations 
are well defined, upon expansion one can see that all terms vanish and
this gauge Jacobiator should vanish. 

This vanishing, however, is not a trivial constraint from the viewpoint
of the $L_\infty$ algebra.  One can compute ${\cal J}$ using the gauge
algebra (\ref{jnhwlvly}) and one finds that the vanishing requires the
$L_\infty$ identities for three and four inputs.   We will do this calculation
below.  
Indeed, we will find  that the gauge Jacobiator is a null transformation and
thus vanishes identically.  Namely, 
\be
\sum_{\hbox{\tiny\rm cyc}}  \bigl[ \delta_{\Lambda_3}\,,  \, [  \delta_{\Lambda_2}\,, 
\, \delta_{\Lambda_1} ]  \bigr] \,   \ = \  0 \,.\ee

\noindent
{\bf Proof:} 
Using the gauge algebra (\ref{jnhwlvly}) we have 
\be
\label{cllckswtss}
\sum_{\hbox{\tiny\rm cyc}}  \bigl[ \delta_{\Lambda_3}\,,  \, [  \delta_{\Lambda_2}\,, 
\, \delta_{\Lambda_1} ]  \bigr] \,   \ = \
\sum_{\hbox{\tiny\rm cyc}}  \bigl[ \delta_{\Lambda_3}\,,  \, \delta_{[ \Lambda_1
\Lambda_2 ]'} \bigr] \,   +   \sum_{\hbox{\tiny\rm cyc}}  \bigl[ \delta_{\Lambda_3}\,,  \, \delta^{{}^T}_{\Lambda_1, \Lambda_2  }   \bigr] \;. 
\ee
For the first term on the right-hand side, we can use the gauge algebra  
noticing, however, 
the presence of an extra term because the gauge parameter $[\Lambda_1 \Lambda_2]'$ is
now field dependent and must be varied using (\ref{var-primed-products}).  Following the same steps as in the derivation of the gauge algebra
one finds 
\be
\sum_{\hbox{\tiny\rm cyc}}  \bigl[ \delta_{\Lambda_3}\,,  \, \delta_{[ \Lambda_1
\Lambda_2 ]'} \bigr] \ = \ \delta_{\sum_{\hbox{\tiny\rm cyc}} 
\bigl(  \bigl[\, [\Lambda_1\Lambda_2]' \Lambda_3 \bigr]' 
+  \bigl[\, \Lambda_1\Lambda_2 Q'\Lambda_3 \bigr]'\bigr)  }  +\  
\sum_{\hbox{\tiny\rm cyc}} 
\delta^{{}^T}_{\   [\Lambda_1\Lambda_2]', \Lambda_3  } \,,
\ee
where the extra term is the one involving $ \bigl[\, \Lambda_1\Lambda_2 Q'\Lambda_3 \bigr]'$. 
We now use the last identity in (\ref{jyndlcsbtt}), with $A_i = \Lambda_i$ 
to simplify  the first term on the above right-hand side:
\be
\label{first_term}
\sum_{\hbox{\tiny\rm cyc}}  \bigl[ \delta_{\Lambda_3}\,,  \, \delta_{[ \Lambda_1
\Lambda_2 ]' } \bigr] 
\ = \ 
\delta_{
\bigl(  - Q'\, [\Lambda_1\Lambda_2\Lambda_3]'
+  \bigl[\, \Lambda_1\Lambda_2 \Lambda_3{\cal F}  \bigr]'\bigr)  }  +\  \sum_{\hbox{\tiny\rm cyc}} \delta^{{}^T}_{\  [\Lambda_1\Lambda_2]', \Lambda_3  } \;. 
\ee
This completes our simplification of the first term on the right-hand side of (\ref{cllckswtss}).
For the second term on that same right-hand side,  acting on $\Psi$,  we get
\be
\label{skstkvg}
\sum_{\hbox{\tiny\rm cyc}}  \bigl[ \delta_{\Lambda_3}\,,  \, 
\delta^{{}^T}_{\Lambda_1, \Lambda_2  }   \bigr] \, \Psi \ = \  
\sum_{\hbox{\tiny\rm cyc}} \Bigl( -\bigl[\Lambda_1\Lambda_2 
\, [ {\cal F}\Lambda_3 ]'  \bigr]'
+ \bigl[\Lambda_1\Lambda_2 {\cal F} \, Q'\Lambda_3  \bigr]' 
- \bigl[\Lambda_3 [\Lambda_1\Lambda_2 {\cal F} ]' \bigr]' \Bigr) \,,
\ee
where we had to use (\ref{jnlvltts}). We now need to use
 the $L_\infty$ identity for four string-fields  $\Lambda_1,\Lambda_2,\Lambda_3,{\cal F}$ 
and primed products.  Although we did not include it in (\ref{jyndlcsbtt}) it is readily
obtained.  
Recalling also that $Q' {\cal F}=0$ and that any product with more than one
${\cal F}$ vanishes we get
\be
\sum_{\hbox{\tiny\rm cyc}}  \bigl[ \delta_{\Lambda_3}\,,  
\, \delta^{{}^T}_{\Lambda_1, \Lambda_2  }   \bigr] \, \Psi \ = \  
-\sum_{\hbox{\tiny\rm cyc}} 
\bigl[ \, [\Lambda_1\Lambda_2]'  \Lambda_3 {\cal F}  \bigr]'  \ 
-  Q' [\, \Lambda_1\Lambda_2 \Lambda_3{\cal F} ]' 
- \bigl[ [\Lambda_1 \Lambda_2\Lambda_3 ]' {\cal F} \bigr]'  \,.
\ee
We can now write the right-hand side in terms of familiar
transformations and there is then no need to keep the string field explicitly:
\be
\label{dkfjheriu}
\sum_{\hbox{\tiny\rm cyc}}  
\bigl[ \delta_{\Lambda_3}\,,  \, \delta^{{}^T}_{\Lambda_1, \Lambda_2  }   \bigr] \,  \ = \  
- \sum_{\hbox{\tiny\rm cyc}} \delta^{{}^T}_{\  [\Lambda_1\Lambda_2]', \Lambda_3  } 
\ - \ \delta_{ [\, \Lambda_1\Lambda_2 \Lambda_3{\cal F} ]'   }
 \ - \ \delta^{{}^T}_{ [\Lambda_1 \Lambda_2\Lambda_3 ]' }\;. 
\ee
This interesting equation  can be thought of as part of the gauge algebra.
It gives the commutator of a standard gauge transformation and an equations-of-motion
transformation of the second type.  The answer is an equations-of-motion
transformation of the first type (last term),  an equations-of-motion
transformation of the second type (first term), and a middle term that could be thought of 
as a new, additional,  equations-of-motion
transformation.

Combining (\ref{dkfjheriu})  and (\ref{first_term})  we see the cancellation of two
equations-of-motion symmetries and two ordinary transformations, leaving:
\be
\sum_{\hbox{\tiny\rm cyc}}  \bigl[ \delta_{\Lambda_3}\,,  \, [  \delta_{\Lambda_2}\,, 
\, \delta_{\Lambda_1} ]  \bigr] \,    =  \,  -\delta_{\, 
Q' [\Lambda_1\Lambda_2\Lambda_3]'    } \,  -
\, \delta^{{}^T}_ {
 [\Lambda_1 \Lambda_2\Lambda_3 ]' } 
\, = \,   -\delta^{\,{}^E}_{Q'\chi, \,\chi } \,, \quad
\hbox{with} \ \  \chi \ = \   [\Lambda_1\Lambda_2\Lambda_3]' \,.   
\ee
The gauge Jacobiator is a null transformation and thus, as claimed, vanishes identically.
\hfill $\square$

\medskip

If a gauge algebra is field independent and closes
off-shell, as in the case of Courant brackets,
we have
\be
\label{jnhsbtfrgs}
\bigl[ \delta_{\Lambda_2} \,, \, \delta_{\Lambda_1} \bigr] 
\ = \  \delta_{[\Lambda_1\, \Lambda_2]} \,. 
\ee
This requires that  
\be
 [\Lambda_1 \Lambda_2 \Psi^n\, ] =  0  \,, \ n \geq 1\,, \quad \hbox{and}  
 \quad
  [\Lambda_1 \Lambda_2 {\cal F}  \Psi^n\, ] = 0 \,,  \  n\geq 0 \,,   
\ee
so that the extra terms in the gauge algebra (\ref{jnhwlvly}) vanish.
In this case the gauge Jacobiator is equal to a gauge transformation with parameter
equal to the standard Jacobiator:
\be
\begin{split}
{\cal J} \ = \ \sum_{\hbox{\tiny\rm cyc}}  \bigl[ \delta_{\Lambda_3}\,,  \, [  \delta_{\Lambda_2}\,, 
\, \delta_{\Lambda_1} ]  \bigr] \,   \ =  \ \sum_{\hbox{\tiny\rm cyc}}  \bigl[ \delta_{\Lambda_3}\,,  \,   \delta_{[\Lambda_1\, \Lambda_2]}   \bigr] \, 
 \ =  \   \delta_{\sum_{\hbox{\tiny\rm cyc}} [ \, [\Lambda_1\, \Lambda_2], \Lambda_3 ]  } \,  \,.
\end{split}
\ee
The third identity in (\ref{clksshrss}) then gives
\be
\label{clclt}
{\cal J} \ = \ -\delta_{ Q [\Lambda_1 \Lambda_2\Lambda_3]+  [Q\Lambda_1 \Lambda_2\Lambda_3]
-  [\Lambda_1 Q\Lambda_2\Lambda_3] +  [\Lambda_1 \Lambda_2 Q\Lambda_3]}\;. 
\ee
The transformation on the right must vanish on fields.  
The way this works (as will be seen
later in section 6.2) is that there are no 3-brackets between the field 
$Q\Lambda$ 
and two gauge parameters.  
Moreover, we also have 
\be
\delta_{Q \chi}  \ = \ 0 \,,
\ee
meaning that such gauge parameters simply generate no transformations. 
These two facts imply with (\ref{clclt}) that ${\cal J}=0$, as required by consistency.

\section{$L_{\infty}$ algebra in
$\ell$ picture and field theory}\setcounter{equation}{0}

In the previous section we reviewed the axioms of $L_\infty$ 
algebras, in the formulation where all products have degree minus one.
We will return to this briefly in a slightly different notation, with elements
$\tilde x_i$ and products written as
\be
[ \tilde x_1, \ldots \,, \tilde x_n ] \quad \to \quad  b_n ( \tilde x_1, \ldots \,, \tilde x_n)  \,. 
\ee  
We will call this the `$b$-picture' of the $L_\infty$ algebra.  The sign conventions
we have described in this  picture  result in a simple action, field equations, and gauge transformations.
The signs for the Jacobi-like identities, however, are a bit unfamiliar.
Shortly after the work in \cite{Zwiebach:1992ie}, 
the axioms of $L_\infty$ algebras  were presented in 
a different convention~\cite{Lada:1992wc} and later reviewed nicely in~\cite{Lada:1994mn}.
In this `$\ell$-picture', the products $\ell_n$ satisfy Jacobi-like identities with more
familiar signs.  The action, field equations, and gauge transformations, however,
have more intricate signs.    These two pictures of the $L_\infty$ algebra 
are related by suspension.

In this section we begin by stating the general identitites for products
in the $\ell$-picture.   Recalling the analogous definitions for the $b$-picture
we explain how suspension relates the two pictures.  We are then able 
to use the familiar $b$-picture results for the field equations, action, and
gauge transformations to obtain the corresponding formulae in the $\ell$-picture.
In the following sections all of our discussions and examples will
be stated in the  
$\ell$-picture.  
To set the stage for these examples,  in the final subsection
we discuss 
general features of field theories in this language.   
We show how to 
read large classes of products from the perturbative setup and 
identify large classes of $L_\infty$ identities that hold 
when the field equations are gauge covariant and gauge 
transformations close.

\subsection{$L_\infty$ algebra identities;  $\ell$-picture}

In an $L_\infty$ algebra we have a vector space $X$ graded by 
a degree:
\be
X = \bigoplus_{n}  X_n \,, \quad  n \in \mathbb{Z} \,. 
\ee
The elements of the vector space $X_n$ are said to be of degree $n$.
We use the notation $x_1, x_2, \ldots$ to denote arbitrary vectors in $X$,
but each one having definite degree;  each $x_k$ belongs to some space
$X_p$.    The degree enters in sign factors where,
for convenience, we omit the `deg' label.  Thus, for example: 
\be
 (-1)^{x_1 x_2}  \ \equiv \  (-1)^{{\small\hbox{deg} }(x_1) \cdot {\small\hbox{deg} } (x_2)}
 \,.  
\ee
In exponents, the degrees are relevant only mod 2.   

In an $L_\infty$ algebra we have multilinear products $\ell_1, \ell_2, \ell_3 , \cdots$.
The multilinear product $\ell_k$ is 
said to have degree $k-2$:
\be
\hbox{deg} \, \ell_k  \ = \ k-2 \,, 
\ee
 meaning that when acting on a collection of inputs we find
\be
\hbox{deg}  \, ( \ell_k (x_1 , \ldots , x_k) ) \ = \   k-2 + \sum_{i=1}^k  \hbox{deg} (x_i) \,. 
\ee 
Thus
 \be
 \hbox{deg} (\ell_1) \ = \   -1 \,, \quad
  \hbox{deg} (\ell_2) \ = \    \  0\,,  \quad
  \hbox{deg} (\ell_3) \ = \   \   1 \,,  \quad {\rm etc.}  
 \ee
The products are defined to be \textit{graded commutative}.  For $\ell_2$, for example, 
 \be
  \ell_2(x_1,x_2) \ = \  (-1)^{1+ x_1 x_2}\,  \ell_2(x_2,x_1)\;. 
   \ee
Note the extra sign added in the exponent, when compared to the $b$-picture formula
in section 2.  
More generally for any permutation $\sigma$ of $k$ labels we have
\be
\ell_k ( x_{\sigma(1)} , \ldots , x_{\sigma(k)} ) \ = \ (-1)^\sigma \epsilon(\sigma;x ) \, 
\ell_k (x_1 , \ldots \,, x_k) \,. 
\ee
Here $(-1)^\sigma$ gives a plus sign if the permutation is even and a minus sign if the permutation is odd.  The Koszul sign $\epsilon (\sigma;x )$ is defined by
considering a graded commutative algebra $\Lambda (x_1, x_2, \cdots )$ with
\be
x_i \wedge x_j \ = \ (-1)^{x_i x_j}   \,  x_j \wedge x_i \,,   \quad \forall i, j , 
\ee
and  reading its value from the relation
\be
 x_1\wedge \ldots  \wedge x_k   =  \epsilon (\sigma; x)  \   x_{\sigma(1)} \wedge \ldots   \wedge \, x_{\sigma(k)} \,. 
\ee

The $L_\infty$ identities given in $b$-language can be 
stated in $\ell$-language and are enumerated by
a positive integer $n$ ~\cite{Lada:1994mn}:
\be
\label{main-Linty-identity}
\sum_{i+j= n+1}  (-1)^{i(j-1)} \sum_\sigma  (-1)^\sigma \epsilon (\sigma; x) \, \ell_j \, \bigl( \, \ell_i ( x_{\sigma(1)}  \,, \, \ldots\,, x_{\sigma(i)} ) \,, \, x_{\sigma(i+1)}, \, \ldots \, x_{\sigma (n)} \bigr) \ = \ 0\,.
\ee
Here $n \geq 1$ is the number of inputs.   The sum over $\sigma$ is a sum over ``unshuffles'' meaning
that we restrict to permutations in which the arguments are partially ordered as follows
\be
\sigma(1) <  \, \cdots \, <  \, \sigma(i) \,,  \qquad 
\sigma(i+1) <  \, \cdots \, <  \, \sigma(n) \,.
\ee
Schematically, the identities are of the form 
\be
\label{main-Linty-identity-schem}
\sum_{i+j= n+1}  (-1)^{i(j-1)}  \ell_j \, \ell_i \ = \ 0\,.
\ee
For $n=1$ we have 
  \be\label{l1-identity}
  \ell_1 ( \ell_1 (x)) \ = \ 0 \,.
  \ee
  This means that the iterated action of $\ell_1$ gives zero. In string field theory
  $\ell_1$ is identified with the BRST operator. 
For $n=2$, the constraint is, schematically,  
  \be
  \ell_1 \ell_2 \ = \ \ell_2 \ell_1\,, 
  \ee
  and in detail it gives
  \be
  \label{L2L1}
 \ell_1(\ell_2(x_1,x_2)) \ = \ \ell_2(\ell_1(x_1),x_2) + (-1)^1
 (-1)^{x_1x_2}\ell_2(\ell_1(x_2), x_1)\;, 
 \ee
where $(-1)^1$ is the sign of $\sigma$ and  $(-1)^{x_1x_2}$ the Koszul sign. 
The arguments in the last term can be exchanged to find
  \be
  \label{L2L1}
 \ell_1(\ell_2(x_1,x_2)) \ = \ \ell_2(\ell_1(x_1),x_2) + (-1)^{x_1}\ell_2(x_1,\ell_1(x_2))\;. 
 \ee
We recognize this as the statement that $\ell_1$ is a derivation of the product $\ell_2$.
 The next identity arises for $n=3$,  
 \be
0 \ = \ \ell_1 \ell_3  + \ell_3 \ell_1  + \ell_2 \ell_2 \, ,
 \ee
 and explicitly reads:
 \be\label{L3L1}
  \begin{split}
0    & =  \ \ell_1(\ell_3 (x_1,x_2, x_3))  \\
&   + \ell_3(\ell_1 (x_1) ,x_2, x_3) 
  +  (-1)^{x_1}  \ell_3( x_1 ,\ell_1(x_2), x_3)
    +  (-1)^{x_1+ x_2}  \ell_3( x_1 ,x_2, \ell_1(x_3)) \\
 & + \ell_2(\ell_2(x_1,x_2),x_3) + (-1)^{(x_1+ x_2) x_3}\ell_2(\ell_2(x_3,x_1),x_2) 
   +(-1)^{(x_2+ x_3) x_1 }\ell_2(\ell_2(x_2,x_3),x_1) \;.  
  \end{split} 
 \ee
The first four terms on the above right-hand side quantify the failure of $\ell_1$ to
be a derivation of the product $\ell_3$.  The last three terms are the Jacobiator for a 
bracket defined by $\ell_2$.  The failure of $\ell_2$ to be a Lie bracket is thus related
to the existence of the higher product $\ell_3$. 

Let us consider one more identity.  For $n=4$ we get, schematically, 
\be
0\ = \ \ell_1 \ell_4  - \ell_2 \ell_3  + \ell_3 \ell_2 - \ell_4 \ell_1 \,. 
\ee
Explicitly we have
\be\label{ell4ell1}
\begin{split}
0 \ = \ & \  \ \ell_1 ( \, \ell_4 ( x_1, x_2, x_3, x_4))  \\[1.0ex]
& - \ell_2 ( \, \ell_3 (x_1, x_2, x_3) , x_4)  
+ \, (-1)^{x_3 x_4}\, \ell_2  ( \, \ell_3 (x_1, x_2, x_4) , x_3) 
 \\ 
 & + (-1)^{(1+x_1)x_2} \ell_2 (x_2, \ell_3 (x_1, x_3, x_4)) 
 \, - (-1)^{x_1} \ell_2 (x_1, \ell_3 (x_2, x_3, x_4) ) \\[1.5ex]
 & +  \ell_3 ( \ell_2 (x_1, x_2 ) , x_3, x_4)   
 \ + (-1)^{1 + x_2 x_3} \   \ell_3 ( \ell_2 (x_1, x_3 ) , x_2, x_4) \\
 &  +  (-1)^{x_4 (x_2 + x_3)} \ell_3 ( \ell_2 (x_1, x_4 ) , x_2, x_3)   
 \ - \ell_3 ( x_1, \ell_2 (x_2, x_3 ) ,  x_4)  \\
 &  + (-1)^{x_3x_4}\ell_3 (x_1,  \ell_2 (x_2, x_4 ) , x_3)  
\  + \ell_3 (x_1,  x_2, \ell_2 (x_3, x_4 ) )  \\[1.5ex]
&  - \ell_4  (\ell_1 (x_1), x_2, x_3, x_4) 
 \ - (-1)^{x_1}  \ell_4 ( x_1, \ell_1 (x_2), x_3, x_4)   \\
 & - (-1)^{x_1+x_2}  \ell_4 ( x_1, x_2, \ell_1 (x_3), x_4)
 \ - (-1)^{x_1+ x_2 + x_4}  \ell_4 ( x_1, x_2, x_3, \ell_1 (x_4))\,. 
 \end{split}
\ee
We now turn to the $b$-picture and the relation between the two pictures.

\subsection{From $b$-picture to $\ell$-picture}

In the $b$-picture of the $L_\infty$ algebra we have a vector space $\tilde X$ graded by 
a degree:
\be
\tilde X = \bigoplus_{n}  \tilde X_n \,, \quad  n \in \mathbb{Z} \,. 
\ee
The elements of the vector space $\tilde X_n$ are said to be of degree $n$.
We use the notation $\tilde x_1, \tilde x_2, \ldots$ to denote arbitrary 
fixed-degree vectors in $\tilde X$.
In the $b$ picture all products have degree minus one:
\be
\hbox{deg} \, b_n \ = \ -1 \,. 
\ee
As we have already explained all 
 products are  \textit{graded commutative}, with no additional factors:
 \be
  b_n(\ldots \,, \tilde x_i, \tilde x_j ,\, \ldots ) \ = \  (-1)^{\tilde x_i\tilde  x_j}\,   b_n( \ldots\,, \,  \tilde x_j, \tilde x_i ,\ldots )\;, 
   \ee
with exponents representing degrees.   The inner product is completely graded commutative
and has a simple exchange symmetry:
 \be
 \label{clmgnfcnttts}
 \begin{split}
 \langle  \tilde x_1\, , \,   b_n(\tilde x_2 \,, \ldots \, ,  \tilde x_n) \rangle \ = \ & \
   (-1)^{\tilde x_1\tilde  x_2}\,     \langle  \tilde x_2\, , \,   b_n(\tilde x_1 \,, \ldots \, ,  \tilde x_n) \rangle
   \\
\langle \tilde x_1 , \tilde x_2 \rangle \ = \ & \ (-1)^{(\tilde x_1 +1)(\tilde x_2+1)}  \langle \tilde x_2 , \tilde x_1 \rangle \,.  
 \end{split} \ee
It follows from these that 
\be
\langle b_1  (\tilde x_1) , \tilde x_2 \rangle =   (-1)^{\tilde x_1}  \langle \tilde x_1 , b_1 
(\tilde x_2) \rangle\, .
\ee
In this new notation, the $L_\infty$ identities are just a simple translation of those
given in (\ref{clksshrss}) and need not be repeated here. 
The gauge transformation and field equations, with $\Lambda \to \tilde \Lambda$ and 
the field denoted by $\tilde\Psi$, take the form (see (\ref{vmlvsagdlck}) and (\ref{calF-def}))   
\be
\begin{split}
\delta_{\tilde \Lambda} \tilde\Psi \ = & \ \ b_1 (\tilde\Lambda)  + b_2 (\tilde\Lambda, \tilde\Psi) + \tfrac{1}{2} b_3 ( \tilde\Lambda , \tilde\Psi, \tilde\Psi) + 
\tfrac{1}{3!} b_3 ( \tilde\Lambda, \tilde\Psi, \tilde\Psi, \tilde\Psi) + \ldots \,, \\
\tilde {\cal F} \ = & \ \  b_1 (\tilde\Psi) +  \tfrac{1}{2} b_2 ( \tilde\Psi, \tilde\Psi) + \tfrac{1}{3!} b_3 ( \tilde\Psi, \tilde\Psi, \tilde\Psi)
+ \ldots 
\end{split}
\ee
The degrees of the various vectors here are
\be
\label{vmmgnfcnttts}
\hbox{deg} \, \tilde \Lambda \ = \ 1 \,,   \quad \hbox{deg} \, \tilde \Psi \ = \ 0 ,   \quad \hbox{deg} \,
\tilde  {\cal F}  \ = \ -1\,. 
\ee

\medskip
\noindent
{\bf Suspension:}  Suspension is a map that starting with a graded vector space $X$ gives us a graded vector
space $\tilde X$.  Acting on $X_n$ suspension gives us the space $\tilde X_{n+1}$.  The map 
simply copies the vectors in $X_n$ into $\tilde X_{n+1}$.  The degree of the elements is
then `suspended', or increased by one unit.   To track properly the various vectors we will
write the suspension map as $s$ or sometimes as $\uparrow$ and say that
\be
\tilde x_i \ = \ s \, x_i  \  =  \ \uparrow\, x_i \,,  
\ee 
leading to
\be
\hbox{deg} \, \tilde x_i  \ = \  \hbox{deg} \, x_i  \ + \ 1 \,. 
\ee
The inverse map is well defined and we will write
\be
x_i \ = \ \downarrow \, \tilde x_i  \,. 
\ee
For gauge parameters, fields and field equations we write, 
\be
\tilde \Lambda =  s\,  \Lambda \ = \ \uparrow \Lambda \,,  \qquad  
\tilde \Psi =  s \, \Psi \ = \ \uparrow \Psi \,, \qquad  \tilde {\cal F} \ = \ s \, {\cal F} \ = \ \uparrow
\, {\cal F} \,. 
\ee
We note that given (\ref{vmmgnfcnttts}) we now have
\be
\label{vmglrsttts}
\hbox{deg} \,  \Lambda \ = \ 0 \,,   \quad \hbox{deg} \,  \Psi \ = \ -1 ,   \quad \hbox{deg} \, {\cal F}  \ = \ -2\,. 
\ee

The products in the two pictures are related as follows.  Up to a sign, 
$b_n (\tilde x_1, \ldots, \tilde x_n)$ is the same as $\ell_n (x_1, \ldots , x_n)$. 
As discussed in \cite{Gaberdiel:1997ia} and \cite{Lada:1994mn}, we  have
\be
\label{clncbtt}
 b_{n+1} ( \tilde x_1, \ldots \tilde x_{n+1}) \ = \ (-1)^{x_1 n + x_2 (n-1) + \ldots + x_n}\ 
 s\, \ell_{n+1} ( x_1, \ldots , x_{n+1} ) \,.
\ee
In the above, the values $x_1 , \ldots , x_n$ in exponents denote the degrees as elements of $X$.
Note that the degree of the right-hand side of (\ref{clncbtt}) is
\be
1 + ((n+1)-2) + \sum_{k=1}^{n+1}  \hbox{deg} \, x_k \ = \ - 1 +   \sum_{k=1}^{n+1}  (\hbox{deg} \,
x_k +1) \ = \ -1 +   \sum_{k=1}^{n+1}  \hbox{deg} \, \tilde x_k\,,
\ee
showing that (\ref{clncbtt}) is  consistent with the stated degrees of $\ell$ and $b$ products.
The first few cases of (\ref{clncbtt}) give 
\be
\begin{split}
b_1 (\tilde x) \ = \ &  s\, \ell_1 (x)\,, \\
b_2 (\tilde x_1, \tilde x_2) \ = \ &  (-1)^{x_1} \,  s\, \ell_2 (x_1, x_2)\,, \\
b_3 (\tilde x_1, \tilde x_2, \tilde x_3) \ = \ &  (-1)^{x_2} \,  s\, \ell_3 (x_1, x_2, x_3)\,, \\
b_4 (\tilde x_1, \tilde x_2, \tilde x_3, \tilde x_4) \ = \ &  (-1)^{x_1+x_3} \,  s\, \ell_4 (x_1, x_2, x_3, x_4)\,.\\
\end{split}
\ee
One can verify with some explicit computation that the Jacobi-like $\ell_n$ identities, upon suspension
become the corresponding $b_n$ identities. 

\medskip
\noindent
It follows from (\ref{clncbtt}), applied to a gauge parameter and $n$ fields,  that 
\be
 b_{n+1} ( \tilde \Lambda, \tilde \Psi^n) \ = \ (-1)^{0 n + (-1) (n-1) + (-1) (n-2) + \ldots + (-1)}\ 
s\, \ell_{n+1} ( \Lambda, \Psi^n ) \,.
\ee
Performing the sum in the exponent and applying $\downarrow$ we get 
\be
\downarrow b_{n+1} ( \tilde \Lambda, \tilde \Psi^n) \ = \ (-1)^{n(n-1)\over 2} \ 
\ell_{n+1} ( \Lambda, \Psi^n ) \,.
\ee
This formula allows us to translate the gauge transformations from the $b$ picture
to the $\ell$ picture.  Consider  
\be
\delta_{\tilde \Lambda} \tilde\Psi \ = \   \sum_{n=0}^\infty  {1\over n!}
b_{n+1} (\tilde \Lambda, \tilde \Psi^n)  \,.  
\ee
Applying $\downarrow$ to the gauge transformation above, 
\be
\downarrow \delta_{\tilde \Lambda} \tilde\Psi \ = \   \sum_{n=0}^\infty
{1\over n!}\   \downarrow\,  b_{n+1}  (\tilde \Lambda, \tilde \Psi^n)  \,, 
\ee
and therefore 
\be
\delta_\Lambda \Psi \  \equiv \ \downarrow \delta_{\tilde \Lambda} \tilde\Psi \ = \   \sum_{n=0}^\infty
{1\over n!}\    (-1)^{n(n-1)\over 2} \,   \ell_{n+1}  (\Lambda, \,  \Psi^n)  \,.  
\ee
Expanding, this gives a series whose signs alternate every two elements
 \be
 \label{jnsqts}
  \delta_{\Lambda}\Psi  \ = \  \ell_1(\Lambda) + \ell_2(\Lambda,\Psi) - \tfrac{1}{2} \ell_3(\Lambda, \Psi,\Psi)
  - \tfrac{1}{3!} \ell_4(\Lambda,\Psi,\Psi,\Psi)  + \ldots \;. 
 \ee 
 
Let us now consider the action.  For this we must consider the inner product. 
The identities for the inner product in the $\ell$ picture arise from the definition
of this inner product in terms of the $b$-picture inner product:
\be
\label{jnlvscm}
\langle x_1, x_2 \rangle \ \equiv  \ \langle \tilde x_1 , \tilde x_2 \rangle \,. 
\ee
Here, $x_1, x_2 \in X$ and, with a  
slight abuse of notation, the inner product on the right-hand side is
in the $b$-picture  and the inner product on the left-hand side is in the $\ell$ picture.
From the properties of the $b$-picture inner product (\ref{clmgnfcnttts}) and the above
definition 
we quickly derive the properties of the
$\ell$-picture inner product:
\be
\label{vmglrsbtt}
\begin{split}
\langle x , \, \ell_n (x_1, \ldots x_{n}) \rangle 
 \  =  \  &  (-1)^{x x_1  +1}  \langle x_1 , \, \ell_n (x, x_2  \ldots x_{n}) \rangle \,,  \\   
\langle x_1, x_2 \rangle \ = \ &    (-1)^{x_1 x_2 } \langle x_2 , x_2 \rangle \,. 
\end{split}
\ee
As we can see the inner product is totally graded symmetric, just as the products are.
A short computation shows that we also have:
\be
\begin{split}
\langle \ell_1(x_1), x_2 \rangle \ = \ &    (-1)^{x_1+1} \langle x_1 , \ell_1(x_2) \rangle \,,\\
\langle \ell_2 (x_1 , x_2) \,,  x_3) \rangle \ = \ & \ \langle x_1 , \, \ell_2 (x_2, x_3) \rangle\,. 
\end{split}
\ee

The translation of the action 
\be
S \ = \sum_{n=1}^\infty  {1\over (n+1)!} \, \langle \tilde \Psi,  b_n (\tilde \Psi^n) \, \rangle \,,
\ee
is done using (\ref{clncbtt}), which gives
\be
\downarrow b_{n} ( \tilde \Psi^{n}) \ = \ (-1)^{n(n-1)\over 2} \ 
\, \ell_{n} ( \Psi^{n} ) \,.
\ee
Indeed, together with (\ref{jnlvscm}) we have the closed form expression
for the action in the $\ell$ picture:
\be
S \ = \sum_{n=1}^\infty  {1\over (n+1)!} \, \langle  \Psi,  \downarrow b_n (\tilde \Psi^n) \, \rangle 
\ = \ \sum_{n=1}^\infty  {(-1)^{n(n-1)\over 2}\over (n+1)!} \, \langle  
\Psi,   \ell_n (\tilde \Psi^n) \, \rangle \,.
\ee
Again, if we expand we get alternating signs:
\be
S \ = \ \tfrac{1}{2} \langle \Psi, \ell_1 (\Psi)  -\tfrac{1}{3!} \langle \Psi, \ell_2 (\Psi^2) \rangle
 - \tfrac{1}{4!} \langle \Psi, \ell_3 (\Psi^3) \rangle  + \tfrac{1}{5!}\langle \Psi, \ell_4 (\Psi^4) \rangle + \ldots
\ee
The field equation takes the form
\be
\label{jngvmkss}
  {\cal F}(\Psi) \ = \ \sum_{n=1}^\infty  
   {(-1)^{n(n-1)\over 2}\over n!}  \ell_n(\Psi^n)\ = \ 
   \ell_1(\Psi)  - \tfrac{1}{2}\ell_2(\Psi^2) - \tfrac{1}{3!} \ell_3(\Psi^3)
  \, +\tfrac{1}{4!}\ell_4(\Psi^4)+\cdots\;. 
  \ee
The gauge transformation of the field equation can be translated starting
from (\ref{jnlvltts}) 
\be
\delta \tilde {\cal F} \ = \ [\, \tilde\Lambda \, \tilde {\cal F}\,  ] ' 
\ee
together with
\be
\downarrow  b_{n+2}  (\tilde \Lambda,\, \tilde {\cal F}\, , \tilde \Psi^n)\ = \ (-1)^{n(n-1)\over 2}  \, \ell_{n+2} (\Lambda, {\cal F} , \Psi^n) 
\ee
This leads to
\be
\label{clwsmbtt}
\delta_{\Lambda}{\cal F}(\Psi) \ = \   \ell_2(\Lambda,{\cal F})\,  +\,  \ell_3(\Lambda,{\cal F}(\Psi),\Psi)
  \, -\, \tfrac{1}{2}\ell_4(\Lambda,{\cal F}(\Psi), \Psi^2)+\cdots
\ee
This expresses the gauge covariance of the field equation in the $\ell$ picture.

For the gauge algebra we had (\ref{jnhwlvly}) stating that $\bigl[ \delta_{\tilde \Lambda_2} \,, \, \delta_{\tilde \Lambda_1} \bigr]$ is a gauge transformation with parameter
\be
\tilde \Lambda_{12} \ \equiv \  [\tilde \Lambda_1\, \tilde\Lambda_2]' \,,
\ee 
in addition to a trivial gauge transformation.  In the $\ell$ picture the commutator 
$\bigl[ \delta_{ \Lambda_2} \,, \, \delta_{ \Lambda_1} \bigr]$ is a gauge transformation
with parameter 
\be\label{lClosure}
\Lambda_{12} \ =  \  \ell_2 (\Lambda_1, \Lambda_2) + \ell_3 (\Lambda_1, \Lambda_2 , \Psi)
- \tfrac{1}{2} \ell_4 ( \Lambda_1, \Lambda_2 , \Psi, \Psi)  - \ldots\,,
\ee
with the by-now-familiar alternating signs.   This translation follows from the identity
\be
\downarrow \, b_{n+2} (\tilde \Lambda_1 , \tilde \Lambda_2, \tilde \Psi^n) \ = \ 
(-1)^{n(n-1)\over 2} \, \ell_{n+2} (\Lambda_1, \Lambda_2, \Psi^n) \,. 
\ee

\subsection{General remarks on the $L_\infty$ algebra of
field theories} 
 \label{fieldSEC}

Let us make a few general remarks  about the extraction of 
products from a gauge invariant perturbative field theory.

We will focus on the part of the theory dealing with gauge parameters,
fields, and field equations. 
We thus consider the graded vector space 
  \be\label{Hextendedvector}
  \begin{split}
  \ldots \ \ \longrightarrow\; & X_{0} 
\longrightarrow \; X_{-1} \longrightarrow \; X_{-2} \,. \\
   &\,  \Lambda\ \ \  \qquad \ \Psi
 \qquad \  \ \  \,  E  \end{split}
 \ee
The arrows are defined as the map $\ell_1$.    
We will assume that there are no spaces $X_{-d}$ with $d\geq 3$.
Recall that the field equations (\ref{jngvmkss}) take the form
\be
\ell_1 (\Psi) - \tfrac{1}{2!} \ell_2 (\Psi, \Psi) - \tfrac{1}{3!} \ell_2 (\Psi, \Psi, \Psi)  + \tfrac{1}{4!} \ell_4 (\Psi, \Psi,\Psi,\Psi) + \ldots   \ = \ 0   \;. 
\ee
It follows that knowledge of the field equations determines explicitly 
all products
\be
\ell_n ( \Psi, \,\ldots \,, \Psi)  \, \in \, X_{-2} \,,  \quad  n \geq 1\;, 
\ee
that involve fields.  Here all arguments are identical, but a general result
(a polarization identity) implies that a multilinear symmetric form is completely determined
by the values on the diagonal.  For example, defining $L_2 (\Psi) = \ell_2 (\Psi, \Psi)$ 
and $L_3(\Psi) =  \ell_3 (\Psi, \Psi, \Psi)$ we have
\be
\begin{split}
2\, 
\ell_2 (\Psi_1, \Psi_2) \ = \ & \   L_2 (\Psi_1 + \Psi_2 ) - L_2 (\Psi_1) - L_2 (\Psi_2)\;,   \\
3!\,
\ell_3 (\Psi_1, \Psi_2, \Psi_3) \ = \ & \  L_3 (\Psi_1 + \Psi_2+ \Psi_3 ) - L_3 (\Psi_1 + \Psi_2)
- L_3 (\Psi_1 + \Psi_3)
- L_3 (\Psi_2 + \Psi_3) \\
&   + L_3 (\Psi_1)+ L_3 (\Psi_2)+ L_3 (\Psi_3) \;.  \\
\end{split}
\ee
More generally, defining $L_n (\Psi) = \ell_n (\Psi, \, \ldots\,, \Psi)$ we have
\be
\begin{split}
n! \ell_n (\Psi_1, \, \ldots , \Psi_n) \ = \ &  \ \ L_n (\Psi_1 + \ldots \, + \Psi_n) \\
& - \bigl[  L_n (\Psi_1 + \ldots \, + \Psi_{n-1}) + \ldots  \bigr]    +  \bigl[ \ldots  \bigr] - \ldots  \\
& + (-1)^{n-k} [ L_n (\Psi_1 + \ldots \Psi_k) + \ldots ]   + \ldots\\
&   + (-1)^{n-1} [ L_n(\Psi_1) + \ldots
+ L_n (\Psi_n) ] \,.   
\end{split}
\ee
The pattern is clear. On the second line we subtract all terms with $L_n$ evaluated on the 
sum of fields leaving one out. 
As we proceed we alternate signs and leave  
out two, three, four, until we leave 
out all fields except one.  This shows we have determined completely the
multilinear products acting on arbitrary fields.

Consider now the $L_\infty$ identities acting on just fields.
The first is 
\be
\ell_1 ( \ell_1 (\Psi)) = 0 \;. 
\ee
Since $\ell_1 (\Psi)$ is an element $E$ of $X_{-2}$ we can satisfy this
constraint by setting 
\be
\ell_1 (E ) = 0 \,.
\ee
For the second identity we have
\be
\ell_1 (\ell_2 ( \Psi, \Psi) ) \,  = \,   2\,  \ell_2 ( \ell_1 (\Psi), \Psi) \,. 
\ee
The left-hand side is of the form $\ell_1 (E)$ and thus vanishes.
Thus the identity holds if we set
\be
\ell_2 ( E, \Psi) = 0\,.
\ee
An inductive argument shows that all $L_\infty$ identities acting
on fields are satisfied if we take
\be
\ell_{n+1} ( E, \Psi_1, \, \ldots\, , \Psi_n) \ = \ 0 \,,   \quad n = 0,1, \ldots 
\ee
This is not surprising, since all of the above are of degree $n+1-2 + (-2)  - n = -3$
and we have not introduced a space $X_{-3}$.  If we did, we could contemplate
setting some of these products to be nonzero:  for example setting $\ell_1 (E)$
to some value such that $\ell_1 ( \ell_1 (\Psi))=0$.

Let us now consider the gauge transformations.  From (\ref{jnsqts}) 
 \be
  \delta_{\xi}{\Psi}  \ = \  \ell_1(\Lambda ) + \ell_2(\Lambda,\Psi) - \tfrac{1}{2} \ell_3(\Lambda, \Psi,\Psi)
  - \tfrac{1}{3!} \ell_4(\Lambda,\Psi, \Psi,\Psi)  + \ldots \;,  
 \ee 
 we are now able to read off the products
 \be
 \ell_{n+1} (\Lambda, \Psi_1, \ldots \Psi_n) \in X_{-1} \,,   \qquad   n \geq 0 \,, 
 \ee
where we can use the polarization identities above to deduce the value
of the product for non-diagonal field entries. 

We can now examine the $L_\infty$ identities when we input a list
$(\Lambda, \Psi, \ldots, \Psi)$ of arguments.   The identity $\ell_1 (\ell_1 (\Lambda)) =0$
is nontrivial but must hold due to gauge invariance of the linearized  
field equation.
The next identity is
\be
\ell_1 (\ell_2 (\Lambda, \Psi))  \ = \ \ell_2 ( \ell_1 (\Lambda), \Psi)  + \ell_2 (\Lambda, \ell_1 (\Psi)) \;. 
\ee
The left-hand side is already determined and so is the first term on the right-hand side.
Thus this identity determines 
\be
\ell_2 ( \Lambda, E)  \, \in \, X_{-2} \;. 
\ee
The next identity can be seen to determine $\ell_3 ( \Lambda, E, \Psi)$.  All in all, 
the set of $L_\infty$ identities acting on $(\Lambda, \Psi, \ldots, \Psi)$ determine
the products
\be
\ell_{n+2} (\Lambda,  E, \Psi_1, \ldots \Psi_n) \in X_{-2} \,,   \qquad   n \geq 0 \,.  
\ee
The identities that lead to this determination are in fact the ones relevant to
the gauge covariance (\ref{jnlvlsckbltts})  
of the field equation.    We can now iterate this process
and consider the $L_\infty$ identities on a list $(\Lambda, E, \Psi, \ldots , \Psi)$.
This time this would lead us to consider products $\ell_{n+3} (\Lambda, E_1, E_2, \Psi, \ldots, \Psi)$.
But these products are all of degree minus three, and thus they vanish with the 
assumption that $X_{-3}$ does not exist. 

The gauge algebra commutator leads to the determination of the following products.
From the field-dependent gauge parameter we read
\be
\ell_{n+2} (\Lambda_1, \Lambda_2 , \Psi_1, \ldots \Psi_n) \in X_{0} \,,   \qquad   n \geq 0 \,.  
\ee
If we have only on-shell closure we then read 
\be
\ell_{n+3} (\Lambda_1, \Lambda_2 , E, \, \Psi_1, \ldots \Psi_n) \in X_{-1} \,,   \qquad   n \geq 0 \,.  
\ee
Using the $L_\infty$ identities for inputs of the form $(\Lambda_1, \Lambda_2, \Psi, \ldots , \Psi)$ 
we get constraints on the  products $\ell_{n+3} (\Lambda_1, \Lambda_2, E, \Psi_1 , \ldots \Psi_n) \in X_{-1}$  
determined by on-shell closure. 
By use of the identities for inputs of the form $(\Lambda_1, \Lambda_2,  E, \Psi, \ldots , \Psi)$ 
we can get information about products of the form $\ell_{n+4} (\Lambda_1, \Lambda_2, E_1, E_2, \Psi_1 , \ldots \Psi_n) \in X_{-2}$. 
Note that the products vanish on $\Lambda$ diagonals and on $E$ diagonals.

\medskip
We want to emphasize an important point.  We have seen in detail how
a consistent set of $L_\infty$ products leads to gauge transformations
under which the field equation transforms covariantly and to a gauge
algebra that closes.    We now want to explain that the reverse is true.
More precisely:
\begin{enumerate}

\item  If we have gauge transformations and 
gauge covariance properties of the field equations 
of a certain standard type, c.f.~(\ref{equall}) below, 
$L_\infty$ identities
acting on inputs 
$$(\Lambda, \Psi, \ldots), $$ with arbitrary numbers
of $\Psi$'s, are all satisfied.

\item  If we have gauge transformations of the standard type and
a standard-form gauge algebra, 
then the $L_\infty$ identities
acting on inputs 
$$(\Lambda_1, \Lambda_2,\Psi\ldots)\,, $$ 
 with arbitrary numbers
of $\Psi$'s, are all satisfied.

\end{enumerate}

Consider the first item above, and work for simplicity in the $b$ picture
where all signs are simple.  We recall the following equalities
\be\label{equall}
\begin{split}
 \delta_{\Lambda}\Psi  \ = \  &\, Q\Lambda + [\Lambda,\Psi] + \tfrac{1}{2} 
 [\Lambda, \Psi,\Psi]
  + \tfrac{1}{3!}  [\Lambda,\Psi,\Psi,\Psi]  + \ldots \;, \\
 {\cal F}(\Psi) \ = \ &   \, 
   Q\Psi  + \tfrac{1}{2} [\Psi, \Psi] + \tfrac{1}{3!} [\Psi, \Psi, \Psi]
  \, +\tfrac{1}{4!}[ \Psi, \Psi, \Psi, \Psi]+\cdots\;, \\
\delta_{\Lambda}{\cal F}(\Psi) \ = \  &\, [\Lambda,{\cal F}] \,  +\,  
[\Lambda,{\cal F},\Psi]
  \, +\, \tfrac{1}{2} [\Lambda,{\cal F}, \Psi, \Psi]+\cdots
\end{split}
\ee
The first equation is what we mean by standard
gauge transformations and the last one what we mean
by a standard-type field-equation covariance. 
Think of the first two equations as definitions. Then, we used some
subset of the $L_\infty$
identities to show that the last one holds.
  But in fact the last one holds
{\em if and only if} that subset of the $L_\infty$ identities hold.
The equation is checked in powers of $\Psi$, and for each power  $\Psi^n$ 
{\em one} $L_\infty$ identity is involved.
It is also clear, because ${\cal F}$ is a sum of 
products of fields, 
that the relevant $L_\infty$ identities are those with one $\Lambda$
and any number of $\Psi$'s.  

For the second item now consider
 the gauge algebra (\ref{jnhwlvly}) acting on a field, 
\be
\label{jnhwlvly77}
\bigl[ \delta_{\Lambda_2} \,, \, \delta_{\Lambda_1} \bigr]  \Psi
\ = \  \delta_{[\Lambda_1\, \Lambda_2]'} \Psi \ 
+ \  [\Lambda_1\Lambda_2 {\cal F}]' \,.
\ee
We call this a standard-form gauge algebra.   We have checked before 
that using the above gauge transformations the gauge algebra
above follows if the collection of $L_\infty$ identities 
that involve inputs $(\Lambda_1,\Lambda_2, \Psi, \ldots)$ hold.
In fact the gauge algebra holds {\em if and only if} those
$L_\infty$ identities hold.  Again,  equation
 (\ref{jnhwlvly77}) is checked in powers
of $\Psi$ and for each power  $\Psi^n$ 
one $L_\infty$ identity with inputs $(\Lambda_1, \Lambda_2, \Psi^n)$ 
is involved.

\medskip
The utility of the above remarks is that if we identify a perturbative
field theory in which we have standard  
gauge transformations, 
field-equation covariance, and gauge algebra, we are guaranteed
that the products  
that can be easily read off from those expressions
will satisfy large subsets of the $L_\infty$ identities.

\section{Non-abelian gauge theories and $L_{\infty}$ algebras}

In this section we formulate Yang-Mills-type gauge theories as $L_{\infty}$ 
algebras. In the first subsection we discuss the Yang-Mills gauge structure in this 
framework. As examples we then consider in the second subsection the dynamical 
theory based on the Chern-Simons action in three dimensions and, in the third subsection, 
the usual Yang-Mills theory that exists in arbitrary dimensions. 
Yang-Mills theories were first   
formulated as $L_{\infty}$ algebras in \cite{Zeitlin:2007vv,Zeitlin:2007fp,Zeitlin:2008cc}
using the BRST complex of open string field theory, which is larger than the complex
we use here.  

\subsection{Generalities on Yang-Mills theory}
Consider a Lie algebra ${\cal G}$ with generators $T_\alpha$:
\be
[ T_\alpha , T_\beta ] \ = \ f_{\alpha\beta}{}^\gamma \, T_\gamma\,,
\ee
where $f_{\alpha\beta}{}^\gamma$ are the structure constants.   We also consider
Lie algebra valued gauge fields $A_\mu (x) = A_\mu^\alpha (x) T_\alpha$ and gauge parameters
$\lambda (x) = \lambda^\alpha(x) T_\alpha$.  
The gauge field transformations are
\be
 \label{a-gauge-transformation}
  \delta_{\lambda}A_{\mu}{}^{\alpha} \ = \ \partial_{\mu}\lambda^{\alpha} +\big[ A_{\mu},\lambda\big]^{\alpha}\; \,, 
  \ee
 and they close according to the Lie algebra structure:
   \be
   \label{gauge-algebra}
  \big[\delta_{\lambda_1},\delta_{\lambda_2}\big] \ = \ \delta_{[\lambda_1,\lambda_2]}\;. 
 \ee
We also have the field strength
  \be
  \label{field-strength-gauge-field}
  F_{\mu\nu}{}^{\alpha} \ = \ \partial_{\mu}A_{\nu}{}^{\alpha}-\partial_{\nu}A_{\mu}{}^{\alpha}
  +\big[A_{\mu},A_{\nu}\big]^{\alpha}\;,  
 \ee
 that transforms covariantly under gauge transformations: 
 \be
 \label{f-gauge-transformation}
   \delta_{\lambda}F_{\mu\nu} \ = \ \big[F_{\mu\nu},\lambda\big]\;. 
 \ee
  Our goal is now to determine the appropriate $L_\infty$ algebra
 for Chern-Simons theory in 3D and for Yang-Mills theory in arbitrary dimensions. 
  For both of these cases the total graded vector space $X$ will be taken to contain three spaces of fixed degrees:  
 \be
  \begin{split}
   &X_0\qquad   X_{-1}\qquad\;\; X_{-2} \\
   &\lambda^{\alpha}\qquad A_{\mu}{}^{\alpha}\ \qquad E_{\mu}{}^{\alpha}
  \end{split}
 \ee
 The gauge parameters $\lambda$ are of degree zero,  
gauge fields $A$  have degree minus one, and equations of motion $E$ have degree minus two.   We write this as
 \be
 \hbox{deg} (\lambda) \ = \ 0\;, \qquad 
  \hbox{deg} (A) \ = \ -1\;, \qquad  \hbox{deg} (E)  \ = \ -2\;. 
 \ee
Recalling that
  $\ell_2(x_1,x_2) \ = \  (-1)^{1+ x_1 x_2}\,  \ell_2(x_2,x_1)$
 we have 
that $\ell_2$ is antisymmetric for gauge parameters, as it befits
a Lie algebra, and symmetric for fields, as it befits the interactions of a bosonic 
field. 

We define the inner product that is non-vanishing only when the total
degree is minus three:    
 \be\label{INNER0}
  \begin{split}
   \langle A, E\rangle \ \equiv  \ 
   \int \, dx\ \kappa_{\alpha\beta}\, \eta^{\mu\nu} A_{\mu}{}^{\alpha} (x) \, E_{\nu}{}^{\beta}(x) \;, \\
  \end{split}
 \ee  
where $\kappa_{\alpha\beta}$ is the Cartan-Killing form and $\eta_{\mu\nu}$ a fixed spacetime metric (say the Minkowski metric) and we include the integration over spacetime, as the inner 
product is supposed to give a number.   

The homotopy Lie algebra implies an infinite number of identities. 
Of course, for polynomial gauge theories we only need to check a finite number of them.
Here is a table of the identities, ordered by total degree of the identity, and showing
 the degrees of total inputs that must be checked given the relevant complex exists at degree zero, minus one
 and minus two. 
 
 \be
 \label{table-gauge-products}
 \begin{split}
 \hbox{deg} = -2\,,  \quad \ell_1 \ell_1 = 0, \qquad    &
 \begin{cases}  \hbox{deg}=0 :\   \lambda  
 \end{cases} \\
 \hbox{deg} = -1\,,   \quad \ell_1 \ell_2 - \ell_2 \ell_1= 0 , \qquad 
 &\begin{cases}  
 \hbox{deg} = \ 0: \  \lambda \lambda \\  \hbox{deg} = -1: \ \lambda A 
 \end{cases} \\
 \hbox{deg} = 0 \,,  \quad \ell_3 \ell_1 + \ell_2 \ell_2  + \ell_1 \ell_3 = 0, \quad 
  &\begin{cases}  
 \hbox{deg} = \ 0:  \  \lambda \lambda \lambda \\  \hbox{deg} = -1:   \ \lambda \lambda A 
 \\  \hbox{deg} = -2 : \ \lambda A A ,  \lambda \lambda E 
 \end{cases} \\ 
 \hbox{deg} = 1 \,,  \quad \ell_1 \ell_4 - \ell_2 \ell_3 + \ell_3 \ell_2  -\ell_4 \ell_1 = 0, \quad 
  &\begin{cases}  
 \hbox{deg} = \ -1 : \  \lambda \lambda \lambda A \\  
 \hbox{deg} = -2  : \ \lambda \lambda AA , \   \lambda\lambda \lambda E \\  
 \hbox{deg} = -3  : \ \lambda A A A ,  \ \lambda \lambda AE 
 \end{cases} \\ 
  \hbox{deg} = 2 \,,  \quad \ell_1 \ell_5  \pm  \ell_2 \ell_4 \pm \ell_3 \ell_3 \pm \ldots   = 0, \quad \quad 
  &\begin{cases}  \ldots
\end{cases}
 \end{split}
 \ee
 For Chern-Simons theory there are only $\ell_1$ and $\ell_2$ products and thus just
 the first three identities must be checked.  Yang-Mills theory has also an $\ell_3$ and 
 thus all identities above must be checked.  As we will see, the last one ends up holding
 trivially, so we did not include the various subcases above. 

Since the gauge structure is the same for Chern-Simons and Yang-Mills theories, we 
can read off some of the basic products.   Comparing the gauge transformation (\ref{a-gauge-transformation})
with the expression
\be
\label{deltaAvm}
\delta_\lambda A  \ = \ \ell_1 (\lambda)  + \ell_2 (\lambda , A)  + \ldots
\ee
we infer:
 \be 
  \begin{split} 
    \ell_1(\lambda) \ & \ = \ \ \ \partial_{\mu}\lambda \ \ \  \in \ X_{-1} \,,\\
    \ell_2(\lambda, A) \ & \ 
 =  \  \big[A,\lambda\big]  \ \ \in  \ X_{-1} \, . 
  \end{split}
 \ee     
All products involving a gauge parameter and two or more fields vanish. 
We can write the indices in these equations explicitly 
\be 
  \begin{split} 
    [\ell_1(\lambda)]_{\mu}{}^{\alpha} \ & \ = \  \ \ \partial_{\mu}\lambda^{\alpha} 
    \ \ \ \in \ X_{-1} \,, \\
    [\ell_2(\lambda, A)]_{\mu}{}^{\alpha} \ & \ 
 =  \  \big[A_{\mu},\lambda\big]^{\alpha} \in  \ X_{-1}\,.  
   \end{split}
 \ee     
Note
that to comply with the graded commutativity we must also define
\be
 \ell_2( A, \lambda ) \ \equiv \ -  \ell_2(\lambda, A) \ = \ -  [ A, \lambda ]  \,. 
\ee
We can now use the gauge algebra to identify the product $\ell_2$ acting on two 
gauge parameters.  From (\ref{deltaAvm}) we quickly find that
\be
\begin{split}
[ \delta_{\lambda_1} , \delta_{\lambda_2} ] \, A \ = \ & \ 
 \delta_{\lambda_1}  ( \ell_1(\lambda_2) + \ell_2 
(\lambda_2 , A) ) - (1 \leftrightarrow 2) \\
\ = \ &  \ \ell_2 (\lambda_2 , \delta_{\lambda_1}A)  - \ell_2 (\lambda_1 , \delta_{\lambda_2}A)  \\
\ = \ &  \ \ell_2 (\lambda_2 , \ell_1(\lambda_1))  - \ell_2 (\lambda_1 , \ell_1(\lambda_2) ) + {\cal O} (A)  \\
\ = \ &   - \ell_2 ( \ell_1(\lambda_1), \lambda_2)  - \ell_2 (\lambda_1 , \ell_1(\lambda_2) ) 
+ {\cal O} (A) 
\end{split}
\ee
We now use the $\ell_1 \ell_2 = \ell_2 \ell_1$ identity to identify the gauge transformation
on the right-hand side:
\be
[ \delta_{\lambda_1} , \delta_{\lambda_2} ] \, A \ = \    \ell_1 (-\ell_2 (\lambda_1, \lambda_2)  )
+ {\cal O} (A) \ = \ \delta_{- \ell_2 (\lambda_1, \lambda_2)}  A \,.    
\ee
The $A$ dependent terms on the right-hand side are not needed for the identification.
Comparing with (\ref{gauge-algebra}) we infer 
\be
\ell_2(\lambda_1,\lambda_2) \ \  = \  - \big[\lambda_1,\lambda_2\big] \ \in  \ X_0\;. 
\ee

\subsection{Chern-Simons Theory}

We now turn to the Chern-Simons theory. In order to define an action we have to assume 
that for the Lie algebra there exists an invariant inner product. 
We write for this inner product
of Lie algebra valued objects  
 \be\label{INNER}
   \ll A_\mu, B_\nu \rr \ = \ \kappa_{\alpha\beta}\, A_{\mu}{}^{\alpha} B_{\nu}{}^{\beta} \,,
 \ee  
where $\kappa_{\alpha\beta}$ is the Cartan-Killing metric.  
With this definition, the full inner product (\ref{INNER0}) becomes 
 \be\label{IntInner}
  \langle A, E\rangle \ = \  \int {\rm d}^3x  \, \eta^{\mu\nu} \, \ll  A_\mu \,, E_\nu \rr  \; . 
 \ee 
Consider now the gauge invariant 3D Chern-Simons action    
  \be
  \label{CS-gauge-invariant}
   S \ = \ \tfrac{1}{2}\int {\rm d}^3x \,\varepsilon^{\mu\nu\rho} \ll A_{\mu},\partial_{\nu}A_{\rho}+\tfrac{1}{3}
   [A_{\nu},A_{\rho}]\rr  \, .  
  \ee
The Chern-Simons action is topological and hence does not depend on the spacetime metric.   
The general variation
of the action is given by 
 \be
 \begin{split}
   \delta S \ = \ & \ \tfrac{1}{2}\int {\rm d}^3x \,\ll\, 
   \delta A_{\mu}\, , \, \varepsilon^{\mu\nu\rho}  ( 
   \partial_{\nu}A_{\rho}-\partial_{\rho}A_{\nu} +\, 
   [A_{\nu},A_{\rho}]) \, \rr  \, \\
    \ = \ & \ \int {\rm d}^3x \, \,  \eta^{\mu\sigma} \ll\, 
   \delta A_{\mu}\, , \, \varepsilon_\sigma{}^{\nu\rho}  ( 
   \partial_\nu A_\rho +\, \tfrac{1}{2} 
   [A_\nu\, ,\, A_\rho]) \, \rr  \, \\
    \ = \ & \ \langle \, 
   \delta A\, , \, \varepsilon_*{}  ( 
   \partial A +\, \tfrac{1}{2} 
   [A,A]) \, \rangle  \,, \\
   \end{split} 
  \ee
where the star denotes the position of the free index on the epsilon symbol
and we used the definition of the inner product.
Comparing with the expected form of the field equation, 
\be
\ell_1 (A ) - \tfrac{1}{2} \, \ell_2 (A , A) \ = \ 0\,,
\ee
we get
 \be\label{CSproducts}
  \begin{split}
   [\ell_1(A)]_{\mu}{}^{\alpha} \ &= \ \varepsilon_{\mu}{}^{\nu\rho}\,\partial_{\nu}A_{\rho}{}^{\alpha} \ \in \ X_{-2}\;, \\
   [\ell_2(A_1,A_2)]_{\mu}{}^{\alpha} \ &= - \varepsilon_{\mu}{}^{\nu\rho} \,\big[A_{1\nu},A_{2\rho}\big]^{\alpha} \ \in \ X_{-2}
   \,.
  \end{split}
 \ee 
 In index free notation we would write
  \be\label{CSproductsIF}
  \begin{split}
   \ell_1(A) \ &= \ \varepsilon_{*}\partial A\ \in \ X_{-2}\;, \\
   \ell_2(A_1,A_2)\ &=  -\varepsilon_{*}\,\big[A_{1},A_{2}\big] \ \in \ X_{-2}\,.
  \end{split}
 \ee
As expected  $\ell_2$ is \textit{symmetric} under the exchange of gauge fields.\footnote{This is the first instance where we derive the general product starting with the product evaluated on diagonals.}  
Note that the inner product in (\ref{IntInner}) 
now  contains the 
spacetime metric, which is also used in $\ell_1(A)$ to lower the index on the epsilon tensor. 
Thus, the $L_n$ formulation obscures the topological nature of the Chern-Simons action, but 
that is unavoidable if we have spaces $X_1$ and $X_2$ with the 
same index structure.
 
We now confirm that the action has the expected form 
 \be\label{LnCS}
  S \ = \  \tfrac{1}{2}\langle A,\ell_1(A)\rangle -\tfrac{1}{3!}\langle A,\ell_2(A,A)\rangle \ = \ 
 \langle A\, ,\ \tfrac{1}{2}\ell_1(A)\, -\tfrac{1}{3!}\ell_2(A,A)  \ \rangle  \,.    
 \ee
 The Chern-Simons action given above can be written as   
 \be
 \begin{split}
   S \ = \ & \  \int {\rm d}^3x \, \ll\,  A_{\mu},\,  \tfrac{1}{2}\varepsilon^{\mu\nu\rho}
   \partial_{\nu}A_{\rho}+\tfrac{1}{3!}\, \varepsilon^{\mu\nu\rho}
   [A_{\nu},A_{\rho}] \ \rr  \,, \\
    \ = \ & \  \langle \,  A,\,  \tfrac{1}{2}\varepsilon_*
   \partial A\, +\tfrac{1}{3!}\, \varepsilon_*
   [A,A] \ \rangle  \,.  \end{split}   
  \ee
Comparing with (\ref{CSproductsIF}) we see that the action is indeed correctly reproduced.

\medskip
\noindent
Let us  verify the $L_n$ axioms. 

\noindent
{\bf Checking 
$\ell_1 \ell_1 = 0$}.  This is only nontrivial at degree zero.  Indeed, we have
 \be
  [\ell_1(\ell_1(\lambda))]_{\mu}{}^{\alpha} \ = \ 
  \varepsilon_{\mu}{}^{\nu\rho}\,\partial_{\nu}[\ell_1(\lambda)]_{\rho}{}^{\alpha} \ = \ 
   \varepsilon_{\mu}{}^{\nu\rho}\,\partial_{\nu}\partial_{\rho}\lambda^{\alpha} \ = \ 0\;. 
 \ee
This is just linearized gauge invariance. 

\noindent
{\bf Checking 
$\ell_1 \ell_2 = \ell_2 \ell_1$}.    This means checking (\ref{L2L1}) at degree zero and minus one.  

\noindent
\underline{Degree zero}.  At this degree we must act on two gauge parameters:
 \be\label{DEGREE0}
  \ell_1(\ell_2(\lambda_1,\lambda_2)) \ = \  \ell_2(\ell_1(\lambda_1),\lambda_2) \ +  \  \ell_2(\lambda_1,\ell_1(\lambda_2))  \;. 
 \ee
 This gives
\be
\begin{split}
  -\partial ( [\lambda_1,\lambda_2]) \ = \ &  \  \ell_2(\partial \lambda_1,\lambda_2) \ +  \  \ell_2(\lambda_1,\partial \lambda_2)\\
 \ = \ &   - [\partial \lambda_1,\, \lambda_2]  \  - \  [  \lambda_1, \partial \lambda_2 ]\,,
 \end{split}
 \ee
which works out correctly.

\medskip 
\noindent
\underline{Degree minus one}.  We must verify
 \be
  \ell_1(\ell_2(A,\lambda)) \ = \  \ell_2(\ell_1(A),\lambda) \ -  \  \ell_2(A,\ell_1(\lambda))  \;. 
 \ee
We then have that the left-hand side is
 \be\label{ell1ell2}
 \begin{split}
  [\ell_1(\ell_2(A,\lambda))]_{\mu}{}^{\alpha} \ &= \ \varepsilon_{\mu}{}^{\nu\rho}\,
  \partial_{\nu}[\ell_2(A,\lambda)]_{\rho}{}^{\alpha} \ = \ -
   \varepsilon_{\mu}{}^{\nu\rho}\,\partial_{\nu}\big[A_{\rho},\lambda\big]^{\alpha} \\
  \ &= - 
   \varepsilon_{\mu}{}^{\nu\rho}\,\big[\partial_{\nu}A_{\rho},\lambda\big]^{\alpha}
  - \varepsilon_{\mu}{}^{\nu\rho}\,\big[A_{\rho},\partial_{\nu}\lambda\big]^{\alpha}\\
  \ &= \ 
  \bigl( -\big[\varepsilon_*\partial A,\lambda\big]
  + \varepsilon_*\,\big[A,\partial\lambda\big]\bigr)_\mu{}^{\alpha}  \,.   
 \end{split}
 \ee  
The right-hand side is
\be
\ell_2(\varepsilon_* \partial A ,\lambda) \ -  \  \ell_2(A,\partial \lambda) 
\ = \ \ell_2(\varepsilon_* \partial A ,\lambda) \ +  \  \varepsilon_* [ A,\partial \lambda] \;. 
\ee
In order for this to agree with the left-hand side we have to define 
for $E\in X_{-2}$, $\lambda \in X_0$
 \be 
 \label{e-lambda}
   \ell_2(E,\lambda) \ = \, - \big[E,\lambda\big] \ \in \ X_{-2}\;. 
 \ee

\noindent
{\bf Checking 
$\ell_3\ell_1+\ell_1\ell_3+\ell_2\ell_2=0$}.   Since $\ell_3$ is assumed to be zero
 this means checking that $\ell_2 \ell_2=0$.  From (\ref{L3L1}) this requires that 
  \be
  \label{L3L1-reduced}
 \ell_2(\ell_2(x_1,x_2),x_3) + (-1)^{(x_1+ x_2) x_3}\ell_2(\ell_2(x_3,x_1),x_2) 
   +(-1)^{(x_2+ x_3) x_1 }\ell_2(\ell_2(x_2,x_3),x_1) \, = \, 0\,.   
 \ee
As indicated in our table this identity can only be nontrivial acting 
on elements whose total degree equals zero, minus one, or minus two.

\noindent
\underline{Degree zero}.  We must act on three gauge parameters.
Since they have degree zero, we have
\be
  \label{L3L1-reduced-lambda}  
 \ell_2(\ell_2(\lambda_1,\lambda_2),\lambda_3) 
 +\ell_2(\ell_2(\lambda_3,\lambda_1),\lambda_2) 
   +\ell_2(\ell_2(\lambda_2,\lambda_3),\lambda_1) \ = \ 0   \;. 
 \ee
Using $\ell_2 (\lambda_1, \lambda_2) = -[\lambda_1 \,, \lambda_2] $ we obtain 
 \be
  \label{L3L1-reduced-lambda-vm}  
 -\ell_2([\lambda_1,\lambda_2],\lambda_3) 
 -\ell_2([\lambda_3,\lambda_1],\lambda_2) 
   -\ell_2([\lambda_2,\lambda_3],\lambda_1) \ = \ 0   \;. 
 \ee
 Since the bracket is another gauge parameter, we use the same expression
 for $\ell_2$ to see that we must have
 \be
  \label{L3L1-reduced-lambda-vmbb}  
 [[\lambda_1,\lambda_2],\lambda_3] 
 +[[\lambda_3,\lambda_1],\lambda_2] 
   +[[\lambda_2,\lambda_3],\lambda_1] \ = \ 0   \,. 
 \ee
This holds because ${\cal G}$ is a  Lie-algebra.
   
\noindent
\underline{Degree minus one}.   Here we have two gauge parameters
and one gauge field 
\be
  \label{L3L1-reduced-vmk}
 \ell_2(\ell_2(\lambda_1,\lambda_2),A) + \ell_2(\ell_2 (A,\lambda_1),\lambda_2) 
   +\, \ell_2( \ell_2 (\lambda_2, A),\lambda_1) \ = \ 0   \,. 
 \ee
Again, first replacing the nested in products   
\be
  \label{L3L1-reduced-vmks}
- \ell_2([ \lambda_1,\lambda_2] ,A) - \ell_2([A,\lambda_1],\lambda_2) 
   -\, \ell_2([\lambda_2, A],\lambda_1) \ = \ 0  \,.  
 \ee
Since $[A, \lambda]\in X_{-1}$ we can now take 
\be
  \label{L3L1-reduced-vmkss}
 [[ \lambda_1,\lambda_2] ,A] + [[A,\lambda_1],\lambda_2] 
   +\, [[\lambda_2, A],\lambda_1] \ = \ 0  \,,  
 \ee
 which holds by virtue of the Jacobi identity of the Lie algebra ${\cal G}$.
       
\noindent
\underline{Degree minus two }. We now act on two gauge fields
and one gauge parameter  $(AA\lambda)$ or two gauge parameters and
one field equation $(\lambda\lambda E)$.  First, for  the former we have 
  \be\label{L2Jacobi}
  \ell_2(\ell_2(A_1,A_2),\lambda) + \ell_2(\ell_2(\lambda,A_1), A_2) - \ell_2(\ell_2(A_2,\lambda),A_1) \ = \ 0\;, 
 \ee
and we compute 
  \be\label{L2Jacobi99}
  -\ell_2(\varepsilon_*[A_1, A_2],\lambda) - \ell_2([\lambda,A_1], A_2) + \ell_2([A_2,\lambda],A_1) \ = \ 0\;. 
 \ee
For the first term we use (\ref{e-lambda}) 
  \be\label{L2Jacobi999}
  \varepsilon_* [ [A_1, A_2],\lambda] + \varepsilon_*[ [\lambda,A_1], A_2]  - \varepsilon_* [[A_2,\lambda],A_1] \ = \ 0\;. 
 \ee
In the first two terms the second and third indices in $\varepsilon$ are
contracted with $A_1$ and $A_2$ respectively.  Not so in the third, so we
can factor out $\varepsilon$ by changing the sign of the last term:
  \be\label{L2Jacobi9999}
  \varepsilon_*\bigl(  [ [A_1, A_2],\lambda] +[ [\lambda,A_1], A_2]  +
   [[A_2,\lambda],A_1] \bigr) \ = \ 0\;. 
 \ee
This holds on account of the Jacobi identity of ${\cal G}$.

\noindent  
Now for the second case   ($\lambda \lambda E$) we have 
\be
\ell_2 (\ell_2 (\lambda_1, \lambda_2), E) 
+ \ell_2 (\ell_2 (E, \lambda_1)  \lambda_2)
 + \ell_2 (\ell_2 (\lambda_2, E)  \lambda_1) \ = \ 0 \,. 
\ee 
This simply gives    
\be
[[ \lambda_1, \lambda_2], E] 
+ [[E, \lambda_1],   \lambda_2]
 + [[ \lambda_2, E],   \lambda_1] \ = \ 0 \,,
\ee 
which again holds by the Jacobi identity. 

\medskip

With all checks done, we list the complete set of {\em nonvanishing}  $L_2$ products: 
  \be\label{Lnproducts}
\boxed{  \begin{split}
\hbox{Chern-Simons:} \qquad   \ell_1(\lambda) \ &= \ \partial\lambda\ \in \ X_{-1} \\
      \ell_1(A)\ &= \ \varepsilon_*\,\partial A \ \in \ X_{-2} \\
    \ell_2(\lambda_1,\lambda_2) \ &= \ -\big[\lambda_1,\lambda_2\big] \ \in  \ X_0\\
    \ell_2(A,\lambda) \ &= \ -\big[A,\lambda\big]\ \in  \ X_{-1} \\
   \ell_2(A_1,A_2) \ &= \ -\varepsilon_*\,\big[A_{1},A_{2 }\big] \ \in \ X_{-2} \ \ \\
   \ell_2(E,\lambda) \ &= \ -\big[E,\lambda\big] \ \in \ X_{-2}\, . 
  \end{split}}
 \ee    
The versions with explicit indices were given above.

 With the identities one can verify that the field equations transform covariantly, 
 \be
  \delta_{\lambda}(\ell_1(A)-\tfrac{1}{2}\ell_2(A,A)) \ = \ \ell_2(\ell_1(A)-\tfrac{1}{2}\ell_2(A,A),\lambda)\,, 
 \ee
which is the correct covariant transformation.

In order to compute the field equations and check gauge invariance 
we need the invariance properties of the inner product: Assuming that we 
can integrate by parts under the integral implicit in the inner product (\ref{IntInner}), we have 
for $A,B\in X_{-1}$
 \be
  \langle A,\ell_1(B)\rangle \ = \ \langle \ell_1(A),B\rangle \;. 
\ee  
Moreover, for $A,B,C\in X_{-1}$ we have explicitly 
 \be
  \langle A , \ell_2(B,C)\rangle \ = \ \int {\rm d}^3 x \,\varepsilon^{\mu\nu\rho}\,\kappa\big(A_{\mu},
  \big[B_{\nu},C_{\rho}\big]\big) \;. 
 \ee 
The invariance of the Cartan-Killing form then implies cyclicity, i.e., 
 \be
  \langle A,\ell_2(B,C)\rangle \ = \ \langle C,\ell_2(B,A)\rangle\;, \quad {\rm etc.}
 \ee   
The general variation of the Chern-Simons action is then 
 \be
 \begin{split}
  \delta S \ &= \ \tfrac{1}{2}\langle \delta A , \ell_1(A) \rangle  + \tfrac{1}{2} \langle A, \ell_1(\delta A) \rangle 
  + \tfrac{1}{3!} \langle\delta A, \ell_2(A,A) \rangle + \tfrac{2}{3!} \langle A, \ell_2(A,\delta A)  \rangle   \\
  \ &= \ \langle \delta A , \ell_1(A) \rangle  + 
  \tfrac{1}{3!} \langle\delta A, \ell_2(A,A) \rangle + \tfrac{2}{3!} \langle \delta A, \ell_2(A,A)  \rangle \\
  \ &= \ \langle \delta A, \ell_1(A)+\tfrac{1}{2} \ell_2(A,A)\rangle\;, 
 \end{split}
 \ee
 implying the correct field equation.

 \subsection{Yang-Mills theory}
 
We now turn to the dynamical Yang-Mills theory, for which we 
keep the general conventions for Yang-Mills  gauge transformations as above. Consider the Yang-Mills Lagrangian 
and its expansion in powers of the gauge field:
 \be
 \begin{split}
  {\cal L} \ &= \ -\tfrac{1}{4}\langle F^{\mu\nu},F_{\mu\nu}\rangle  \\
  \ &= \ \tfrac{1}{2}
  \langle A^{\mu},\partial^{\nu}(\partial_{\nu}A_{\mu}-\partial_{\mu}A_{\nu})\rangle 
\   - \langle \partial^{\mu}A^{\nu},[A_{\mu},A_{\nu}]\rangle
 \  -\tfrac{1}{4}\langle [A^{\mu},A^{\nu}],[A_{\mu},A_{\nu}]\rangle \,.
 \end{split}  
 \ee    
To derive a few of the products we consider the field equations: 
  \be
  \begin{split}
   0 \  & = \ D^{\mu}F_{\mu\nu} \ = \ \partial^{\mu}F_{\mu\nu}+[A^{\mu},F_{\mu\nu}] \\
   \ &= \ \partial^{\mu}(\partial_{\mu}A_{\nu}-\partial_{\nu}A_{\mu}+[A_{\mu},A_{\nu}])
   +[A^{\mu},\partial_{\mu}A_{\nu}-\partial_{\nu}A_{\mu}+[A_{\mu},A_{\nu}]] \\
    \ &= \ \square A_{\nu}-\partial_{\nu} \, \partial \cdot A\ \  
    +\partial^\mu [A_{\mu},A_{\nu}]
   +[A^{\mu},\partial_{\mu}A_{\nu}-\partial_{\nu}A_{\mu}] \ \ +[ A^\mu\,, [A_{\mu},A_{\nu}]]
   \,.
  \end{split}
 \ee  
We now compare with the expectation for the gauge transformations and the equations of motion 
 \be
 \begin{split}
 \delta_{\lambda}A \ &= \ \ell_1(\lambda) + \ell_2(\lambda, A)\;, \\
  {\cal F}(A) \ &\equiv \ \ell_1(A)-\tfrac{1}{2}\ell_2(A,A)-\tfrac{1}{3!}\ell_3(A,A,A) \ = \ 0 \;, 
 \end{split}
 \ee     
and we read off
 \be
 \begin{split}
 \ell_1(A) \ = & \ \ \square A-\partial(\partial\cdot A) \,, \\ 
  [\ell_2(A_1,A_2)]_{\mu} 
  \ =  & \ \ -\partial^{\nu}[A_{1\nu}, A_{2\mu}] - [\partial_{\mu}A_{1\nu}-\partial_{\nu}A_{ 1 \mu}, A_{2}^{\nu}]
  +(1\leftrightarrow 2)\,,  \\
  \ell_3(A_1,A_2,A_3)_{\mu} \ = & \ 
  -[A_1^{\nu},[A_{2\nu},A_{3\mu}]]  - [A_2^{\nu},[A_{3\nu},A_{1\mu}]]
   -[A_3^{\nu},[A_{1\nu},A_{2\mu}]]\\
 & \;\, -[A_2^{\nu},[A_{1\nu},A_{3\mu}]] - [A_1^{\nu},[A_{3\nu},A_{2\mu}]]
 -[A_3^{\nu},[A_{2\nu},A_{1\mu}]] \, . 
\end{split}
 \ee
Since the gauge field has degree minus one, the above products are
symmetric under the exchange of any two gauge fields.
We can confirm that $\ell_2$, so defined, gives the correct cubic terms in the action: 
\be
 \begin{split}
  -\tfrac{1}{3!}\langle A,\ell_2(A,A)\rangle \ &= \ \tfrac{2}{3!}\langle A^{\mu}, 
  \partial^{\nu}[A_{\nu}, A_{\mu}] + 
  [\partial_{\mu}A_{\nu}-\partial_{\nu}A_{  \mu}, A^{\nu}]\rangle \\
  &= \ -\tfrac{2}{3!}\langle \, \partial^{\mu}A^{\nu}, 
  [A_{\mu}, A_{\nu}] \, \rangle  -\tfrac{2}{3!} \langle \, A^\mu, [ A^{\nu}\, , 
  \partial_{\mu}A_{\nu}-\partial_{\nu}A_{  \mu} ]\rangle \\
   &= \ -\tfrac{1}{3}\langle \, \partial^{\mu}A^{\nu}, 
  [A_{\mu}, A_{\nu}] \, \rangle  -\tfrac{2}{3} \langle \, [A^\mu,  A^{\nu}]\, , 
  \partial_{\mu}A_{\nu} ]\rangle \\
   \ &= \ -\,\langle \, \partial^{\mu}A^{\nu}\, , [A_{\mu}, A_{\nu}] \rangle \;, 
 \end{split} 
 \ee  
where from the first to second line we integrated by parts and  used the invariance of the Cartan-Killing metric.

In the following we verify the $L_{\infty}$ relations:  \\[0.5ex]
\noindent
{\bf Checking 
$\ell_1 \ell_1 = 0$}.  This must only be checked at degree zero, and it works out immediately: 
 \be
  \ell_1(\ell_1(\lambda)) \ = \ 
 \ell_1 ( \partial \lambda)   \ = \  \square \partial_* \lambda - \partial_* ( \square \lambda) \ = \ 0 \,. 
  \ee
\noindent
{\bf Checking 
$\ell_1 \ell_2 = \ell_2 \ell_1$}.    At degree zero the computation is identical to
that for Chern-Simons.  At degree minus one we must verify
 \be\label{Degree-1}
  \ell_1(\ell_2(A,\lambda)) \ = \  \ell_2(\ell_1(A),\lambda) \ -  \  \ell_2(A,\ell_1(\lambda))  \;. 
 \ee
All terms are calculable except for the first one on the right-hand side.
This identity works out correctly if, again, we choose  \be 
 \label{e-lambda-ym}
   \ell_2(E,\lambda) \ = \ - \big[E,\lambda\big] \ \in \ X_{-2}\;. 
 \ee
There are no more cases to check here.

\noindent
{\bf Checking 
$\ell_3 \ell_1 + \ell_2 \ell_2  + \ell_1 \ell_3 = 0$}. 
  
Since the products on the identity do not change degree, this identity is nontrivial
only in degrees zero, minus one and minus two.

\noindent
We set the following combinations to zero:  
 \be\label{AAlambdaiszero}
\ell_3(\lambda_1,\lambda_2,\lambda_3) \ = \ 0\;,  \quad  \ell_3(\lambda_1,\lambda_2,A) \ = \ 0\;, 
\quad \ell_3(A,A,\lambda) \ = \ 0\;, 
  \quad     \ell_3 ( \lambda_1,\lambda_2,E) \ = \ 0 \,, 
 \ee 
because there is no Lie algebra Jacobiator,  no $\ell_3(A,A,\lambda)$ term in  $\delta_{\lambda}A$, 
no field dependent structure constants $\ell_3(\lambda_1 ,\lambda_2,A)$ in the gauge algebra, and no 
$\ell_3 ( \lambda_1,\lambda_2,E)$ because the algebra closes on shell.

At degree zero we act on $(\lambda\lambda\lambda)$ and the $\ell_3$ terms in the identity
will vanish because the $\ell_3$ acts on $(\lambda\lambda\lambda)$ or $(\lambda\lambda A)$.    At degree minus one we act on $(\lambda\lambda A)$ and the $\ell_3$ terms in the identity
will vanish because $\ell_3$  acts on $(\lambda\lambda A)$,
$(\lambda AA)$ or $(\lambda \lambda E)$.  
Thus both at degree zero and minus one the computation reduces to 
$\ell_2 \ell_2 =0$ and is the same as in CS.

\noindent
At degree minus two we have $AA\lambda$ and $\lambda\lambda E$.   
 For the first one,
(\ref{L3L1}) requires
 \be
 \begin{split}
  & \ell_2(\ell_2(A_1,A_2),\lambda) + \ell_2(\ell_2(\lambda,A_1),A_2)-\ell_2(\ell_2(A_2,\lambda),A_1)\\
  &\; \ = \ -\ell_1(\ell_3(A_1,A_2,\lambda)) -\ell_3(\ell_1(A_1),A_2,\lambda) + \ell_3(A_1,\ell_1(A_2),\lambda)
  -\ell_3(A_1,A_2, \ell_1(\lambda))\,. 
 \end{split} 
 \ee 
A computation of the l.h.s.~gives 
 \be
  {\rm l.h.s.} \ = \ [[\partial_{\mu}\lambda, A_{1\nu}],A_{2}^{\nu}] 
  - [[\partial_{\nu}\lambda, A_{1\mu}],A_{2}^{\nu}]
  -[[A_{1\nu},A_{2\mu}],\partial^{\nu}\lambda]
  +(1\leftrightarrow 2) \,. 
 \ee 
On the r.h.s.~the first term is zero because of (\ref{AAlambdaiszero}). The final term on the r.h.s.~is 
 \be
  -\ell_3(A_1,A_2, \ell_1(\lambda))_{\mu} \ = \ [[A_1^{\nu},[A_{2\nu},\partial_{\mu}\lambda]]
  +[A_1^{\nu},[\partial_{\nu}\lambda,A_{2\mu}]]
  +[\partial^{\nu}\lambda,[A_{1\nu},A_{2\mu}]]+(1\leftrightarrow 2) \,. 
 \ee 
This agrees precisely with the l.h.s., and so the identity holds in the form 
 \be\label{Degree-2}
  \ell_2(\ell_2(A_1,A_2),\lambda) + \ell_2(\ell_2(\lambda,A_1),A_2)-\ell_2(\ell_2(A_2,\lambda),A_1)
  \ = \ - \ell_3(A_1,A_2, \ell_1(\lambda))\;, 
 \ee
implying that we can satisfy the equation by setting
 \be
   \ell_3(E, A, \lambda) \ = \ 0\;.  
 \ee 
The second one is  $\lambda \lambda E$.  The $\ell_3$ terms in the identity 
will find $\ell_3$ acting on $\lambda \lambda E$ and $\lambda A E$, both of which vanish.
Thus we again have to check only $\ell_2\ell_2=0$ and this is the same check as in Chern-Simons.
At degree minus three there is nothing to check.
 
 \noindent
{\bf Checking 
$\ell_1 \ell_4 - \ell_2 \ell_3 + \ell_3 \ell_2  -\ell_4 \ell_1 = 0$}. 
Since we will take $\ell_4 =0$ we just need to check
\be
\ell_2 \ell_3 = \ell_3 \ell_2\,. 
\ee
Since $\ell_2 \ell_3$ has degree plus one,  this identity
must only be checked on four arguments adding to 
degree $-1, -2,$ or  $-3$.  Since $\ell_3$ is only non-vanishing on three $A$'s,
we claim we need three $A$'s and the last one must be a $\lambda$, giving total degree $-3$. 
Indeed, if there are two or less $A$'s no term survives: when $\ell_3$ acts first there are
no three $A$'s to give something nonzero.  When it acts after $\ell_2$ there are no three $A$'s either, since $\ell_2$
does not change degree.   Thus the only nontrivial check is on $AAA\lambda$:
The explicit form given in (\ref{ell4ell1}) then requires 
 \be\label{Degree-3}
  \ell_2(\ell_3(A_1,A_2,A_3),\lambda) \ = \ \ell_3(\ell_2(A_1,\lambda),A_2,A_3)
  +\ell_3(A_1,\ell_2(A_2,\lambda),A_3)+\ell_3(A_1,A_2,\ell_2(A_3,\lambda)) \,. 
 \ee 
This relation can be proved by multiple use of the Jacobi identity.

 \noindent
{\bf Checking 
$\ell_1 \ell_5 \pm  \ell_2 \ell_4 \pm  \ell_3 \ell_3  \pm \ell_4\ell_2  \pm \ell_5 \ell_1   = 0$}. 

\noindent
Here we only need to check the
\be
\ell_3 \ell_3 = 0\,.
\ee
Conceivably, having degree $+2$, this equation should
be tested on degrees $-2, -3$, and $-4$.  But acting on $AAA$, the product $\ell_3$ produces an $E$
and then the second $\ell_3$ will always give zero.  So this equation is trivially
satisfied.

 \noindent
{\bf Checking $\sum_{i,j} \ell_i \ell_j = 0$ with $i+j \geq 7$.}
In here every term has an $\ell_k$ with $k\geq 4$ therefore these are trivially satisfied.

    \be\label{LnproductsNEW}
\boxed{  \begin{split}
\hbox{Yang-Mills:} \qquad   \ell_1(\lambda) \ &= \ \partial\lambda\ \in \ X_{-1} \\
      \ell_1(A)\ &= \ \square A-\partial(\partial\cdot A) \ \in \ X_{-2} \\
    \ell_2(\lambda_1,\lambda_2) \ &= \ -\big[\lambda_1,\lambda_2\big] \ \in  \ X_0\\
    \ell_2(A,\lambda) \ &= \ -\big[A,\lambda\big]\ \in  \ X_{-1} \\
   \ell_2(A_1,A_2)_* \ &= \  -\partial[A_{1}, A_{2*}] - [\partial_{*}A_{1}-\partial A_{1*}\, , \, A_{2}]
  +(1\leftrightarrow 2) \ \in \ X_{-2} \ \ \\
   \ell_2(E,\lambda) \ &= \ -\big[E,\lambda\big] \ \in \ X_{-2}\,   \\
 \ell_3(A_1, A_2, A_3)_* \ &= \  -[A_1,[A_{2},A_{3*}]]\  + \  \hbox{sym.} \  \ \in \ X_{-2}\,   \\
   \end{split}}
 \ee

\medspace 

\section{Double field theory and $L_{\infty}$ algebras}
In this section we discuss double field theory (DFT) in the framework 
of $L_{\infty}$ algebras. In the first subsection, following 
the notation and setup of  
Roytenberg and Weinstein, 
we discuss the subalgebra 
corresponding to the pure gauge structure, given by the C-bracket algebra, 
which in turn is the $O(D,D)$ covariantization of the Courant algebroid. 
The results in this subsection were obtained by Deser and Saemann~\cite{Deser:2016qkw} in a geometrical setup that involves
 symplectic N$Q$-manifolds and a derived bracket construction~\cite{Deser:2014mxa}.
In the second subsection we extend this to the $L_{\infty}$ algebra that also encodes  
fields and their off-shell gauge transformations. Finally, in the third subsection, 
we discuss the full  $L_{\infty}$ algebra describing the complete DFT symmetries and 
dynamics, using perturbation theory around flat space.

\subsection{DFT C-bracket algebra as an $L_3$ algebra}
We begin by recalling a few generalities of DFT, which is manifestly $O(D,D)$ covariant. 
We denote $O(D,D)$ indices by $M,N=1,\ldots, 2D$, and the group-invariant inner product is defined on vectors
by 
\be
\langle  V_1, V_2 \rangle \ = \   \eta_{MN} \, V_1^M V_2^N\;, \qquad
\eta_{MN} \ \equiv \  \begin{pmatrix}   0 & {\bf 1}\\[0.5ex]
  {\bf 1} & 0 \end{pmatrix}\,.  
\ee
The role of infinitesimal gauge transformations will be played by 
generalized Lie derivatives w.r.t.~to a gauge parameter $\xi^M$, 
\be
 {\cal L}_{\xi}V^M \ = \ \xi^N\partial_NV^M + \big(\partial^M\xi_N-\partial_N\xi^M\big)V^N\;. 
\ee
The generalized Lie derivatives form an algebra, 
$[{\cal L}_{\xi_1},{\cal L}_{\xi_2}]={\cal L}_{[\xi_1,\xi_2]_c}$, 
which is governed by the antisymmetric C-bracket
\be
[\xi_1, \xi_2]_c^M \ \equiv \  \xi_1^K\partial_K \xi_2^M  -  \half \xi_1^K  \partial^M \xi_{2K}
 - (1 \leftrightarrow 2) \,, 
\ee
where in the following we will sometimes leave out the sub-index $c$ in order not to clutter 
the equations. 
Both the inner product and the C-bracket are covariant under the action of the generalized
Lie derivative.   
Moreover, the generalized Lie derivative and the C-bracket are equal up to a total 
derivative of the inner product, 
\be
{\cal L}_{V}W \ = \ [V,W]+\tfrac{1}{2} \partial \langle V,W\rangle \;. 
\ee
The generalized Lie derivative w.r.t.~a gauge parameter that is a total derivative, $\xi^M=\partial^M\chi$, 
acts trivially on fields as a consequence of the strong constraint 
($\partial^M\partial_M A =0$ and $\partial^M A\, \partial_M B = 0$ for all $A, B$). 
Moreover, when one of the two gauge parameters (vectors) inside the C-bracket is trivial, i.e., $\xi_2 = \partial \chi$ for some function
$\chi$, one finds 
\be
\label{c-bracket-trivial-vector}
[\xi, \partial \chi] \ =  \   \half  \partial ( \xi^K \partial_K \chi)  \ = \ 
\partial \, \half  \langle \xi , \partial \chi \rangle 
\,.  
\ee
The C-bracket satisfies a Jacobiator identity
\be\label{CJACO}
J (\xi_1, \xi_2, \xi_3 ) \ \equiv  \ 3\bigl[ \, [\xi_{[1} , \xi_2 ] \,, \xi_{3]} \bigr] \ = \   \ \bigl[ \, [\xi_1, \xi_2 ] \,, \xi_3 \bigr]  + \hbox{c.p.} 
\ = \   \, \partial \,  T(\xi_1, \xi_2, \xi_3) \,, 
\ee
where the antisymmetrization is over all three indices, c.p.~denotes `cyclic permutation', and 
$T$ is defined by 
\be
T(\xi_1, \xi_2, \xi_3) \ \equiv \ \tfrac{1}{2} \bigl\langle
 \, [\xi_{[1}, \xi_2 ] \,, \xi_{3]} \bigr\rangle \ = \   \ \tfrac{1}{6}  \,  \bigl( \, \bigl\langle
 \, [\xi_1, \xi_2 ] \,, \xi_3 \bigr\rangle  + \hbox{c.p.}  \bigr)  \,. 
\ee

\medskip

Since the above identities take the same form as for the Courant algebroid, the setup of
Roytenberg and Weinstein applies here, and we can next 
reformulate this as an $L_3$ homotopy Lie algebra.  The total graded vector space $X$ will be taken to contain three spaces of fixed degrees:  
 \be
  \begin{split}
0 \longrightarrow \   &X_2 \longrightarrow   X_{1}\longrightarrow\;  X_{0} \\
   &\, c  \hskip10pt \qquad \chi \ \  \qquad \xi^M
  \end{split}
 \ee
The space of degree zero contains the gauge parameters, the space of degree one
contains functions, and the space of degree zero contains the constants.  The above
arrows define the $\ell_1$ map. From $X_2$ to $X_1$ it is the inclusion map $\iota: c \to \iota c$, which is the same constant, now viewed as a trivial function in $X_1$.
From $X_1$ to $X_2$ is the partial derivative $\partial$.  Acting on $X_0$ the map 
$\ell_1 $ is defined to give zero
\be
\ell_1 (\xi) \ = \ 0 \,, 
\ee
in agreement with the fact that we are not taking fields into account. 

The non-vanishing multilinear maps are
\be\label{Courantprod}
\boxed{
\begin{split}
\ell_1 (\chi) \ = \ &  \ \partial \chi  \  \in X_0\;, \\
\ell_1 (c)  \ = \ &  \  \iota\,  c  \ \in X_1\;, \\
\ell_2 ( \xi_1, \xi_2) \ = \ & \ [\xi_1, \xi_2] \ \in X_0\;,   \\
\ell_2\,  ( \xi,\,  \chi ) \ = \ & \ 
\half \, \langle \xi \,, \partial \chi \rangle \ = \ \half \xi^K \partial_K \chi \ \in X_1  \;,  \\
\quad \ell_3 (\xi_1, \xi_2, \xi_3) \ = \ &  - T(\xi_1, \xi_2, \xi_3)\ \in X_1 \,.  \ \ 
\end{split} }
\ee
Note that no product, except for $\ell_1$, involves an input from the  
space $X_2$ nor does any product give an element of $X_2$.   
Additionally, we have the following interesting relations:
\be
\label{note-nice}
  \begin{split}
  \partial \ell_3 (\xi_1, \xi_2 , \xi_3)  \ = \ & \  - J(\xi_1, \xi_2 , \xi_3) \,, \\
  \ell_2  ( \xi , \partial  \chi )  \ = \ & \   \partial  \ell_2 ( \xi , \chi) \, . 
  \end{split}
  \ee  
 The first relates the Jacobiator (\ref{CJACO}) to the derivative of $\ell_3$.  The second encodes the
 behavior  (\ref{c-bracket-trivial-vector})  of the C-bracket when one of the inputs is
 a trivial vector.

\noindent
{\bf Step by step construction.}     
We now sketch how the construction of the above algebra is done, step by step,
starting with the C-bracket.  
\begin{enumerate}

\item We  begin with the space $X_0$ of gauge parameters.
From the C-bracket one sets the product $\ell_2 (\xi_1, \xi_2)$
equal to the bracket itself.  At this stage one does not know if any other products are needed or not.

\item
With just $\ell_2 \not= 0$ the only nontrivial identity would be $\ell_2 \ell_2 = 0$, acting on three
$\xi$'s.   But $\ell_2 \ell_2$  gives the Jacobiator $J(\xi_1, \xi_2, \xi_3)$, which  does not vanish.   This implies we have to introduce both $\ell_3$ and 
$\ell_1$ to fix this identity.  

\item Since the Jacobiator can be written as $\partial T(\xi_1, \xi_2, \xi_3)$ this suggests 
setting $\ell_3$ on three gauge parameters equal to the function $T(\xi_1, \xi_2, \xi_3)$.
Since $\ell_3$ has degree plus one, we now need a space $X_1$ of functions.  On this space
of functions $\ell_1$ acts as a derivative and this fixes the homotopy identity
$\ell_1 \ell_3 + \ell_2 \ell_2 + \ell_3 \ell_1 = 0$, assuming the last term is zero.

\item   The 
presence of $\ell_1$ forces one to reconsider the lower identities. In order to 
guarantee that $\ell_1 \ell_1 =0$ acting on $X_1$ we now set $\ell_1 : X_0 \to 0$.  
This now confirms the last term in the previous item vanishes.

\item 
We then consider  $\ell_1 \ell_2 = \ell_2 \ell_1$  which is only nontrivial acting on
a $\chi\in X_1$ and a gauge parameter $\xi\in X_0$.  That identity determines $\ell_2 (\xi, \chi)$.

\item
At this point all nontrivial products have been determined and one must verify that 
all homotopy identities hold without the need of additional products.  

\item If one wishes to have
an exact sequence of spaces then one can introduce the space $X_2$ of constants, and 
$\ell_1$ acting on it simply gives the same constant, now as an element of the space of functions
$X_1$.   This completes the construction.

\end{enumerate}

\bigskip     
As indicated above, the only nontrivial computation is checking that
no $\ell_4$ is needed because the identity $\ell_3 \ell_2 - \ell_2 \ell_3= 0$ 
holds when acting on four gauge
parameters.  Indeed, using (\ref{ell4ell1}) we see that
\be
\label{eofi}
\ell_3 \ell_2 - \ell_2 \ell_3  \ = \  \  
 6\, \ell_3   ([ \xi_{[1}, \xi_2] , \xi_3, \xi_{4]}) \ - \ 
  4 \, \ell_2  (\ell_3 (\xi_{[1}, \xi_2, \xi_3) , \xi_{4]} )  \,. 
\ee
We will show that $\ell_3 \ell_2 - \ell_2 \ell_3$ must be a constant by 
proving that its derivative vanishes.  But this means 
that $\ell_3 \ell_2 - \ell_2 \ell_3$ actually vanishes, 
because it is a local function of arbitrary space-dependent gauge parameters;  
if it did not vanish it
would have to have space dependence and could not be a constant.    
Taking the derivative of the above equation and using both lines  in (\ref{note-nice}), 
we compute  
\be
\partial (\ell_3 \ell_2 - \ell_2 \ell_3 ) \ = \  \  
 -6\, J ([ \xi_{[1}, \xi_2] , \xi_3, \xi_{4]}) \ + \ 
  4 \, \ell_2  (J (\xi_{[1}, \xi_2, \xi_3) , \xi_{4]} )  \,. 
\ee
Rearranging the inputs on both terms and recalling the definition of $\ell_2$ on 
two vectors we get
\be
\partial (\ell_3 \ell_2 - \ell_2 \ell_3 ) \ = \  \  
 -6\, J (\, \xi_{[1}, \xi_2\,  , [\xi_3, \xi_{4]}] ) \ + \ 
  4 \, [ \, \xi_{[1} \,, \, J (\xi_{2}, \xi_3, \xi_{4]} )  \, ] \,. 
\ee
It is straightforward to see that the right-hand side vanishes.  It does so trivially, just
upon using the definition $J (\xi_1, \xi_2, \xi_3 )  \equiv  3\bigl[ \, [\xi_{[1} , \xi_2 ] \,, \xi_{3]} \bigr]
= \bigl[ \, [\xi_1, \xi_2 ] \,, \xi_3 \bigr]  + \hbox{c.p.}$.  Thus, as claimed,  the derivative of
$\ell_3 \ell_2 - \ell_2 \ell_3$ is guaranteed to vanish by our definitions.  As argued above, this
means that $\ell_3 \ell_2 - \ell_2 \ell_3=0$.

We can contemplate the possibility
that in some other scenario   $\ell_3 \ell_2 - \ell_2 \ell_3\not= 0$ is a non-vanishing constant and we 
would  require an $\ell_4$ product  that would contribute, for example,
a term $\ell_1 \ell_4$ to the identity.   Note that acting on four gauge parameters $\ell_4 \in X_2$,
which is correctly identified as the space of constants.  That constant in $X_2$ would be mapped
by $\ell_1$ to the same constant in $X_1$, allowing the possibility of cancellation
of the constant $\ell_3 \ell_2 - \ell_2 \ell_3\not= 0$.  The space $X_2$
would then play an important role.  This, however, does not happen for the C-bracket.

A more conventional proof of the identity $\ell_3 \ell_2 - \ell_2 \ell_3= 0$ 
is just by direct computation:
Indeed,  starting from (\ref{eofi})  we can show that
\be
\begin{split}
\ell_3 \ell_2 - \ell_2 \ell_3  \ = \ & \  
 6\, \ell_3   ([ \xi_{[1}, \xi_2] , \xi_3, \xi_{4]}) \ - \ 
  4 \, \ell_2  (\ell_3 (\xi_{[1}, \xi_2, \xi_3) , \xi_{4]} )\\
  \ = \ &  
- 6\, T ([ \xi_{[1}, \xi_2] , \xi_3, \xi_{4]}) \ + \ 
  4 \, \ell_2  ( T (\xi_{[1}, \xi_2, \xi_3) , \xi_{4]} )\\
  \ = \ &   -\langle \,   [\,  [ \xi_{[1} , \xi_2 ] , \xi_3 ] , \, \xi_{4]} \rangle - \,
  \langle \    [ \xi_{[3} , \xi_4 ] \ ,\  [ \xi_1 \, \xi_{2]} ] \rangle 
   -\langle \  [\, \xi_{[4}\, ,  [ \xi_{[1} , \xi_2 ] \, ],  \, \xi_{3]} \rangle \\
  &   - 2 \langle \, \xi_{[4} \,, \, \partial T (\xi_{1}, \xi_2, \xi_{3]}) \ \rangle \\
 \ = \ &   -2\langle \  [\,  [ \xi_{[1} , \xi_2 ] , \xi_3 ] , \, \xi_{4]} \rangle - \,
  \langle \    [ \xi_{[1} , \xi_2 ] \, ,\,  [ \xi_3 \, \xi_{4]} ] \rangle \    - 2 \langle \, \xi_{[4} \,, \, J (\xi_{1}, \xi_2, \xi_{3]}) \ \rangle \\
  \ = \ &   -\tfrac{2}{3} \langle \  J ( \xi_{[1} , \xi_2 , \xi_3 ) , \, \xi_{4]} \rangle - \,
  \langle \    [ \xi_{[1} , \xi_2 ] \, ,\,  [ \xi_3 \, \xi_{4]} ] \rangle \    + 2 \langle \, J (\xi_{[1}, \xi_2, \xi_{3}) \, , \, \xi_{4]}\ \rangle \\
    \ = \ &   -\tfrac{8}{3} \langle \  J ( \xi_{[1} , \xi_2 , \xi_3 ) , \, \xi_{4]} \rangle - \,
  \langle \    [ \xi_{[1} , \xi_2 ] \, ,\,  [ \xi_3 \, \xi_{4]} ] \rangle \     \\
    \ = \ &   -\tfrac{2}{3}  {\bf J}  - \, \tfrac{1}{3} {\bf K}    \ = \   -\tfrac{1}{3} (2 {\bf J} + {\bf K} )  \,.
\end{split}
\ee
Here, following Roytenberg-Weinstein, we have defined the scalars 
\be
 \begin{split}
  {\bf J}(\xi_1,\xi_2,\xi_3,\xi_4) \ &\equiv \ 4\langle J(\xi_{[1},\xi_2,\xi_3),\xi_{4]}\rangle
  \ = \ -2\xi_{[1}\langle [\xi_{2},\xi_3],\xi_{4]}\rangle\,. 
  \\
  {\bf K}(\xi_1,\xi_2,\xi_3,\xi_4) \ &\equiv \ 3\langle [\xi_{[1},\xi_2],[\xi_3,\xi_{4]}]\rangle\,. 
 \end{split}
 \ee  
Writing the antisymmetrizations out one has
\be
\begin{split}
{\bf J} \ \equiv \ & \langle  J(\xi_1, \xi_2, \xi_3)\,, \xi_4 \rangle  
- \langle  J(\xi_1, \xi_2, \xi_4)\,, \xi_3 \rangle   
+ \langle  J(\xi_1, \xi_3, \xi_4)\,, \xi_2 \rangle  
-\langle  J(\xi_2, \xi_3, \xi_4)\,, \xi_1 \rangle \,,  \\
{\bf K} \ \equiv \ & \langle  [\xi_1, \xi_2] , [\xi_3, \xi_4] \rangle
- \langle  [\xi_1, \xi_3] , [\xi_4, \xi_4] \rangle
+\langle  [\xi_1, \xi_4] , [\xi_2, \xi_3] \rangle\,. 
\end{split}
\ee
The requisite identity is satisfied because 
\be
\label{KJident}
{\bf K} + 2{\bf J}= 0\,,  
\ee
as we show now.
 Using the covariance of the inner product and bracket, leaving the total antisymmetrization 
in the four arguments implicit from now on, we compute 
 \be
 \begin{split}
  \xi_{1}\langle [\xi_{2},\xi_3],\xi_{4}\rangle \ = \ {\cal L}_{\xi_{1}}\langle [\xi_{2},\xi_3],\xi_{4}\rangle
  \ &= \ 2\, \langle [{\cal L}_{\xi_{1}}\xi_{2},\xi_3],\xi_{4}\rangle + \langle [\xi_{2},\xi_3],{\cal L}_{\xi_{1}} \xi_{4}\rangle\\
  \ &= \ 2\, \langle [[\xi_{1},\xi_{2}],\xi_3],\xi_{4}\rangle + \langle [\xi_{2},\xi_3],[{\xi_{1}}, \xi_{4}] \rangle\\
  \ &= \ \tfrac{2}{3} \langle J(\xi_1,\xi_2,\xi_3),\xi_4\rangle + \langle [\xi_{1},\xi_2],[{\xi_{3}}, \xi_{4}] \rangle\\
  \ &= \ - \tfrac{1}{3}\xi_1 \langle [\xi_{2},\xi_3],\xi_{4}\rangle+ \langle [\xi_{1},\xi_2],[{\xi_{3}}, \xi_{4}] \rangle\;, 
 \end{split}
 \ee
where we used  ${\cal L}_{X}Y=[X,Y]+\frac{1}{2} \partial \langle X,Y\rangle$, noting that   under the total antisymmetrization 
the symmetric terms drop out. Thus, bringing the first term on the r.h.s.~to the l.h.s., 
 \be
  \tfrac{4}{3}\, \xi_1 \langle [\xi_{2},\xi_3],\xi_{4}\rangle \ = \ \langle [\xi_{1},\xi_2],[{\xi_{3}}, \xi_{4}] \rangle\;, 
 \ee 
and thus 
 \be
   2\,{\bf J}(\xi_1,\xi_2,\xi_3,\xi_4)  \ = \ -4\, \xi_{1}\langle [\xi_{2},\xi_3],\xi_{4}\rangle 
   \ = \ -3 \, \langle [\xi_{1},\xi_2],[{\xi_{3}}, \xi_{4}] \rangle \ = \ -{\bf K}(\xi_1,\xi_2,\xi_3,\xi_4)\;, 
 \ee  
proving  (\ref{KJident}). Thus, we have proved that all $L_{\infty}$ identities are satisfied.

\subsection{Off-shell DFT as extended $L_3$ algebra}

Here we extend the  $L_3$ algebra describing the C-bracket algebra 
to include the fields and gauge transformations of DFT, but still without 
taking  dynamics into account. In other words, we consider the off-shell gauge structure 
of the DFT fields and build the algebra 
\be
{L_\infty^{\phantom{0}}}^{\hskip-7pt \rm gauge+fields}
\ee 
described in the introduction. 
We discuss two alternative formulations.   
In the first we  include only the generalized   
metric ${\cal H}_{MN}$ and its gauge transformations. 
In the second we include the non-symmetric metric ${\cal E}_{ij}$
and its gauge transformations.   
Both formulations are background independent
and non-perturbative:
there is no need to consider expansions around some specific backgrounds.

\subsubsection*{Gauge structure in terms of  ${\cal H}_{MN}$}  

We start by extending the total graded vector space as follows: 
  \be\label{Hextendedvector}
  \begin{split}
0 \longrightarrow \   &X_2 \longrightarrow   X_{1}\longrightarrow\;  X_{0} 
\longrightarrow \; X_{-1} \\
   &\, c  \hskip10pt \qquad \chi \ \  \qquad \xi^M\ \  \qquad {\cal H}_{MN}
  \end{split}
 \ee
Since there is no vector space of degree minus two, we have 
\be
\ell_n ({\cal H}^n) \ = \ 0\,, \ \ \hbox{for} \ \ n \geq 1 \,.  
\ee
This also implies that there are no field equations and no dynamics.  
The gauge transformation of ${\cal H}_{MN}$ is given by the generalized Lie derivative 
 \be\label{genLieH}
  \delta_{\xi}{\cal H}_{MN} \ \equiv \ {\cal L}_{\xi}{\cal H}_{MN} \ = \ 
  \xi^K\partial_K {\cal H}_{MN}
  + K_{M}{}^{K}{\cal H}_{KN}
  + K_{N}{}^{K}{\cal H}_{MK}\;, 
 \ee
where we defined $K_{MN}\equiv \partial_M\xi_N-\partial_N\xi_M$.   
The generalized metric satisfies ${\cal H}\eta{\cal H}=\eta$, and this constraint is 
preserved by the generalized Lie derivative. 
In this background independent 
formulation 
the gauge transformations are homogenous in the fields. 
We now compare the above with the general form (\ref{jnsqts}) of the
gauge transformations. This comparison shows 
that we require one 
new non-trivial product:    
 \be\label{newL2}
  \ell_2(\xi,{\cal H}) \ \equiv \ {\cal L}_{\xi}{\cal H}\;.
 \ee 
The gauge transformation then reads, by definition, $\delta_{\xi}{\cal H}=\ell_2(\xi,{\cal H})$. 
The lack of an inhomogeneous term and of terms nonlinear in the field imply that  
\be
\ell_1(\xi)\ =\ 0\,, \quad   \ell_{n+1} ( \xi, \Psi^n) \ = \ 0 \,, \ \   \hbox{for} \ n \geq 1 \,. 
\ee
We claim that with the addition of the product (\ref{newL2})    
to the list of products (\ref{Courantprod})  we have a consistent $L_3$ algebra 
structure on the vector space (\ref{Hextendedvector}). 

In order to prove this claim we have to verify the $L_{\infty}$ relations. 
The relation $\ell_1^2=0$ does not need to be re-checked because the $\ell_1$ product 
is not modified. The relation $\ell_1\ell_2=\ell_2\ell_1$, c.f.~(\ref{L2L1}), 
tells us that $\ell_2 (\chi, {\cal H})$ and $\ell_2 (c, {\cal H})$ can be taken 
to be zero.  The first one is a bit nontrivial:  take $x_1=\chi$, $x_2={\cal H}$ for a
scalar function $\chi$:
 \be\label{secondL3}
 \ell_1 \ell_2(\chi,{\cal H}) \ = \ \ell_2(\partial\chi,{\cal H})  \;, 
 \ee
using $\ell_1(\chi)=\partial\chi$.  The left-hand side is zero because 
$\ell_1$ is acting on
a vector in $X_0$. But the right-hand side is also zero, 
 \be\label{secondL3p}
  \ell_2(\partial\chi,{\cal H}) \ = \ {\cal L}_{\partial\chi}{\cal H} \ = \ 0 \;, 
 \ee
because $\partial^M\chi$ 
is a trivial parameter and the associated generalized Lie derivative (\ref{genLieH}) vanishes 
by the strong constraint.   The consistency of setting $\ell_2 (c , {\cal H}) =0$ now follows 
quickly.

We now turn to the relation $0=\ell_1\ell_3+\ell_3\ell_1+\ell_2\ell_2$, c.f.~(\ref{L3L1}), which requires three inputs.  We must check it on 
inputs that include at least one ${\cal H}$.  If we introduce no new $\ell_3$ products,
the $\ell_1 \ell_3$ and $\ell_3 \ell_1$ terms must vanish and the identity reduces
to $\ell_2 \ell_2 =0$.   To get something non-vanishing when having an ${\cal H}$
we must then have two $\xi$'s.  Thus we find that  
for arguments $\xi_1, \xi_2, {\cal H}$,  the identity reads  
 \be
  \ell_2(\ell_2(\xi_2,\xi_1),{\cal H}) \ = \ \ell_2(\xi_2,\ell_2(\xi_1,{\cal H}))-(1\leftrightarrow 2)\;. 
 \ee
It is easy to see that 
this is precisely the closure condition of $\delta_{\xi}$ on ${\cal H}$ and hence satisfied 
by the general DFT results. 

Finally, the $L_{\infty}$ relation (\ref{ell4ell1}) and all higher ones are trivially satisfied, because 
they involve products like $\ell_3$ or higher. Since all higher products vanish under our assumption, we need only concern ourselves with the appearance of $\ell_3$.  The $\ell_3$ product is only non-zero evaluated 
for three gauge parameters and  takes values in the space $X_1$ of scalar functions. But 
there is no non-zero product for ${\cal H}$ and an argument in $X_1$. 
This proves that the $L_{\infty}$ or $L_3$ relations remain 
valid after the extension in (\ref{Hextendedvector}) and the addition of the new product (\ref{newL2}).

\subsubsection*{Gauge structure in terms of  ${\cal E}_{ij}$}

Let us now turn to a similar but somewhat more intriguing extension of the $L_3$ algebra. 
We still work with non-perturbative off-shell fields, in this case the `non-symmetric metric' 
${\cal E}=g+b$, so that the graded vector space is still given by (\ref{Hextendedvector}), 
but with the elements of $X_{-1}$ now being fields ${\cal E}$. 
The products (\ref{Courantprod}) encoding the pure C-bracket algebra 
are unchanged, 
but we decompose the gauge parameter as $\xi^M=(\tilde{\xi}_i, \xi^i)$,
with $i = 1, \ldots , D$.  
Despite being non-perturbative, 
the gauge transformations of ${\cal E}$ have inhomogeneous 
and higher order terms. Specifically, the gauge transformation can be written as 
 \be\label{deltaE}
  \delta_{\xi}{\cal E}  \ = \  \ell_1(\xi) + \ell_2(\xi,{\cal E}) - \tfrac{1}{2} \ell_3(\xi, {\cal E},{\cal E}) \;,  
 \ee 
where according to eqn.~(2.32) in \cite{Hohm:2010jy} 
 \be
 \label{sgbtfltts}
 \begin{split}
  [\ell_1(\xi)]_{ij} \ &= \ \partial_{i}\tilde{\xi}_{j} -  \partial_{j}\tilde{\xi}_{i} \;, \\
  [\ell_2(\xi,{\cal E})]_{ij} \ &= \ { L}_{\xi}{\cal E}_{ij}+\tilde{ L}_{\tilde{\xi}}{\cal E}_{ij}\;, \\
  [\ell_3(\xi, {\cal E}_1,{\cal E}_{2})]_{ij} \ &= \ {\cal E}_{1ik}(\tilde{\partial}^k\xi^l -\tilde{\partial}^l\xi^k)
  {\cal E}_{2lj}+(1\leftrightarrow 2)\;, 
 \end{split}
 \ee 
using the notation of \cite{Hohm:2010jy} for Lie derivatives $L_{\xi}$ and dual 
Lie derivatives $\tilde{L}_{\tilde{\xi}}$,  defined by 
 \be\label{LLIIEE}
  \begin{split}
     L_{\xi}{\cal E}_{ij} \ &= \ \xi^k\partial_k {\cal E}_{ij} +\partial_i\xi^k\, {\cal E}_{kj}
   +\partial_j\xi^k\, {\cal E}_{ik}\;, \\
   \tilde{L}_{\tilde{\xi}}{\cal E}_{ij}\ &= \ \tilde{\xi}_k\tilde{\partial}^k {\cal E}_{ij} -\tilde{\partial}^k\tilde{\xi}_i\,{\cal E}_{kj}
   -\tilde{\partial}^k\tilde{\xi}_j\,{\cal E}_{ik}\;. 
  \end{split}
 \ee
Since the gauge algebra is field independent  and we have no dynamics, 
 \be\label{zeroproducts}
  \ell_{n+2}(\xi_1,\xi_2,{\cal E}^n) \ = \ 0\;, \;\;n\geq 1\;, 
  \qquad 
  \ell_n({\cal E},\ldots, {\cal E}) \ = \ 0 \quad 
    \forall\, n\geq 1 \;. 
   \ee 
We claim that with the addition of the products (\ref{sgbtfltts})    
to the list of products (\ref{Courantprod})  we have a consistent $L_3$ algebra 
structure on the vector space (\ref{Hextendedvector}) (with ${\cal E}$ replacing ${\cal H}$). 
 
Let us now verify the $L_{\infty}$ relations. 
First, the relations involving only gauge parameters $\xi$ and functions $\chi$ still hold as 
in the first subsection, since we merely changed the notation by splitting $\xi^M$ into $\tilde{\xi}_i$ 
and $\xi^i$. For instance, $\ell_1^2=0$ on $X_1$ holds for ${\ell_1(\chi)}_i=\partial_i\chi$: 
 \be
  \ell_1(\ell_1(\chi))_{ij} \ = \ \partial_i{\ell_1(\chi)}_j   -\partial_j{\ell_1(\chi)}_i \ = \ 0\;.  
 \ee
Second, the relation $\ell_1\ell_2=\ell_2\ell_1$ 
of the same form as in (\ref{secondL3}), with ${\cal H}$ replaced by ${\cal E}$, follows again because 
$\tilde{\xi}_i=\partial_i\chi$, $\xi^i=\tilde{\partial}^i\chi$ are trivial parameters of (\ref{deltaE}), 
which can be easily verified by use of (\ref{LLIIEE}) and the strong constraint.  
So again we have that $\ell_2 (\chi, {\cal E})$ and $\ell_2 (c, {\cal E}) $ can be set to zero.

Under our assumption that there are no additional products, the $L_\infty$ identities of the form
$\ell_i \ell_j + \ldots =0$ with $i+j \geq 7$ are trivially satisfied, since we have no products
$\ell_n$ with $n\geq 4$.   We must consider the identities for $i+j = 4,5,6$, having 3,4, and 5 inputs,
respectively.  In each case we must have at least one field ${\cal E}$ among the inputs,
otherwise the identity was checked before (note that the products in (\ref{Courantprod}) 
can never generate an object in $X_{-1}$).

We sketch the procedure now.  
Take $i+j=4$, which corresponds to the identity $\ell_1 \ell_3 + \ell_3 \ell_1 + \ell_2 \ell_2 =0$,
of intrinsic degree zero.
This requires three inputs.  Three ${\cal E}$ works trivially because of degree.  The cases $(\star, {\cal E}{\cal E})$
for $\star = \xi, \chi, c$ also work trivially.  From the six cases $(\star, \star, {\cal E})$ the only nontrivial
one happens to occur for $(\xi\xi {\cal E})$.  For $i+j=5$ the identity is $\ell_3 \ell_2 = \ell_2 \ell_3$ 
and it requires four inputs.  An enumeration shows that the only nontrivial case occurs for inputs
$(\xi \xi {\cal E} {\cal E})$.  Finally, for $i+j =6$  the identity is $\ell_3 \ell_3 =0$ and requires 5 inputs,
the only nontrivial one being $(\xi \xi {\cal E} {\cal E} {\cal E})$.  
It follows that the remaining non-trivial $L_{\infty}$ relations are  
 \be\label{Linftyidentities}
  \begin{split}
   \ell_2(\ell_2(\xi_2,\xi_1),{\cal E}) \ &= \ \ell_2(\xi_2,\ell_2(\xi_1,{\cal E})) - \ell_3(\xi_2,\ell_1(\xi_1),{\cal E}) 
   - (1\leftrightarrow 2)\;, \\
   \ell_3(\ell_2(\xi_2,\xi_1),{\cal E},{\cal E}) \ &= \ \ell_2(\xi_2,\ell_3(\xi_1,{\cal E},{\cal E}))
   +2\,\ell_3(\xi_2,\ell_2(\xi_1,{\cal E}),{\cal E}) - (1\leftrightarrow 2)\;, \\
   0 \ &= \ \ell_3(\xi_2,\ell_3(\xi_1,{\cal E},{\cal E}),{\cal E})- (1\leftrightarrow 2)\;, 
  \end{split}
 \ee  
with diagonal arguments for fields ${\cal E}$, which by the general discussion  
is sufficient for the validity of the $L_{\infty}$ relations.\footnote{These relations were applied recently in~\cite{Hohm:2016yvc}.} 
These relations, 
together with those in (\ref{Courantprod}),  
are sufficient for closure of the gauge transformations, 
 \be
  \begin{split}
   \big[\delta_{\xi_1},\delta_{\xi_2}\big]{\cal E} \ = \ 
   \ell_1(\xi_{12})+\ell_2(\xi_{12},{\cal E})-\tfrac{1}{2}\ell_3(\xi_{12},{\cal E},{\cal E})\;, 
  \end{split}
 \ee   
where $\xi_{12}  \equiv [\xi_2,\xi_1]_{c}  \equiv  \ell_2(\xi_2,\xi_1)$, c.f.~(\ref{lClosure}).  As explained at the end of section 3.3,  
the closure of the gauge algebra implies that the identities  
(\ref{Linftyidentities})  hold. 
We have also verified (\ref{Linftyidentities}) by direct 
computations.

\subsection{Perturbative DFT as $L_{\infty}$ algebra}

We finally take dynamics into account, employing a perturbative formulation obtained by 
expanding DFT around a background. The fundamental fields are the dilaton $\phi$ 
and the generalized metric ${\cal H}$,   
which is expanded around a constant background as follows:  
 \be\label{expandHAAA}
  {\cal H}_{MN} \ = \ \bar{\cal H}_{MN} + h_{\,\nin{M}\bar{N}}+ h_{\,\nin{N}\bar{M}}
   -\tfrac{1}{2} h^{\,\nin{K}}{}_{\bar{M}}\, h_{\,\nin{K}\bar{N}} + \tfrac{1}{2} h_{\,\nin{M}}{}^{\bar{K}}\,
  h_{\,\nin{N}\bar{K}}  \ +  \ {\cal O}(h^3)\;.  
 \ee 
This expansion is  compatible with the constraint   
${\cal H}_{M}{}^{K}{\cal H}_{K}{}^{N}=\delta_M{}^{N}$.  
We use projected
 $O(D,D)$ indices defined for a vector by $V_{\,\nin{M}}=P_M{}^{N}V_N$, $V_{\bar{M}}=\bar{P}_M{}^{N}V_N$, 
with the projectors 
 \be\label{projectorsINTRO}
  P_M{}^{N} \ =  \ \tfrac{1}{2}\big(\delta_M{}^{N} - \bar{\cal H}_M{}^N\big)\;, \qquad 
  \bar{P}_M{}^{N} \ =  \ \tfrac{1}{2}\big(\delta_M{}^{N} + \bar{\cal H}_M{}^N\big)\;. 
 \ee
Indeed, 
$P^2=P$, $\bar{P}^2=\bar{P}$, and $P\bar{P}=0$ as a consequence of the constraint on the background 
generalized metric, $\bar{\cal H}_{M}{}^{K}\bar{\cal H}_{K}{}^{N}=\delta_M{}^{N}$. 
 
Let us now discuss the $L_{\infty}$ algebra encoding the symmetries and dynamics of the 
above perturbative field variables. 
To this end we extend the above sequence once more to 
 \be
  \begin{split}
0\; \longrightarrow \;   &X_2\; \longrightarrow  \;\; X_{1}\longrightarrow\;  X_{0} \;
\longrightarrow\;\;  X_{-1}\; \longrightarrow\;\;  X_{-2}\\
   &\, c  \hskip18pt \qquad \chi \ \ \  \qquad \xi^M  \qquad (h_{\,\nin{M}\bar{N}},\phi) 
    \quad ({\cal R}_{\,\nin{M}\bar{N}}, {\cal R}) 
  \end{split}
 \ee
where $X_{-1}$ encodes the fields and $X_{-2}$ the field equations. More precisely, 
since we have the fundamental fields $h_{\,\nin{M}\bar{N}}$ and 
$\phi$ we have a further decomposition into direct sums: 
 \be  
  X_{-1} \ = \ X_{-1,t}  \ \oplus \ X_{-1,s} \;, \qquad X_{-2} \ = \ X_{-2,t}  \ \oplus \ X_{-2,s}\;, 
 \ee 
with subscript `$s$' denoting the dilaton (scalar) component and subscript `$t$' the tensor component. 
Collectively, we denote fields by $\Psi=(h_{\,\nin{M}\bar{N}},\phi)$ and field equations by $E=({\cal R}_{\,\nin{M}\bar{N}}, {\cal R})$, so that 
 \be
 [\Psi_t]_{\,\nin{M}\bar{N}}  \ = \ 
  h_{\,\nin{M}\bar{N}}\,, \quad  \Psi_s \ = \ \phi\;, 
\qquad  [E_t]_{\,\nin{M}\bar{N}} \ = \ {\cal R}_{\,\nin{M}\bar{N}}\;, \quad  
 E_s \ = \ {\cal R}\;. 
 \ee 
We will leave out the subindex of the grading if the tensor character is 
evident from the index structure.

Before starting the construction we must digress.  
The gauge parameters are $\xi^M$ as above, 
and the gravitational field is  $h_{\,\nin{M}\bar{N}}$,  but 
the conventional perturbative 
DFT expressions are defined in terms of 
gauge parameters $\lambda_i$ and $\bar{\lambda}_i$  
and closed string field theory (CSFT) variables $e_{ij}$ and $d=-\tfrac{1}{2}\phi$~\cite{Hull:2009mi}.
By means of background 
frame fields it is straightforward to 
translate the indices $i, j$ of the original perturbative DFT expressions to the projected 
$2D$ valued $O(D,D)$ indices. 
Indeed,  we assume a constant background frame field $E_{A}{}^{M}$
(an anchor map), and use \cite{Hohm:2011dz,Hohm:2014xsa} 
 \be\label{eandh}
  {E}_{a}{}^i\,{E}_{\,\bar{b}}{}^{\,j}\, e_{ij} \ = \ \tfrac{1}{2}\,{E}_{a}{}^{M}\, {E}_{\,\bar{b}}{}^{\,N} \, 
  h_{\,\nin{M}\bar{N}}\;, 
 \ee
and 
 \be \label{barredunbarred}
  \begin{split}
   \lambda_i \ &\equiv \ - E_{i}{}^{a} E_{a}{}^{M}\xi_M \;, \qquad 
   \bar{\lambda}_i \ \equiv \ E_{i}{}^{\bar{a}} E_{\bar{a}}{}^{M} \xi_{M}\;, \\
   D_i \ &\equiv \ E_{i}{}^{a} E_{a}{}^{M}\partial_M\;, \qquad \;\; 
   \bar{D}_i \ \equiv  \ E_{i}{}^{\bar{a}} E_{\bar{a}}{}^{M}\partial_M\;, 
  \end{split}
 \ee
where the sign in the first line is introduced for convenience, in order to comply with the CSFT conventions for 
$\lambda, \bar{\lambda}$.  Note that we could also replace the $O(D,D)$ indices 
appearing above immediately as projected indices, so that 
e.g.~$\bar{\lambda}_i = E_{i}{}^{\bar{a}} E_{\bar{a}}{}^{M} \xi_{\bar{M}}$. 
Moreover, 
one must recall 
that the flattened $O(D,D)$ metric ${\cal G}_{AB}={E}_A{}^M {E}_{BM}$ is related to the 
background metric $G$ via 
\be
 G_{ij} \ = \ -\tfrac{1}{2}\, {E}_{i}{}^{a}\, {E}_{j}{}^{{b}}\, {\cal G}_{ab} \ = \ 
 \tfrac{1}{2}\, {E}_{i}{}^{\bar{a}}\, {E}_{j}{}^{\bar{b}}\, {\cal G}_{\bar{a}\bar{b}}\;,  
\ee
where ${E}_{i}{}^{a}$ is the inverse of ${E}_{a}{}^{i}$ and similarly for the other fields. 

The $L_{\infty}$ products governing the C-bracket algebra 
are given by (\ref{Courantprod}) and still apply in this construction.  
If desired, they could be rewritten in terms of 
projected 
gauge parameters using (\ref{barredunbarred}). 
We must now determine what are the extra products that make up the complete  
dynamical ${L_\infty^{\phantom{0}}}^{\hskip-7pt \rm full}$
of the interacting DFT.

We begin by inspecting 
the perturbative gauge transformations for the CSFT variable $e_{ij}$, given by \cite{Hohm:2010jy}
 \be
 \begin{split}
  \delta e_{ij} \ = \ &\, D_i\bar{\lambda}_j+\bar{D}_{j}\lambda_i 
   +\tfrac{1}{2}(\lambda\cdot D+\bar{\lambda}\cdot \bar{D})e_{ij} \\
   &\, +\tfrac{1}{2}(D_i\lambda^k-D^k\lambda_i) e_{kj} 
   +\tfrac{1}{2}(\bar{D}_j\bar{\lambda}^k-\bar{D}^k\bar{\lambda}_j) e_{ik}
   +\tfrac{1}{4}e_{ ik}(D^l\bar{\lambda}^k - \bar{D}^k\lambda^l)
    e_{ lj}\;. 
 \end{split}   
 \ee
Converting to $O(D,D)$ indices by means of (\ref{eandh}), this implies 
 \be\label{exactHgauge}
 \begin{split}
  \delta_{\xi} h_{\,\nin{M}\bar{N}} \ = \ &\,
  2(\partial_{\,\nin{M}}\xi_{\bar{N}}-\partial_{\bar{N}}\xi_{\,\nin{M}})
   + \xi^{P}\partial_P h_{\,\nin{M}\bar{N}}+
   K_{\,\nin{M}}{}^{\,\nin{K}}
   h_{\,\nin{K}\bar{N}} + K_{\bar{N}}{}^{\bar{K}} h_{\,\nin{M}\bar{K}}\\
   &\, +\tfrac{1}{8}\,h_{\,\nin{M}\bar{K}} \, K^{\,\nin{L}\bar{K}}\, h_{\,\nin{L}\bar{N}}
   \;. 
 \end{split}   
 \ee
This form of the gauge transformation can be taken to be 
exact. More precisely, there is a choice for the higher order terms in 
the expansion (\ref{expandHAAA}) of ${\cal H}$   
so that $\delta_\xi {\cal H} = {\cal L}_\xi {\cal H}$   
yields (\ref{exactHgauge}) exactly. 
Finally, the gauge transformation of the dilaton reads
 \be
  \delta_{\xi}\phi \ = \ \xi^N\partial_N\phi +\partial_N\xi^N\;. 
 \ee 
The products read off from this and (\ref{exactHgauge}) are 
 \be\label{fieldparametERRR}
 \begin{split}
  [\ell_1(\xi)]_{\,\nin{M}\bar{N}} \ = \ &\,
  2(\partial_{\,\nin{M}}\xi_{\bar{N}}-\partial_{\bar{N}}\xi_{\,\nin{M}})\;, \\
   [\ell_2(\xi,h)]_{\,\nin{M}\bar{N}} \ = \ &\,
   \xi^{P}\partial_P h_{\,\nin{M}\bar{N}}+
   K_{\,\nin{M}}{}^{\,\nin{K}}
   h_{\,\nin{K}\bar{N}} + K_{\bar{N}}{}^{\bar{K}} h_{\,\nin{M}\bar{K}}\;,  \\
    [\ell_3(\xi,h_1,h_2)]_{\,\nin{M}\bar{N}} \ = \ &\,
    -\tfrac{1}{2}\,h_{1\,\nin{M}\bar{K}} \, K^{\,\nin{L}\bar{K}}\, h_{2\,\nin{L}\bar{N}} 
    +(1\leftrightarrow 2)\;, \\[0.5ex]
   \ell_1(\xi)_s \ = \ &\,   \partial_N\xi^N\;, \\
   \ell_2(\xi,\phi)_s \ = \ &\, \xi^N  \partial_N \phi \,. 
 \end{split}
 \ee 
All other products involving only gauge parameters $\xi$ and fields 
are zero. 

Let us now turn to the field equations. The full field equations can be written in terms of the 
generalized metric and the dilaton, 
 \be\label{DFTFIELDeq}
  {\cal R}_{MN}({\cal H},\phi) \ = \ 0\;, \qquad
  {\cal R}({\cal H},\phi) \ = \ 0\;. 
 \ee 
Expanding around a constant background and taking `off-diagonal' projections, 
one obtains  
 \be
   0 \ = \ {\cal R}^{(1)}_{\,\nin{M}\bar{N}}(h,\phi)+ {\cal R}^{(2)}_{\,\nin{M}\bar{N}}(h,\phi)+\cdots 
    \;, 
    \qquad
   0 \ = \ {\cal R}^{(1)}(h,\phi)+  {\cal R}^{(2)}(h,\phi) +\cdots
   \;, 
 \ee    
where the superscript denotes the number of fields.  
The linearized equations can be read off from eq.~(6.7) in \cite{Hohm:2014xsa}
(defining $\square=\partial^{\,\nin{M}}\partial_{\,\nin{M}}$) 
 \be
 \begin{split}
  {\cal R}_{\,\nin{M}\bar{N}}^{(1)} \ &= \ 
  \square h_{\,\nin{M}\bar{N}}-\partial^{\,\nin{K}} \partial_{\,\nin{M}} h_{\,\nin{K}\bar{N}} 
  +\partial^{\bar{K}} \partial_{\bar{N}} h_{\,\nin{M}\bar{K}}
  -2\partial_{\,\nin{M}}\partial_{\bar{N}} \phi  \;, \\
  {\cal R}^{(1)} \ &= \ \partial^{\,\nin{M}}\partial^{\bar{N}} h_{\,\nin{M}\bar{N}} -2 \square \phi \;. 
 \end{split} 
 \ee    
We can now read off the $\ell$ products taking values in the space of field equations $X_{-2}$. 
Specifically, to lowest order we find the two projections 
 \be\label{ell1EOM}
 \begin{split}
  [\ell_1(\Psi)_t]_{\,\nin{M}\bar{N}} \ &= \ {\cal R}_{\,\nin{M}\bar{N}}^{(1)}\;, \\
  \ell_1(\Psi)_s \ &= \ {\cal R}^{(1)}\;. 
  \end{split}
 \ee 
The higher products taking values in $X_{-2}$ can be determined algorithmically by expanding (\ref{DFTFIELDeq}) 
to the desired order and using the polarization identities of sec.~\ref{fieldSEC}
 to determine the product for arbitrary different 
arguments in the space of fields $X_{-1}$.  For instance, the correction to second order in fields 
for ${\cal R}$ has been given explicitly in \cite{Hohm:2010jy} in terms of the original 
CSFT variables. Writing 
\be\label{R2isell2}
 {\cal R}^{(2)} \ = \ -\tfrac{1}{2}\ell_2(\Psi,\Psi)_s\;, 
\ee
we read off from eq.~(4.28) in \cite{Hohm:2010jy} for the (diagonal) product 
 \be\label{ell2psipsi}
 \begin{split}
   \ell_2(\Psi,\Psi)_s \ = \ &\, 2\,D^i\phi\,D_i\phi 
   -4e^{ij}D_i\bar{D}_{j}\phi -2D^ie_{ij}\, \bar{D}^j\phi  - 2\bar{D}^je_{ij}\, D^i \phi
   +\tfrac{1}{2}D^pe^{ij}\, D_pe_{ij} \\
  &\, + e^{ij}\big(D_iD^k e_{kj} +\bar{D}_j\bar{D}^ke_{ ik}\big)
  +\tfrac{1}{2}\big(D_le^{li}\, D^k e_{ki} +\bar{D}_le^{il}\, \bar{D}^k e_{ ik}\big)   \;, 
 \end{split} 
 \ee
 or, translating into $O(D,D)$ indices by means of the anchor map, 
\be\label{ell2psipsi}
 \begin{split}
   \ell_2(\Psi_1,\Psi_2)_s \ = \ &\, -2\,\partial^{\,\nin{M}}\phi_1\,\partial_{\,\nin{M}}\phi_2 
   +4\,h_1^{\,\nin{M}\bar{N}}\partial_{\,\nin{M}}\partial_{\bar{N}}\phi_2 
   +2\,\partial^{\,\nin{M}}h_{1\,\nin{M}\bar{N}}\, \partial^{\bar{N}}\phi_2  \\
   &\, +2\, \partial^{\bar{N}}h_{1\,\nin{M}\bar{N}}\, \partial^{\,\nin{M}} \phi_2 
    +\tfrac{1}{2}\,\partial^{\,\nin{K}}h_1^{\,\nin{M}\bar{N}}\, \partial_{\,\nin{K}}h_{2\,\nin{M}\bar{N}} \\
   &\, +h_1^{\,\nin{M}\bar{N}}\big(\partial_{\,\nin{M}}\partial^{\,\nin{K}} h_{2\,\nin{K}\bar{N}} -
  \partial_{\bar{N}}\partial^{\bar{K}}h_{2 \,\nin{M}\bar{K}}\big)\\
  &\, +\tfrac{1}{2}\big(\partial_{\,\nin{L}}h_1^{\,\nin{L}\bar{N}}\, \partial^{\,\nin{K}} h_{2\,\nin{K}\bar{N}} 
  -\partial_{\bar{L}}h_1^{\,\nin{M}\bar{L}}\, \partial^{\bar{K}} h_{2 \,\nin{M}\bar{K}}\big) 
  + (1\leftrightarrow 2)  
  \;, 
 \end{split} 
 \ee
where we restored the two arbitrary field arguments. 
In general, the field equations contain arbitrary powers of the fields and hence are non-polynomial. 
Thus, all higher products for fields are expected to be 
non-vanishing, 
 \be
  \ell_n(\Psi, \ldots, \Psi) \ \neq \ 0 \quad \text{for}  \  n\geq 1 \,.
 \ee

Finally, we employ the gauge covariance of the field equations in order to determine 
the products involving arguments in the space of field equations $X_{-2}$. The full field equations 
(\ref{DFTFIELDeq}) transform covariantly according to generalized Lie derivatives: 
 \be\label{EOMCOV}
 \begin{split}
  \delta_{\xi}{\cal R}_{MN} \ & = \ {\cal L}_\xi {\cal R}_{MN} \  = \ \xi^K\partial_K {\cal R}_{MN}  +K_M{}^K {\cal R}_{KN} + K_N{}^K {\cal R}_{MK} \;, \\   
  \delta_{\xi}{\cal R} \ &= \   {\cal L}_\xi {\cal R}\ \ = \ \  \xi^K\partial_K {\cal R}\;. 
 \end{split} 
 \ee 
 Note that the right-hand sides are linear in the field equations and do not contain bare fields.
 On the other hand, as we showed in (\ref{clwsmbtt}) one has a general formula for the gauge variation of
 the field equations
 \be
\label{clwsmbtty}
\delta_{\xi}{\cal F} \ = \   \ell_2(\xi,{\cal F})\,  +\,  \ell_3(\xi,{\cal F}(\Psi),\Psi)
  +\dots
\ee
Comparing the two equations above we see that 
only the first term on the right-hand side of the second equation is present.  We learn therefore that
\be
\label{jnwsmtts}
   \ell_2(\xi,E) \ = \ {\cal L}_{\xi}E \ \in \ X_{-2}\;, 
 \ee 
and higher products vanish
 \be\label{onemoreZERO}
  \ell_{n+2}(\xi, E, \Psi^n ) \ = \ 0 \;, \quad n\geq 1\;. 
 \ee

 As a consistency check,  
 we can show that (\ref{jnwsmtts}) is required by the $L_\infty$
 relation $\ell_1\ell_2=\ell_2\ell_1$.
We focus on the dilaton and thus  expand the  second equation in (\ref{EOMCOV})  
to first order in fields, 
 \be\label{Rcovariance}
  \xi^K\partial_K
  {\cal R}^{(1)} \ = \ 
  \delta^{(1)}{\cal R}^{(1)} + \delta^{(0)}{\cal R}^{(2)} \;. 
 \ee
Here $\delta^{(n)}$ refers to the terms in the gauge variation with $n$ powers of the fields, i.e., 
the terms encoded in the product $\ell_{n+1}(\xi, \Psi^n)$.   
For the dilaton component $\ell_1\ell_2=\ell_2\ell_1$, 
gives
 \be\label{ellIDENTTY}
  \ell_2(\xi, \ell_1(\Psi))_s \ = \ \ell_1(\ell_2(\xi, \Psi))_s - \ell_2(\ell_1(\xi), \Psi)_s\;.  
 \ee
To this end, we rewrite the two terms on the right-hand side of (\ref{Rcovariance}) as follows.  
First, using $\delta^{(1)}\Psi=\ell_2(\xi,\Psi)$, we compute  with (\ref{ell1EOM}) 
 \be
  \delta^{(1)}{\cal R}^{(1)} \ = \ \delta^{(1)}\big(\ell_1(\Psi)_s\big) 
  \ = \ \ell_1\big(\delta^{(1)}\Psi \big)_s
  \ = \ \ell_1(\ell_2(\xi,\Psi))_s\;. 
 \ee 
Second, using $\delta^{(0)}\Psi=\ell_1(\xi)$, we compute with (\ref{R2isell2}) 
 \be
  \delta^{(0)}{\cal R}^{(2)} \ = \ -\tfrac{1}{2}\delta^{(0)}\big(\ell_2(\Psi,\Psi)\big)_s 
  \ = \ -\ell_2(\delta^{(0)}\Psi,\Psi)_s \  = \ -\ell_2(\ell_1(\xi),\Psi)_s\;, 
 \ee
where we used the symmetry of $\ell_2$ for two arguments in the space of fields.   
We have thus shown that the right-hand side of (\ref{ellIDENTTY}) equals the 
right-hand side of (\ref{Rcovariance}). Therefore, (\ref{ellIDENTTY}) is satisfied 
if the left-hand sides are also equal, i.e., 
 \be
    \ell_2(\xi, \ell_1(\Psi))_s \ = \  \xi^N\partial_N {\cal R}^{(1)}\;. 
 \ee
This relation is satisfied provided we define the product for a general $E\in X_{-2}$ to be 
   \be\label{scaltrans}
   \ell_2(\xi,E)_s \ = \ \xi^N\partial_N E_s\;. 
  \ee  
This is what we wanted to show.  The above derivation goes through for the tensor 
component in the same way.

So far we have determined the non-trivial $\ell_2$ product between gauge parameters 
and field equations, and it is easy to see that there are no higher products involving one 
gauge parameter and several field equations. For instance, an $\ell_3$ product like 
 \be
  \ell_3(\xi, E, E) \ \in  \ X_{-3}\;, 
 \ee
has to vanish because there is no space $X_{-3}$, and similarly for higher products. 
Moreover, there is no product $\ell_3(\xi_1, \xi_2,E)$ of two gauge parameters and a field equation, 
because this would imply that closure of the gauge algebra holds only on-shell, 
while in DFT 
we have off-shell closure. 
Similarly, there is no need for higher products with two gauge parameters and an arbitrary number of field equations, so those are also zero.

 We now claim that the products we have identified so far are all the products
 that are non-zero.  In summary, the non-vanishing products for the $L_{\infty}$ algebra describing  
 the full (perturbative) DFT are the following: 
 \begin{itemize} 
 \item[\textit{i)}] the products governing the 
 pure gauge structure, i.e., gauge parameters $\xi$, `trivial' functions $\chi$ and constants $c$, which are 
 non-vanishing for 
  \be
  \label{jnfntstbtt}
   \ell_1(\chi)\;, \quad \ell_1(c)\;, \quad \ell_2(\xi_1,\xi_2)\;, \quad \ell_2(\xi,\chi)\;, \quad 
   \ell_3(\xi_1,\xi_2,\xi_3)\;, 
  \ee 
  and given explicitly in (\ref{Courantprod});  
 \item[\textit{ii)}] the products involving gauge parameters and fields describing the 
 full gauge transformations of fields, which are non-vanishing for 
  \be
  \label{jnfntstbtt1}
   \ell_1(\xi)\;, \quad \ell_2(\xi, \Psi)\;, \quad \ell_3(\xi,\Psi_1,\Psi_2)\;, 
  \ee
 and given explicitly in (\ref{fieldparametERRR}); 
 \item[\textit{iii)}] products $\ell_n$ for arbitrary $n$ involving only fields $\Psi$, 
  \be
  \label{jnfntstbtt2}
  \ell_n(\Psi_1, \ldots, \Psi_n)  \qquad \text{for}\qquad \Psi_1, \ldots, \Psi_n \ \in \ X_{-1}\;, 
 \ee
 such as given to 
 lowest order in (\ref{ell1EOM}), (\ref{ell2psipsi}); we did not attempt to write these products in a closed form, but we 
 explained how they can be determined systematically from the field equations to any desired order $n$; 
 \item[\textit{iv)}] the product  
 between gauge parameter and field equation, 
  \be
  \label{jnfntstbtt3}
   \ell_2(\xi,E) \ = \ {\cal L}_{\xi}E \;. 
  \ee 
 \end{itemize} 
 
 \medskip
 \noindent
 {\bf Proof.} We now explain why all $L_\infty$ identities are satisfied.  
 For this we will consider the possible lists of inputs for the  identities.  
 The inputs, of course, can be various numbers of $c$'s, $\chi$'s,
 $\xi$'s, $\Psi$'s, and $E$'s.   
 Note that given a list of $n$ inputs there is a single $L_\infty$ identity that
 must be cheched, the identity with 
 sums of products of the form $\ell_i \ell_j$, with $i+j = n+1$.
For any term $\ell_i \ell_j$ we will call $\ell_j$ the {\em first} product and $\ell_i$
the {\em second} product,  as they act first and second, respectively, on any
list of inputs. 
 
  As discussed in section 3.3 we do not need to consider any identity on only
 pure $\Psi$ inputs as they hold if all products involving one $E$ and any number of $\Psi$'s vanish,
 as they do here.    Moreover it is explained there as well that we need not consider
 identities acting on input sets $(\xi, \Psi^n)$ with one gauge parameter and multiple $\Psi$'s as those identities hold when the field equations transform covariantly, as we have checked before.   Finally, we need not consider
 identities acting on input sets $(\xi_1\xi_2\Psi^n)$ with two gauge parameters and multiple $\Psi$'s as those identities hold when the gauge algebra
 takes a standard form.

\noindent
{\em Claim 1: Any list of inputs that includes one or more $E$'s leads to 
correctly satisfied identities.}
Suppose there is one $E$.  If the first product involves the $E$ then it must be
an $\ell_2$ that couples it to a $\xi$ and gives another $E$.  The second product
must then couple the new $E$ to another $\xi$. 
The inputs in this case are $\xi_1 \xi_2 E$ and 
are  relevant to the $\ell_1 \ell_3 + \ell_2 \ell_2 + \ell_3 \ell_1=0$ identity.  
The identity holds: terms with
$\ell_3$ vanish and $\ell_2 \ell_2=0$ because Lie derivatives form a Lie algebra. 
If the first product does not involve $E$ then the second product must, and 
has to be of the form $\ell_2 (E, \xi)$.  This means that the first product must have
taken some inputs and given a gauge parameter.  The options for this are 
$\ell_1(\chi)$ and $\ell_2 (\xi_1, \xi_2)$.  Thus the possible lists to check
are $E \chi$ and $E\xi_1\xi_2$.  The second list was dealt with a few lines 
above.   The first list is relevant to the identity $\ell_1 \ell_2 = \ell_2 \ell_1$
and holds because $\ell_2 (E, \ell_1(\chi))$
vanishes as it is a generalized Lie derivative along a trivial parameter.  
Now consider the case when there are two $E$'s in the original list. 
Since there is no product that includes two $E$'s
the first product must involve an $E$.  But the product gives another $E$,
and therefore the second product is faced with two $E$'s and it must vanish.
The case of more than two $E$'s works for 
analogous reasons.   \hfill $\square$

\medskip
\noindent
{\em Claim 2: Any list of inputs that includes one or more $c$'s leads to 
correctly satisfied identities.}
If there are two or more $c$'s any sequence of products will give zero because
the only product involving $c$ is $\ell_1$.  
If there is one $c$ and no other inputs, this is trivially satisfied.  If there
is one $c$ and some other inputs, the first product cannot act on the other
inputs because then there would be no suitable second product.
The only possibility is that the first product is $\ell_1$ 
and acts on $c$ to give a constant $\chi$.  That $\chi$ can be acted by another $\ell_1$, in which case the identity is trivial, or appear in $\ell_2 (\chi,  \xi)$,
which vanishes because $\chi$ is constant.  \hfill $\square$

\medskip
\noindent
{\em Claim 3: Any list of inputs that includes one or more $\chi$'s leads to 
correctly satisfied identities.}
Consider first the case when we have two $\chi$'s.
Assume the first product is not acting on any of the $\chi$'s.  Then there
is no available second product that acts on a list with at least two $\chi$'s.
If the first product acts on one of the $\chi$'s it could be 
in the form $\ell_1(\chi)$ or $\ell_2 (\chi, \xi)$, because of (\ref{jnfntstbtt}).  Since the latter gives
another $\chi$, it must be the former.  The second product is then
faced with at least a $\chi$ and a $\xi$.  But there can be no more inputs, 
so that we can use $\ell_2$.  In summary, the only possibility
for the original list is $\chi\chi$.  This is certainly a trivially satisfied identity. 
The same argument holds for more than two $\chi$'s.

Consider now the case of a single $\chi$ in the list of inputs.
If the first product does not act on $\chi$ then it must act on all
of the other inputs to produce a $\xi$.  Looking again at the list
of products (\ref{jnfntstbtt}), the only option is $\ell_2 (\xi_1, \xi_2)$
showing that the original list must have been $(\chi \xi\xi)$.  The
associated identity was already checked in the Courant algebroid.
If the first product acts on the $\chi$ it may act with $\ell_2$ or $\ell_1$.
If  it acts with $\ell_2$, it must be
$\ell_2(\xi, \chi)$ giving another $\chi$-type input. The second product can
be $\ell_1$ if the list is $(\chi\xi)$ or another $\ell_2$ in which case the
list is $(\chi\xi\xi)$.  Both lists were checked in the Courant algebroid.
If the first product acts on the $\chi$ with $\ell_1$
 it turns it into a $\xi$ and for nontrivial products one must then have at most
 two $\Psi$'s.  So the only possibilities are inputs $(\chi \Psi)$ or $(\chi \Psi\Psi)$.   In both cases these vanish because the $\xi$ obtained as 
 $\ell_1(\chi)$ is a trivial gauge parameter.    
 \hfill $\square$
 
 \medskip
Because of our earlier observations and the three claims above
 we need only check identities with inputs that 
 have three or more $\xi$'s and any number of $\Psi$'s.    
 The case when all inputs are $\xi$ need not be 
 considered because this was part of the analysis in the Courant algebroid.

Consider the lists $(\xi\xi\xi \Psi^n)$   
  with three $\xi$'s and a number $n\geq 1$ of 
 $\Psi$'s.  
 With three $\xi$'s we cannot have both the first and the second product
 arise from the list $\ell_1 (\xi),  \ell_2 (\xi, \Psi),
\ell_3 (\xi,\Psi,\Psi)$ that use one $\xi$.  One of them must be a $\ell_3 (\xi,\xi,\xi)$.
Suppose the first product is $\ell_3 (\xi,\xi,\xi)$.  Then the second product
is acting on $(\chi\Psi^n)$ and vanishes, as $\chi$ never couples to a field.  
Suppose the first product is
not $\ell_3 (\xi,\xi,\xi)$, then 
the chosen product uses one $\xi$ 
and returns a $\Psi$,
leaving for the second product inputs $(\xi\xi\Psi^k)$ with $k\geq 1$. But 
no product is non-zero for this list.  

The lists $(\xi\xi\xi\xi \Psi^n)$
with four $\xi$'s and a number of $n\geq 1$ of $\Psi$'s also leads
to no constraints.  If the first product is one of $\ell_1 (\xi),  \ell_2 (\xi, \Psi),
\ell_3 (\xi,\Psi,\Psi)$, then the second product is facing the list $(\xi\xi\xi\Psi^k)$
with $k\geq 1$ gives zero.  If the first product is $\ell_3 (\xi,\xi,\xi)$
the second product faces the list $(\chi\xi\Psi^n)$ and gives zero.  It is clear 
that more than four $\xi$'s and a number $n\geq 1$ of $\Psi$'s will also
give trivially satisfied identities.   This concludes our proof that the products
listed in (\ref{jnfntstbtt} --\ref{jnfntstbtt3}) define a consistent $L_\infty$ algebra for DFT.
\hfill $\square$

\subsection{Comments on Einstein gravity}  
We close this section be briefly commenting on the description of Einstein gravity 
as $L_{\infty}$ algebra. Einstein gravity is contained in DFT, so this is a special case 
of our results above, but it is instructive  to see how the   $L_{\infty}$ algebra simplifies 
for pure gravity. 
As before, we consider perturbative gravity, in which the Einstein-Hilbert theory is expanded 
around flat space, writing $g_{mn}=\eta_{mn}+h_{mn}$. The diffeomorphism symmetry acts on 
the massless spin-2 fluctuation $h_{mn}$ as 
 \be
  \delta_{\xi}h_{mn} \ = \ \partial_m\xi_n+\partial_n\xi_m + L_{\xi}h_{mn}\;, 
 \ee
where $L_{\xi}$ is the conventional Lie derivative, defined like in the first line of (\ref{LLIIEE}). 
These transformations close according to the Lie bracket $[\,,]$ of vectors fields, which in turn 
satisfies the Jacobi identity. Thus, there is no need for a space $X_1$ of `trivial' functions, 
and it is sufficient to consider the graded vector space   
  \be
  \begin{split}
 & X_{0} \;
\longrightarrow\;\;  X_{-1}\; \longrightarrow\;\;  X_{-2}\\
   &\;  \xi^m  \qquad \;\; h_{mn}
    \qquad\quad  R_{mn}
  \end{split}
 \ee
The non-zero products are the following: 
 \begin{itemize} 
 \item[\textit{i)}] the product governing the 
 pure diffeomorphism Lie algebra for  $\xi$, 
  \be
    \ell_2(\xi_1,\xi_2) \ = \ [\xi_1,\xi_2]
  \ee 
 \item[\textit{ii)}] the products involving gauge parameters and fields describing the 
 gauge transformations, 
  \be
   \ell_1(\xi)_{mn} \ = \ \partial_m\xi_n+\partial_n\xi_m \;, \qquad 
   \ell_2(\xi, h) \ = \ L_{\xi}h 
  \ee
 \item[\textit{iii)}] products $\ell_n$ for arbitrary $n$ involving only the field $h$, 
  \be
  \ell_n(h, \ldots, h)  \qquad \text{for}\qquad h  \ \in \ X_{-1}\;, 
 \ee
 as can be determined from the Einstein equations to any desired order; 
 \item[\textit{iv)}] the product  
 between gauge parameter and field equation, 
  \be
   \ell_2(\xi,E) \ = \ {L}_{\xi}E \;. 
  \ee 
 \end{itemize}

\section{$A_{\infty}$ algebras and revisiting Chern-Simons}

In this final section we briefly contrast the $L_{\infty}$ constructions of this paper with 
the $A_{\infty}$ formulation of Chern-Simons theory. 
The $A_\infty$ axioms relevant to the construction of a theory include a set of products
$m_n$, with $n= 1,2,3, \ldots$.  The product $m_n$, with $n$ inputs, is of degree $n-2$.
The first couple of  identities are~\cite{Gaberdiel:1997ia}:   
\be
\label{first-two-A-infty}
\begin{split}
m_1 ( m_1 (x) ) \ = \ &\, 0 \,, \\
m_1 (m_2 (x_1, x_2) \ = \ &\,  m_2 (m_1 (x_1) , x_2)  + (-1)^{x_1} m_2 ( x_1, m_1 (x_2))\,.  \\
\end{split}
\ee
For the example we want to discuss, and for Witten's open string field theory, 
the product $m_{3}$ and all higher ones vanish.   In this case, the remaining identity
in the algebra is the associativity condition for $m_2$:
\be
\label{A-infty-associativity}
m_2 ( \, m_2 (x_1, x_2) , x_3 ) \ = \ m_2 ( \, x_1 , \, m_2 (x_2, x_3)\,  )  \,. 
\ee 
If we supply an inner product one can also write an action.  The inner product must
satisfy
\be
\begin{split}
\langle  x_1 , x_2 \rangle \ = \ &\, (-1)^{x_1x_2} \langle x_2, x_1 \rangle\,, \\
\langle m_1 (x_1), x_2 \rangle \ = \ &\, - (-1)^{x_1} \, \langle x_1, \, m_1 (x_2) \, \rangle \,, \\
\langle x_1 , m_2 (x_2, x_3) \rangle \ = \ &\,  \langle m_2 (x_1, x_2) , \, x_3  \rangle \,. 
\end{split}
\ee

In writing a field theory with a field $A \in X_{-1}$ 
we have   an action
\be
S \ = \ \tfrac{1}{2}  \langle  A \,, m_1 (A) \rangle  +  \, \tfrac{1}{3}  \langle A , \, m_2 ( A, A) \, \rangle\,.
\ee
The field equation takes the form ${\cal F}=0$  with
\be
{\cal F}(A) \ \equiv \ m_1 (A)  +  m_2 (A, A) \,.
\ee 
With a gauge parameter $\lambda \in X_0$ the gauge transformations leaving the action
invariant take the form
\be
\label{ainfty-gt} 
\delta_\lambda  A \ = \ m_1 (\lambda)  \, +\,  m_2 (A, \lambda) - m_2 (\lambda , A)\,.
\ee
The gauge algebra takes the form
\be
\label{a-infty-gauge-algebra} 
\bigl[ \, \delta_{\lambda_1} \, , \,  \delta_{\lambda_2} \, \bigr] \ = \ \delta_{ m_2 (\lambda_1, \lambda_2) 
- m_2 (\lambda_2 , \lambda_1) } \,. 
\ee 
 The field equation is gauge covariant:  we have 
 \be
 \delta_\lambda {\cal F} \ = \ m_2\, ({\cal F} , \lambda) \, - \, m_2 (\lambda, {\cal F} ) \,. 
 \ee
 
 \medskip
\noindent  
In order to formulate the Chern-Simons theory we 
consider the graded vector space 
  \be
  \begin{split}
   &X_0\qquad   X_{-1}\qquad\;\; X_{-2} \\
   &\, \lambda \qquad \,\,   A_{\mu}\ \ \  \qquad E_{\mu\nu}
  \end{split}
 \ee
Here we think of these objects as {\em matrix valued} fields:
\be
\lambda \, \equiv \, \lambda^\alpha\, t_\alpha \,, 
\qquad A_\mu \, \equiv \, A_\mu^\alpha \, t_\alpha \,, 
\qquad E_{\mu\nu} \, \equiv \, E_{\mu\nu}^\alpha \, t_\alpha\,,  \ee
where the $t_\alpha$ can be chosen as the adjoint representation
 of the generators $T_\alpha$ of the 
Lie algebra.   
We have the commutator $[t_\alpha, t_\beta] = f_{\alpha\beta}{}^\gamma t_\gamma$ and the relation $\kappa_{\alpha\beta}  =  - \hbox{tr} ( t_\alpha t_\beta)$.

The inner product is given by 
\be
\langle A , E \rangle \ \equiv \  \int d^3 x  \, \varepsilon^{\mu\nu\rho}  \, \hbox{tr} ( A_\mu  E_{\nu\rho} ) \ = \ \int d^3 x  \, \varepsilon^{\mu\nu\rho}  \kappa_{\alpha\beta} 
\,  A_\mu^\alpha\, 
   E_{\nu\rho}^\beta\,. 
\ee
 
We list the complete set of {\em nonvanishing}  $A_\infty$ products: 
  \be\label{Lnproducts}
\boxed{  
\begin{split}
A_\infty \, \hbox{Chern-Simons:} \qquad   
m_1(\lambda)_\mu \ &= \ \partial_\mu\lambda\ \in \ X_{-1} \\
     m_1(A)_{\mu\nu} \ &= \ \partial_\mu A_\nu - \partial_\nu A_\mu  \ \in \ X_{-2} \ \ \ \\
    m_2(\lambda_1,\lambda_2) \ &= \ \lambda_1 \lambda_2\ \in  \ X_0\\
    m_2(A,\lambda)_\mu \ &= \ A_\mu\,\lambda \ \in  \ X_{-1} \\
    m_2(\lambda, A )_\mu \ &= \ \lambda\, A_\mu  \ \in  \ X_{-1} \\
   m_2(A_1,A_2)_{\mu\nu}  \ &= \ A_{1\mu}\, A_{2 \nu}- A_{1\nu}\, A_{2 \mu} \ \in \ X_{-2} \ \ \\
   m_2(E,\lambda) \ &= \ E\, \lambda\ \in \ X_{-2}\, , \\
   m_2(\lambda, E) \ &= \ \lambda\, E \ \in \ X_{-2}\, .
  \end{split}
  }
 \ee 
 Note that the products $m_2$ have no exchange property: they are neither symmetric nor
 antisymmetric;   
 they are intrinsically non-commutative products, 
 which are associative, however.

 Let us briefly go over the derivation of such products and the check that they satisfy the 
 relevant identities.   Comparing the gauge transformation 
$\delta_\lambda  A =  m_1 (\lambda)  +  m_2 (A, \lambda) - m_2 (\lambda , A)$
with $\delta_\lambda A_\mu  = \partial_\mu \lambda  + A_\mu \lambda - \lambda A_\mu$ 
(which follows from (\ref{a-gauge-transformation})) we read off expressions
for $m_1 (\lambda) $, $m_2 (A, \lambda)$ and $m_2 (\lambda , A)$.    Next, we compare the 
field equation $m_1 (A ) + m_2 (A, A)= 0$ to the explicit field equation $F=0$, which using
(\ref{field-strength-gauge-field}) reads $ \partial_{\mu}A_{\nu}-\partial_{\nu}A_{\mu}
  +A_{\mu}A_{\nu} - A_{\nu}A_{\mu}=0$.  This allows us to read off $m_1(A)$ and $m_2 (A, A)$.   
  Comparing the gauge algebra (\ref{a-infty-gauge-algebra})  to the gauge algebra
 $[ \delta_{\lambda_1}  ,  \delta_{\lambda_2}]   = \delta_{\lambda_1 \lambda_2 - \lambda_2 \lambda_1}$
 we obtain the value of $m_2 (\lambda_1, \lambda_2)$.    The last two entries in the above
 table are obtained from the identities themselves. 
 
 The list (\ref{table-gauge-products})  of gauge theory inputs to products 
 applies here.  The identity $m_1 m_1 =0$ need only
 be checked acting on $X_0$ and holds trivially.  The identity $m_1 m_2 = -m_2 m_1$ 
 must be checked on $(A, \lambda)$, $(\lambda, A)$ and $(\lambda_1, \lambda_2)$.
 The first two determine $m_2 (E, \lambda)$ and $m_2 (\lambda, E)$, respectively.
 The last holds trivially.  The associativity condition (\ref{A-infty-associativity}) 
 must be checked on $\lambda\lambda \lambda$,  $\lambda\lambda A$, $\lambda A A$
 and $\lambda \lambda E$.  All those are readily verified.  
 
The Chern-Simons action is also reproduced correctly:
\be
\begin{split}
 \langle  A \,, \tfrac{1}{2}  m_1 (A) + \tfrac{1}{3}  m_2 ( A, A) \, \rangle  =  &
\int d^3 x \, \varepsilon^{\mu\nu\rho}   \, \hbox{tr} \bigl( \, A_\mu, \, \tfrac{1}{2} (\partial_\nu A_\rho
-\partial_\rho A_\nu) + \tfrac{1}{3} (A_\nu A_\rho - A_\rho A_\nu ) \bigr) \\
= &   \int d^3 x \, \varepsilon^{\mu\nu\rho}   \, \hbox{tr} \bigl( \, A_\mu, \, \partial_\nu A_\rho
+ \tfrac{1}{3} (A_\mu A_\nu - A_\nu A_\mu ) \bigr)\,,
\end{split} 
\ee
 which agrees with (\ref{CS-gauge-invariant}).   As discussed before neither the inner product
 nor the products refer to 
 a spacetime metric.  In this sense the $A_\infty$ formulation
 seems more natural than the $L_\infty$ formulation for Chern-Simons theory.

 \section{Conclusions and outlook}

Homotopy Lie algebras or $L_{\infty}$ algebras are generalizations
of Lie algebras  
that describe the underlying algebraic structure of 
classical closed string field theory. 
It might appear that $L_{\infty}$ algebras are somewhat exotic, because the gauge symmetries
of ordinary field theories, properly extended to include equations-of-motion symmetries,
form a Lie algebra. 
We argued that, on the contrary,   
$L_{\infty}$ algebras are the underlying algebraic structure for any consistent classical field theory. 
We illustrated this with examples and outlined a general algorithm to determine the  $L_{\infty}$ structures 
for a given field theory. 
It must be emphasized that 
this is not in conflict with the fact that in conventional classical field theories 
field products are naturally associative and 
symmetry variations always satisfy a Jacobi identity.

One possible application is to formulate the `Wilsonian effective actions' recently 
described by Sen \cite{Sen:2016qap}. 
In principle, these can be obtained by integrating out all modes
except for some specific sub-sectors that, along with massless fields, can also
include arbitrarily massive fields.   A particularly interesting   
case is that of double field theory as envisioned in~\cite{Hull:2009mi}, where 
one would include the Kaluza-Klein and winding modes associated with
the massless fields of string theory in toroidal backgrounds.   
While in this paper we have made no 
attempt to construct such theory,   
 the results here should be the proper 
starting point for any such endeavor.

Other possible applications are in M-theory, for which  
we have exceptional field theory \cite{Hohm:2013pua},
a formulation analogous to double field theory that makes the 
U-duality groups E$_{d(d)}$, $d=2,\ldots, 8$, manifest. 
Unlike double field theory, these theories require a `split-formulation'  
in which the coordinates of $D=11$ supergravity are decomposed into external and internal coordinates in 
analogy to Kaluza-Klein.  The internal coordinates are then enlarged to 
transform in the fundamental representation of E$_{d(d)}$. 
The theory features 
$p$-forms of various ranks with respect to the \textit{external} space, 
transforming as generalized tensors 
under the \textit{internal} symmetries. The gauge structure of these $p$-forms is governed 
by so-called tensor hierarchies, which were originally introduced in 
gauged supergravity \cite{deWit:2005hv} and 
have various features in common with $L_{\infty}$ algebras. 
Notably, the gauge algebra structure does not satisfy the Jacobi identity; 
rather, the failure of the Jacobi identity 
is `absorbed' by higher-form gauge symmetries, with a `generalized Cartan structure' emerging 
naturally \cite{Hohm:2015xna,Wang:2015hca}. 
It thus appears likely that there is an 
$L_{\infty}$ description of the tensor hierarchy, which in turn could shed light on a more fundamental 
formulation of exceptional field theory. Indeed, so far exceptional field theory has only been constructed on a 
case-by-case basis, for each duality group E$_{d(d)}$ separately. One might hope that eventually there will 
be a formulation based on a larger algebraic structure, realizing the U-duality groups as sub-structures. 
This algebraic structure might well be an $L_{\infty}$ algebra.

Finally, $L_{\infty}$ algebras are important for higher-spin gravity. 
Indeed, the early investigation of the consistency of non-linear higher-spin symmetries in \cite{Berends:1984rq}
naturally led to a structure that can be interpreted as a homotopy Lie algebra. 
It would therefore be interesting to reformulate or extend higher-spin theories such as constructed 
by Vasiliev (see \cite{Vasiliev:2014vwa} for a recent review) in terms of $L_{\infty}$ algebras.  
Aspects of this relation have already been discussed in \cite{Vasiliev:2005zu}.  
Specifically, in the formulation of higher-spin theories in \cite{Vasiliev:1992av}
the gauge symmetries are governed by a Lie algebra (albeit infinite-dimensional), 
but this is achieved thanks to additional unphysical coordinates. Upon `integrating them out' one should 
recover an $L_{\infty}$ algebra. It would be interesting to see if other theories whose gauge symmetries 
need $L_{\infty}$ structures can be reformulated with pure Lie algebras by using additional coordinates. 
Further illuminating the  $L_\infty$ description of higher-spin symmetries 
may also shed a new light on the open problem of constructing an action for higher-spin gravity, 
which would be important for holographic applications.    
Perhaps the $L_\infty$ algebra can be naturally constructed by adding sets of free fields, in the
way that the difficulties in constructing actions
for superstring field theories were overcome in~\cite{Sen:2015uaa}.

\section*{Acknowledgments}

We would like to thank Martin Rocek,  Ashoke Sen, and Anton Zeitlin   
for useful discussions,
and Andreas Deser, Christian Saemann, Jim Stasheff, 
and Misha Vasiliev for comments on the first version of this paper. 

O.H.~is supported by a DFG Heisenberg Fellowship 
of the German Science Foundation (DFG).  The work of 
B.Z.~is supported by the U.S. Department of Energy under grant Contract Number de-sc0012567.

\end{document}